\documentclass{article}

\usepackage{mathpazo}
\usepackage[english]{babel}
\usepackage{pifont}
\usepackage{paralist}
\usepackage[numbers]{natbib}
\usepackage{amsmath}
\usepackage{bm}
\usepackage{empheq}
\usepackage{comment}

\addto\captionsenglish{
  \renewcommand{\contentsname}%
    {Table of contents}%
}

\usepackage[usenames,dvipsnames,table]{xcolor}
\usepackage[pdftex,plainpages=false,colorlinks=true,
linkcolor=Red, citecolor=NavyBlue, urlcolor=NavyBlue, filecolor=Magenta]{hyperref}

\usepackage[margin=1.3in, headheight=14pt, paper=letterpaper]{geometry}

\usepackage[nottoc,notlot,notlof]{tocbibind}

\usepackage{titlesec}
\newcommand\basecolor{NavyBlue} 
\newcommand\basecolorsec{ForestGreen} 

\usepackage[most]{tcolorbox}
\newcommand{\cbeq}[1]{\tcboxmath[enhanced, colback=\basecolor!15, colframe=\basecolor, sharp corners]{#1}}

\newenvironment{sidenote}[1]
	{\begin{tcolorbox}[breakable, enhanced jigsaw, #1]}
	{\end{tcolorbox}}

\newenvironment{mainpoint}[1]
	{\begin{tcolorbox}[breakable, enhanced jigsaw, colback=\basecolorsec!15, colframe=\basecolorsec, #1]}
	{\end{tcolorbox}}
	
\usepackage[customcolors, norndcorners]{hf-tikz}
\hfsetfillcolor{\basecolor!15}
\hfsetbordercolor{\basecolor}

\renewcommand{\thesection}{\arabic{section}}
\renewcommand{\thesubsection}{\thesection.\arabic{subsection}}
\newcommand\sectionfont{\Large\sffamily\bfseries}
\newcommand\sectionheadfont{\Large\sffamily\bfseries}
\newcommand\subsectionfont{\large\sffamily\bfseries}

\titleformat{name=\section}[hang]%
	{\vglue 0cm plus 1.5cm\addpenalty{-2000}{\color{\basecolorsec}\titlerule[2pt]\medskip}\nobreak\sectionfont\color{\basecolorsec}\filright}
	{\fcolorbox{\basecolorsec}{\basecolorsec!20!white}{\sectionheadfont\thesection}}
	{1em}
	{}
\titlespacing*{\section}{0mm}{10mm}{5mm}

\titleformat{name=\subsection}[hang]%
	{\vglue 0cm plus 4mm\addpenalty{-500}{\color{\basecolor}\titlerule[1pt]\medskip}\nobreak\subsectionfont\color{\basecolor}\filright}
	{\thesubsection}
	{1em}
	{}
\titlespacing*{\subsection}{0mm}{6mm}{2mm}

\renewcommand{\exp}{\mathrm{e}}

\begin{document}

\thispagestyle{empty}

\vspace*{\fill}

{\sffamily\color{\basecolor}
\hrule height 1pt\medskip
{\LARGE\bfseries \noindent Advanced encoding methods in diffusion MRI}\\
\medskip
\begin{flushright}
{\Large\bfseries \noindent Alexis Reymbaut \& Maxime Descoteaux}\\
{\Large\bfseries \noindent Sherbrooke Connectivity Imaging Lab}
\end{flushright}
}

\vspace*{\fill}

\newpage

\pagenumbering{roman}
\setcounter{page}{1}

{\hypersetup{linkcolor=black}
\tableofcontents
}

\newpage

\vglue 0cm plus 1.5cm\addpenalty{-2000}{\color{NavyBlue}\titlerule[2pt]\medskip} {\LARGE \sffamily \textcolor{NavyBlue}{\textbf{Preamble}}}\\
\addcontentsline{toc}{section}{Preamble}

Discovering a new field is not usually an easy process, especially when you choose magnetic resonance imaging (MRI). I (A. Reymbaut) was trained as a theoretical physicist, did not know much about imaging or diffusion MRI, but was eager to try it on. Indeed, MRI is one of the few non-invasive medical procedures, if not the only one, that someone can receive in modern day hospitals. Its high sensitivity is a true prowess that is due to three things: subtle quantum physics, great engineering and clever imaging techniques from mathematics and computing science. However, the mathematics and physics of it are already daunting tasks by themselves, which makes MRI difficult to comprehend. To put things into perspectives, let us just quote Pr. Martin Lepage, from the research center of the Université de Sherbrooke hospital (CR-CHUS): 
\begin{center}
\begin{minipage}[c]{0.75\linewidth}
``Everybody knows how X-ray imaging works. One shines this light onto a given body, then lets it be stopped by dense tissue and mark a sensitive sheet so that to create a footprint of the bone structure. To reach the same extent of explanation with MRI, it will take me six hours."
\end{minipage}
\end{center}
I initially wrote this document to pave a rather short, yet solid and logical, path from the subtle concept of spin to the current limitations of diffusion MRI. Then I went on to write about more advanced encoding methods proper to diffusion MRI, such as tensor-valued encoding, which interested my advisor at the time (Pr. M. Descoteaux) as it is designed to overcome diffusion MRI's limitations.\\

Needless to say that the reader should possess a reasonable mathematical background to begin with. However, the introductory content (Secs.~\ref{Sec_NMR} to \ref{Section_sym_tensors}) will never be as thorough as a combination of certain reviews in the field. Also, for the sake of their mental health, readers are advised not to try to understand the fundamental rules of quantum mechanics that might be encountered in Sec.~\ref{Sec_NMR}. The reason behind this advice is that there is absolutely nothing to understand there. These rules are completely counter-intuitive but have been experimentally verified countless times. The only thing brought by a solid education in physics in that matter is the acceptance of these experiments as forming a cornerstone on which a new intuition can be built. \\

Finally, we would like to thank the authors of the references used to write this document.


\newpage

\clearpage
\pagenumbering{arabic}
\setcounter{page}{1}

\section{Nuclear magnetic resonance (NMR)}
\label{Sec_NMR}

\subsection{A bit of quantum mechanics}

\subsubsection{Spin}

Nowadays, pretty much anyone knows that particles such as electrons, protons and neutrons bear a mass and an electrical charge, even without knowing their approximate values. These quantities are actually defining, intrinsic, properties of these particles. For instance, a proton always have a mass equal to $m_\text{p} \simeq 1,673\times 10^{-27} \; \mathrm{kg}$ and an electrical charge equal to $q_\text{p} = +e \simeq 1,602\times 10^{-19}\;\mathrm{C}$. In fact, all particles are defined by a much larger array of quantities than just their mass and electrical charge. Among this array is the spin, a quantity that grants additional magnetic properties to particles (in the sense that these properties cannot be explained by the electrical charge alone).\\

At the beginning of the 20$^\text{th}$ century, various experiments were attempted in order to establish the nature of the light shone by a heated gas and the effects of a magnetic field on this light. In both cases, the spectrum of the emitted light (its wavelength content) was studied in varying conditions so as to unveil the great traits of the atom's schematics. While Niels Bohr's model of the atom enabled a rough understanding of the atom at the time, the devil was once again in the details as substructures in the light spectrum, appearing once the heated gas was submitted to a magnetic field, could not be explained by it. The concept of spin was born in the 1920's, after many false experimental interpretations (such as the Stern and Gerlach experiment of 1922) and forced theoretical introductions (such as Wolfgang Pauli's attempt at it in 1924). \\

\begin{sidenote}{title = Food for thought}
In a letter sent to fellow physicist Ralph Kronig in 1925, Wolfgang Pauli wrote ``Physics is very muddled again at the moment; it is much too hard for me anyway, and I wish I were a movie comedian or something like that and had never heard anything about physics".
\end{sidenote}\bigskip

Spin is usually a vector $\mathbf{S}$ that represents a quantized intrinsic magnetic moment. The ``intrinsic magnetic moment" part encapsulates the fact that spin can be seen as a compass directly borne by a particle. To say that this compass is ``quantized" means that it can \textit{a priori} only points in a finite number of directions (this adjective actually comes from quantum mechanics). This quantization manifests itself in the different ``projections" that the spin can have along an applied magnetic field. Denoting $\Vert\mathbf{S}\Vert = S$ (an integer or half-integer number), a spin $\mathbf{S}$ (or $S$) possesses $2S+1$ ``projections" of spin along a magnetic field: $-S,-(S-1),\dots,0,\dots , S-1, S$. \\

To make this clearer, let us take the example of the proton, whose spin is equal to $S=1/2$. Once in a magnetic field $\mathbf{B}_0$, the proton spin can \textit{a priori} point in $2S+1 = 2$ directions: 
\begin{itemize}
\item[$\bullet$] aligned with the magnetic field (``spin up" state $\vert \!\uparrow \rangle$ by convention);
\item[$\bullet$] or anti-aligned with the magnetic field (``spin down" state $\vert \!\downarrow \rangle$ by convention).
\end{itemize}
These two configurations are called ``stationary states", meaning that if the spin is in one of these configurations when the magnetic field is switched on, it remains in the same configuration through time. In a magnetic field, the two spin configurations are not on an equal footing: they have different energies, given by
\begin{align}
E_{\uparrow} = - \frac{1}{2}\,\hbar\gamma B_0   \qquad \qquad E_{\downarrow} = + \frac{1}{2}\,\hbar\gamma B_0\, , \label{Eq_energy_spin}
\end{align}
where $\hbar = h/(2\pi) \simeq 1.05\times 10^{-34}\;\mathrm{J}\cdot \mathrm{s}$ is the reduced Planck constant and $\gamma \simeq 267,513\times 10^6 \; \mathrm{rad}\cdot \mathrm{s}^{-1}\cdot \mathrm{T}^{-1}$ is the gyromagnetic ratio of the proton (an intrinsic property of the proton). In other words, the configuration in which the spin is aligned with the magnetic field has the lowest energy (is physically favored), just like the case of an actual compass aligning with Earth's magnetic field.\\

\begin{sidenote}{title = ``Spin" is a very misleading name}
Spin was first interpreted as the consequence, through Amp\`{e}re's law of electrical induction, of the rotation of any charged particle on itself. However, it only took a few months for Wolfgang Pauli to show that this would imply a faster-than-light rotation at the particle's ``equator", which does not make any sense. Therefore, spin has to be accepted as an intrinsic property of particles!
\end{sidenote}

\subsubsection{The Bloch sphere}

From now on, let us remain within the case of the spin-$1/2$ proton. The reader may have noticed the use of the Latin expression ``\textit{a priori}" when the link between spin quantization and spin orientation was discussed. This has to do with the fact that the spin up state $\vert \!\uparrow \rangle$ and spin down state $\vert \!\downarrow \rangle$ are just two particular states of the system (the stationary states). A general spin state $\vert S \rangle$ can actually be written as a superposition of these two states:
\begin{equation}
\vert S \rangle = \alpha \vert \!\uparrow \rangle + \beta \vert \!\downarrow \rangle \, ,
\label{Eq_physical_spin_superposition}
\end{equation}
where $(\alpha, \beta) \in \mathbb{C}^2$. Indeed, within quantum mechanics, massive particles such as protons obey the Schrödinger equation
\begin{equation}
i\hbar\,\frac{\partial\;}{\partial t} \,\vert \psi \rangle = -\frac{\hbar^2}{2m}\, \mathbf{\nabla}^2\vert \psi \rangle + V \vert \psi \rangle \, ,
\end{equation}
which is a differential equation for the general particle state $\vert \psi \rangle$ bathing in the potential $V$ (describing the effect of an electromagnetic field or gravity for instance). This implies that if $\vert \!\uparrow \rangle$ and $\vert \!\downarrow \rangle$ are true physical solutions of the Schrödinger equation (which they are when spin is the only relevant degree of freedom in the problem), any linear combination Eq.~\eqref{Eq_physical_spin_superposition} of these states is a physical solution. In order to be able to compare all states with one another, one imposes the following normalization condition:
\begin{equation}
\vert \alpha \vert^2 + \vert\beta\vert^2 = 1 \, .
\label{Eq_spin_state_normalization}
\end{equation}
With this normalization, the numbers $\alpha$ and $\beta$ acquire an important physical meaning: if one measures a particle that is initially in the superposition Eq.~\eqref{Eq_physical_spin_superposition}, the state $\vert \! \uparrow\rangle$ will be measured with probability $\vert \alpha \vert^2$ and the state $\vert \! \downarrow\rangle$ will be measured with probability $\vert \beta \vert^2$.\\

\begin{figure}[h!]
\begin{center}
\includegraphics[width=0.52\textwidth]{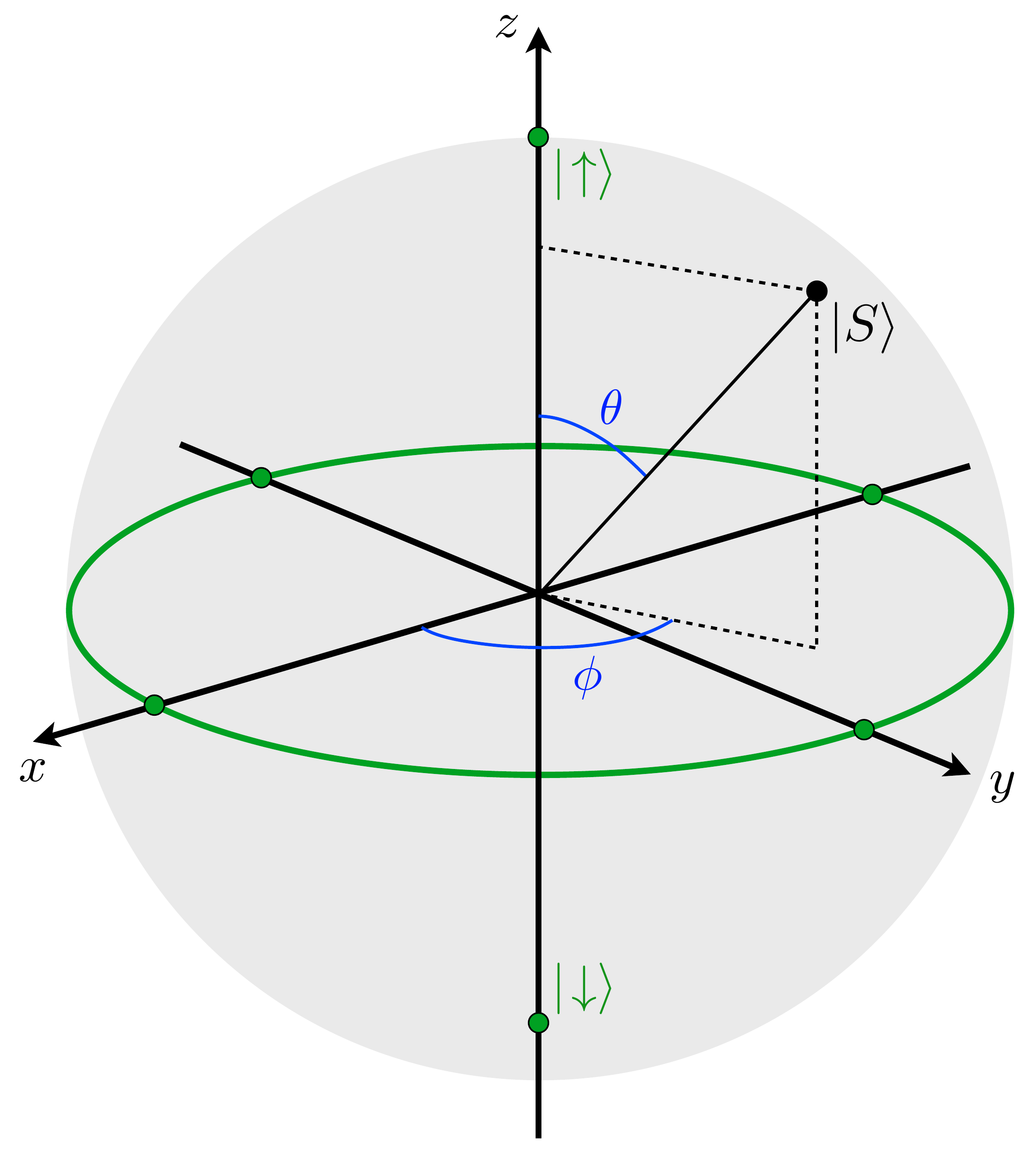}
\caption{Bloch sphere graphically representing all normalized $1/2$-spin states $\vert S \rangle$ as quantum superposition of the eigenstates $\vert \! \uparrow\rangle$ and $\vert \! \downarrow\rangle$ (the poles of the sphere). All states with equal probability of measuring either $\vert \! \uparrow\rangle$ or $\vert \! \downarrow\rangle$ lie on the green equator of the sphere.}
\label{Fig_Bloch_sphere}
\end{center}
\end{figure}

How can one represent all these spin states graphically? Fortunately, it is quite simple in the case of a spin $1/2$. Indeed, let us write the complex numbers $\alpha$ and $\beta$ under their exponential form,
\begin{align}
\vert S \rangle & = \rho_\alpha\, \exp^{i\phi_\alpha} \vert \!\uparrow \rangle + \rho_\beta\, \exp^{i\phi_\beta} \vert \!\downarrow \rangle \nonumber \\
 & = \exp^{i\phi_\alpha}\left[ \rho_\alpha \vert \!\uparrow \rangle + \rho_\beta\, \exp^{i(\phi_\beta - \phi_\alpha)} \vert \!\downarrow \rangle \right] \, ,
\end{align}
and use one of the postulates of quantum mechanics, which states that physical states that differ only by a phase factor ($\exp^{i\phi}$) cannot be distinguished (any measurement is only sensitive to the squared norm of a given state and $\vert \exp^{i\phi} \vert = 1$), to obtain
\begin{equation}
\vert S \rangle \equiv \rho_\alpha \vert \!\uparrow \rangle + \rho_\beta\, \exp^{i(\phi_\beta - \phi_\alpha)} \vert \!\downarrow \rangle\, .
\end{equation}
Now, using the normalization Eq.~\eqref{Eq_spin_state_normalization} written as 
\begin{equation}
\rho_\alpha^2 + \rho_\beta^2 = 1\, ,
\end{equation}
one can define two angles, $\theta$ and $\phi$, such that
\begin{equation}
\cbeq{\vert S \rangle \equiv \cos\left( \frac{\theta}{2} \right)\vert \!\uparrow \rangle + \sin\left( \frac{\theta}{2} \right) \exp^{i\phi}\vert \!\downarrow \rangle} \; ,
\end{equation}
where $\theta \in [0,\pi]$ and $\phi\in [0,2\pi[$. Such definitions for the angles $\theta$ and $\phi$ match the usual angular components of spherical coordinates in geometry. The normalization condition Eq.~\eqref{Eq_spin_state_normalization} setting a natural and constant radius of $1$ for any spin state, such a state can be represented on a sphere of radius $1$, called the Bloch sphere (after Felix bloch, that will be mentioned again later) and illustrated in Fig.~\ref{Fig_Bloch_sphere}.\\

\begin{mainpoint}{title = Link between the Bloch sphere representation and the average spin $\langle \mathbf{S} \rangle$}
It so happens (the reader may look for a proof in any good quantum mechanics textbook) that the orientation of the state $\vert S \rangle$ in the Bloch sphere coincides with the actual orientation of the average of the spin $\langle \mathbf{S} \rangle$ within this state! One can now visualize the average spin associated to any spin superposition in the lab frame of reference!
\end{mainpoint}

\newpage

\subsubsection{Magnetization and Larmor precession}

One could wonder why the average spin $\langle \mathbf{S} \rangle$ is so important. As a matter of fact, this quantity is related to the concept of magnetization. To understand this relation, one has to write down the actual magnetic moment borne by the spin of a given particle:
\begin{equation}
\bm{\mu} = \gamma \mathbf{S}\, .
\end{equation}
The magnetic moment is exactly what one would associate to a classical bar magnet since it defines its magnetic poles and how strongly it responds to an applied magnetic field. The magnetization is a quantity that characterizes an ensemble of spins at a given temperature. Let us imagine an ensemble of $N$ non-interacting spins at temperature $T$ in the magnetic field $\mathbf{B}_0$. On the one hand, thermal energy $k_\text{B} T$ (where $k_\text{B} \simeq 1,38\times 10^{-23}\;\mathrm{J}\cdot\mathrm{K}^{-1}$ is the Boltzmann constant) favors disorder and thus randomness of spin orientation. On the other hand, magnetic energy, of order $\hbar \gamma B_0$ as shown in Eq.~\eqref{Eq_energy_spin}, favors spin alignment with the magnetic field. The magnetization $\mathbf{M}$ is the subtle compromise found between these antagonistic effects. In the aforementioned case, it writes
\begin{equation}
\cbeq{
\mathbf{M} = N \langle \bm{\mu} \rangle = N\gamma \langle \mathbf{S}\rangle
}\; .
\end{equation} 
In the case of the proton, where the gyromagnetic ratio $\gamma$ is strictly positive, the magnetization points in the same direction as the average spin borne by the ensemble of spins.\\

\begin{sidenote}{title = The brain's magnetization}
One has to realize how weakly magnetic the brain is. Considering a brain temperature $T \simeq 310\;\mathrm{K}$, a magnetic field $B_0 = 1\;\mathrm{T}$ and the gyromagnetic ratio of the hydrogen atom's nucleus (its proton) $\gamma \simeq 267,513\times 10^6\;\mathrm{rad}\cdot \mathrm{s}^{-1}\cdot \mathrm{T}^{-1}$, one obtains the following statistical ratio of down to up spins:
\begin{equation}
\frac{N_\downarrow}{N_\uparrow} \sim \exp^{-\hbar\gamma B_0/(k_\text{B} T)} \simeq 1 - 10^{-6}\, .
\end{equation}
In other words, there's in average only one spin in a million contributing to the magnetic resonance (MR) signal! Fortunately, this signal can be measured thanks to two things: first, the brain contains roughly trillions of trillions of water molecules and second, quantum mechanics offers measurements an unrivaled level of precision.
\end{sidenote}\bigskip

The previous treatment of a system of $N$ non-interacting spins is ``statistical" in the sense that it tells how the overall spin behaves in average without giving any insight in the actual behavior of any spin once submitted to the magnetic field. Let us consider a magnetic field $\mathbf{B}_0$ applied in the $z$ direction of the Bloch sphere presented in Fig.~\ref{Fig_Bloch_sphere}. Without delving into the quantum dynamics of spin orientation, it so happens that a spin superposition state $\vert S \rangle$ precesses around the applied magnetic field at an angular frequency (or pulsation) given by
\begin{equation}
\cbeq{
\omega_0 = \gamma B_0
}\; ,
\label{Eq_Larmor_frequency}
\end{equation}
commonly called the ``Larmor frequency", even though the actual frequency is $f_0 = \omega_0/(2\pi)$. In this simple case, this corresponds to precessing at constant angle $\theta$ (the initial angle of the spin superposition) around the $z$ axis in a counterclockwise fashion. If the magnetic field $\mathbf{B}_0$ is not applied in the $z$ direction, the same precession occurs at a fixed angle with the magnetic field. However, one usually talks of ``Rabi oscillations" in that case. Both types of precession are mathematically described by the Bloch equation \cite{Bloch:1946}:
\begin{equation}
\frac{\mathrm{d}\mathbf{M}}{\mathrm{d}t} = -\gamma\, \mathbf{B}_0 \mathbf{\times} \mathbf{M}\, .
\label{Eq_Bloch_equation}
\end{equation}
For an ensemble of $N$ non-interacting spins, all spins precesses around the $z$ axis, so that the average spin $\langle \mathbf{S}\rangle$ remains along the $z$ axis, just as the magnetization $\mathbf{M}$. This means that spins are not generally aligned with the magnetic field, there simply is a vast majority of spins that are globally aligned with it. Others can be globally anti-aligned with it. The larger the temperature, the larger the fraction of globally anti-aligned spins. In other words, applying a magnetic field on an ensemble of spins generates a magnetization along this magnetic field!

\subsubsection{Selectivity through resonance}
\label{Sec_resonance}

It is possible to target and excite a specific population of spins in a given system by taking advantage of the concept of resonance. Before addressing the magnetic resonance of spins, the reader should note that resonance extends to classical phenomena as it only relies on two components:
\begin{itemize}
\item[$\bullet$] a natural frequency driving the system's dynamics; 
\item[$\bullet$] an external periodic perturbation that can match this natural frequency. 
\end{itemize} 
For instance, the position of a mass $m$ attached to a spring of stiffness $k$ oscillates through time at a natural (angular) frequency $\omega_0 = \sqrt{k/m}$. If the spring is attached to a wall oscillating at frequency $\omega$, the system enters a resonance when $\omega \simeq \omega_0$, meaning that the amplitude of the position's oscillations would increase continuously if no drag were present. In other words, the wall oscillation acts as an external perturbation that can drive the system out of a steady oscillation. The same thing can happen in an electrical circuit containing an inductance $L$ and a capacitance $C$ driven by a generator of alternative current, when the current's frequency $\omega$ closely matches the natural frequency of the system $\omega_0 = 1/\sqrt{LC}$. To put things into perspective, resonance is an inherent part of any oscillating behavior, whether it is classical or quantum mechanical!\\

For a spin system, one first needs to apply a magnetic field $\mathbf{B}_0$ in order to set a natural frequency in the system, the Larmor frequency $\omega_0$ Eq.~\eqref{Eq_Larmor_frequency} (by convention, the direction of $\mathbf{B}_0$ defines the $z$ direction). Once this frequency is set, one can apply a magnetic pulse $\mathbf{B}_1(\omega)$ of amplitude $B_1$ and duration $\tau_1$ in the $x$-$y$ plane (perpendicular to the $\mathbf{B}_0$ field) at the frequency given by the resonance condition 
\begin{equation}
\cbeq{
\omega \simeq \omega_0 = \gamma B_0
}\; .
\end{equation}
The effect of this RF pulse ($\omega_0$ usually corresponds to a radiofrequency) is to tip the initial magnetization with respect to the $z$-axis by the tipping angle $\theta_1$
\begin{equation}
\theta_1 = \gamma B_1 \tau_1
\label{Eq_tipping_angle}
\end{equation}
in radians. This tipping is specific because its efficiency depends on the nature of the targeted spin through the gyromagnetic ratio $\gamma$.\\ 

\begin{sidenote}{title = The Nobel Prize for NMR}
If the targeted spin is part of an atom's nucleus, one obtains the ``nuclear induction" independently discovered by Felix Bloch \cite{Bloch:1946} and Edward Purcell in the 1940's. They were both awarded the Nobel Prize in Physics in 1952 for this effect, called ``nuclear magnetic resonance" (NMR) nowadays. 
\end{sidenote}\bigskip

A pulse of duration
\begin{equation}
\tau_1^{90^\circ} = \frac{\pi}{2\gamma B_1}
\label{Eq_duration_90_RF_pulse}
\end{equation}
is called a 90$^\circ$ RF pulse, as it tips the initial magnetization into the $x$-$y$ plane. The magnetization then precesses in this plane under the effect of the stationary $\mathbf{B}_0$ field. Let us proceed with an incredibly simplistic picture of the magnetization readout. Once the RF pulse tips the magnetization into the $x$-$y$ plane, the rotating (precessing) magnetization generates a varying magnetic field. By magnetic induction, electrical currents are then generated in coils contained in the MR scanner. This electrical current is the actual signal being measured. \\


\begin{sidenote}{title = Slice-select gradients}
Resonance enables highly specific excitation of spin populations. However, magnetic resonance imaging (MRI) gives images of brain slices for a given spin population. So, where is this ``slice" aspect coming from? As a matter of fact, during the 90$^\circ$ RF pulse, a linear slice-select field gradient $\mathbf{G}_\text{slice}$ is applied so that the $\mathbf{B}_0$ field actually writes
\begin{equation}
\mathbf{B}_0^\text{slice}(\mathbf{r}) = \mathbf{B}_0 + (\mathbf{G}_\text{slice}\cdot \mathbf{r})\, \mathbf{u}_G\, ,
\end{equation}
where $\mathbf{u}_G$ is the unit vector along the direction of the gradient. In other words, a linear gradient only changes the value of the $\mathbf{B}_0$ magnetic field along a single direction, which designs perpendicular slices of uniform magnetic field, although each slice bathes into a different magnetic field than the others. That way, one can apply the right slice-select gradient so that a slice of interest, bathing in the field $\mathbf{B}_0^\text{target}$, is the only one being targeted by the RF pulse of frequency 
\begin{equation}
\omega_\text{target} = \gamma B_0^\text{target}\, .
\end{equation}

To be a bit more subtle, the resonant response is not perfectly peaked at frequency $\omega_0$, it has a width whose order of magnitude compared to $\omega_0$ is given by the ratio $B_1/B_0$. The technical limitations of magnetic resonance make their appearance in this very statement. Indeed, Eq.~\eqref{Eq_tipping_angle} makes clear that larger $B_1$ values enable shorter RF pulses, which is useful for many reasons related to relaxation (less loss of signal, see Sec.~\ref{Sec_relax}) and because the slice-select gradient, if turned on for too long, induces significant diffusion dephasing (this last concept will be thoroughly dealt with in Sec.~\ref{Sec_dMRI}). However, larger $B_1$ values also increase the thickness of the imaging slices, so that larger values for $B_0$ are also required to counteract this disadvantage.
\end{sidenote}\bigskip

\begin{mainpoint}{title = Magnetic resonance}
The Larmor frequency $\omega_0 = \gamma B_0$ plays the role of a natural frequency for a spin system. Therefore, applying an RF magnetic pulse in the transverse plane at a frequency $\omega \simeq \omega_0$ induces a resonant response of the system. This response corresponds to an efficient tipping of the system's magnetization away from the axis of the $B_0$ field. With the right pulse duration Eq.~\eqref{Eq_duration_90_RF_pulse}, the magnetization is effectively sent into the transverse plane, where it can start precessing under the remaining effect of $B_0$.
\end{mainpoint}

\subsection{Relaxation and spin echo}

\subsubsection{$T_1$ and $T_2$ relaxation}
\label{Sec_relax}

\begin{figure}[h!]
\begin{center}
\includegraphics[width=0.7\textwidth]{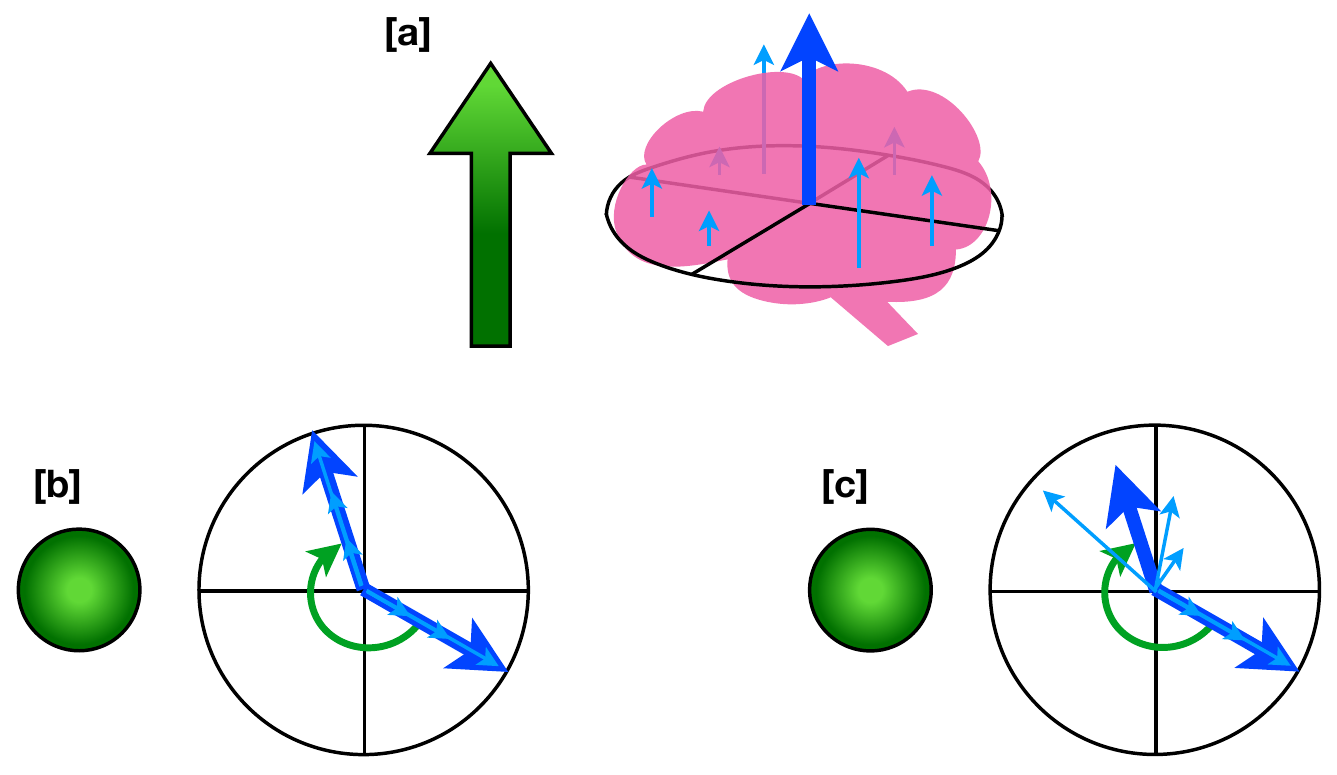}
\caption{\textbf{[a]} Depending on the local density of spins, different parts of the brain exhibit different total spins (sky blue arrows). Once aligned in average by the $B_0$ field (green arrow), the magnetic moments borne by these spins sum up to a magnetization for the whole brain (dark blue arrow). \textbf{[b]} View from above after the 90$^\circ$ RF pulse in the case of a perfect medium. All spins precess at the same rate, maintaining the whole in-plane magnetization. \textbf{[c]} View from above after the 90$^\circ$ RF pulse in the case of a magnetically heterogeneous medium. Different local environments of the brain precess at different rates, which weakens the magnetization over time.}
\label{Fig_Spin_echo_1}
\end{center}
\end{figure}

The combination of resonance and slice selection does not solve all problems when it comes to measuring a MR signal. Indeed, right after the 90$^\circ$ RF pulse, the magnetization is not in its lowest energy configuration anymore (the lowest energy configuration corresponds to the magnetization aligned with the applied $B_0$ field). In that case, quantum physics predicts that any process lowering the energy of the system will be favored. This family of processes is called ``relaxation processes". These processes weakens the $x$-$y$ magnetization by default:
\begin{itemize}
\item[$\bullet$] One calls $T_1$-processes the magnetic field disturbances that send the magnetization out of the $x$-$y$ plane, effectively reducing the in-plane component of the magnetization. To be efficient, such processes have to occur close to the Larmor frequency $\omega\simeq \omega_0$ so as to satisfy the resonance condition and to induce an efficient rotation of the magnetization.
\item[$\bullet$] One calls $T_2$-processes the local stationary ($\omega \simeq 0$) magnetic field disturbances that make certain spins precess at different rates compared to the Larmor frequency, hence weakening the transverse magnetization, as shown in Fig.~\ref{Fig_Spin_echo_1}.
\end{itemize}
Here, the terms $T_1$ and $T_2$ denote timescales so that the magnetization's components write
\begin{align}
M_{z}(t) & \propto M_0 \left[ 1 - \exp^{-t/T_1} \right] \\
M_{x,y}(t) & \propto M_0\, \exp^{-t/T_2} \, ,
\end{align}
where $M_0 = \Vert \mathbf{M}(t=0)\Vert$. The shorter the time $T_1$ ($T_2$), the stronger the $T_1$ ($T_2$) relaxation. 

\subsubsection{Physical origins in the human body}

Here is a table giving a few orders of magnitude for $T_1$ and $T_2$:\\

\begin{figure}[h!]
\centering
\begin{tabular}{|c||c|c|}
\hline
Tissue & $T_1$ (ms) & $T_2$ (ms) \\
\hline
\hline
Water & 4000 & 2000 \\
Grey matter & 900 & 90 \\
Muscle & 900 & 50 \\
Liver & 500 & 40 \\
\hline
\end{tabular}
\hspace*{2cm}
\begin{tabular}{|c||c|c|}
\hline
Tissue & $T_1$ (ms) & $T_2$ (ms) \\
\hline
\hline
Fat & 250 & 70 \\
Tendons & 400 & 5 \\
Proteins & 250 & 0.1 - 1 \\
Ice & 5000 & 1 \\
\hline
\end{tabular}
\end{figure}

While most organs present a $T_1$ 5 to 10 times longer than $T_2$, pure liquids have very long values of $T_1$ and $T_2$ and macromolecules and solids have very short values of $T_2$. Why is that so?\\

Let us focus on natural relaxation processes in biological tissues, where the dipole-dipole interaction is the single most important relaxation mechanism. This interaction is the one that exists classically between two bar magnets. In the quantum mechanical setting, the strength of this interaction is globally given by
\begin{equation}
V_\text{dipole-dipole} \propto \gamma_1\gamma_2\, \frac{3\cos^2\theta_{12}-1}{r_{12}^6}\, ,
\label{Eq_dipole_dipole_interaction}
\end{equation}
where $\gamma_1$ and $\gamma_2$ are the gyromagnetic ratios of each interacting spins, $\theta_{12}$ is the angular difference between these spins' orientations, and $r_{12}$ is the spatial distance separating them. The $1/r_{12}^6$ radial dependency already tells us that only dipole-dipole interaction between intramolecular spins effectively matters in biological systems. Also, since the gyromagnetic ratio of the electron is far larger than the one of the proton ($\gamma_\text{e} \simeq 1000\, \gamma_\text{p}$), a proton-electron dipolar interaction is much larger than a proton-proton one, which explains the use of Gd$^{3+}$, that has a lot of unpaired electrons, as a contrast agent to enhance relaxation contrast.\\

\begin{figure}[h!]
\begin{center}
\includegraphics[width=0.5\textwidth]{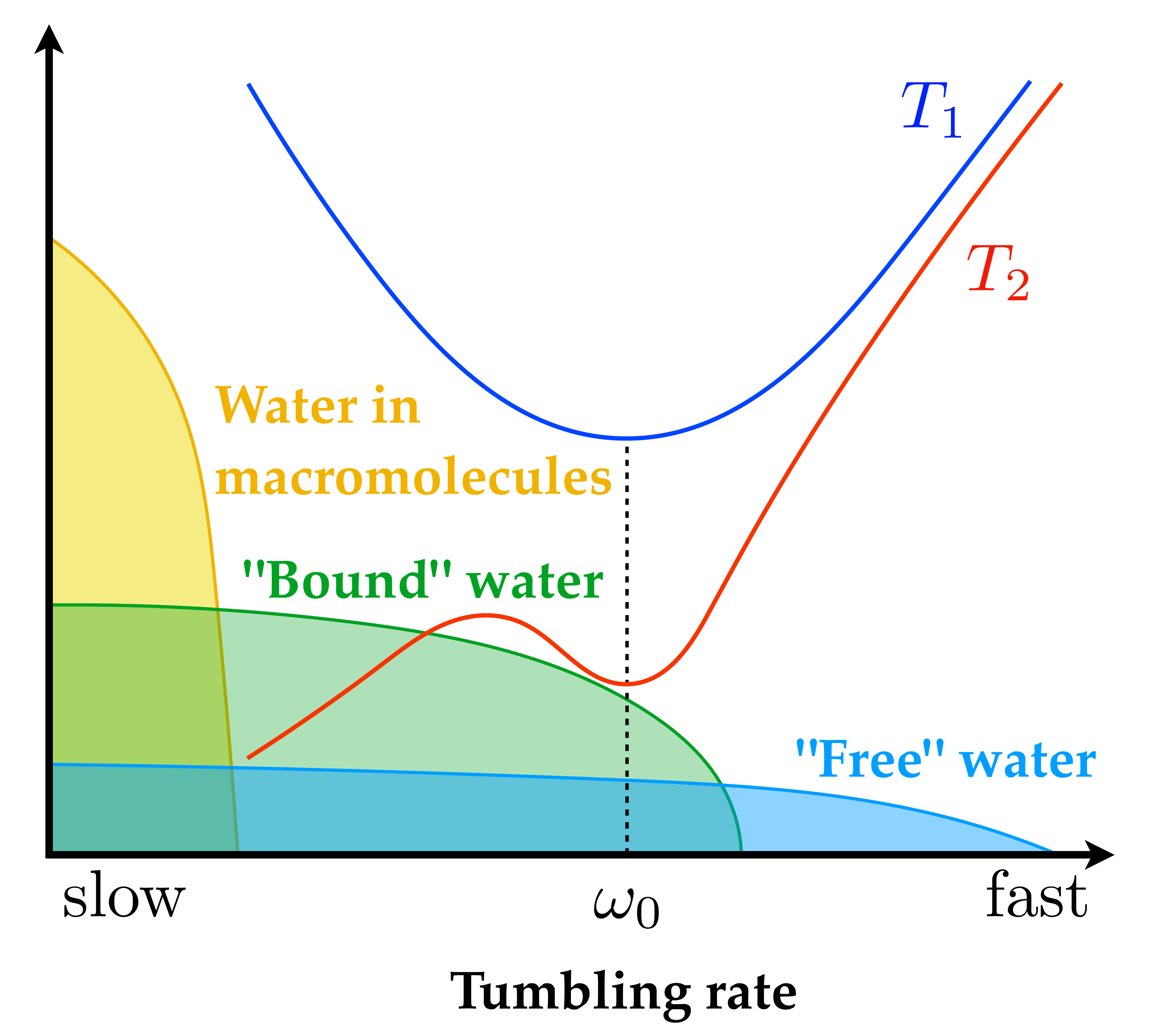}
\caption{Evolution of $T_1$ and $T_2$ as a function of the tumbling rate $\omega_\text{tumbling}$. The colored areas correspond to the distribution of spins in given media along the tumbling rate axis. The small dip in $T_2$ relaxation around $\omega_0$ happens because of the boost in $T_1$ processes that also generate some additional $T_2$ relaxation.}
\label{Fig_relaxation}
\end{center}
\end{figure}

Let us give a very naive example of how the dipole-dipole interaction and relaxation processes are linked in water molecules. In a water molecule, the hydrogen protons are separated by a distance of around 0.2 to 0.3 nanometers. At this distance, each spin creates a magnetic field going from $7\;\mathrm{G} = 7\times 10^{-4}\;\mathrm{T}$ (along the spin) to $-7\;\mathrm{G}$ (perpendicular to the spin). This means that each spin will see an extra varying magnetic field from the other spin, with a frequency roughly given by the tumbling frequency $\omega_\text{tumbling}$ of the water molecule! If $\omega_\text{tumbling} \simeq \omega_0$, the dipole-dipole interaction generates $T_1$ relaxation. If $\omega_\text{tumbling} \simeq 0$, the dipole-dipole interaction generates $T_2$ relaxation. One then realizes that $T_2$ should correlate with structure: the more restricted the structure ($\omega_\text{tumbling} \to 0$), the shorter the $T_2$ (more $T_2$ processes, so more efficient $T_2$ relaxation). This is summed up in Fig.~\ref{Fig_relaxation}.

\subsubsection{Spin echo}

What about unnatural sources of relaxation? For instance, one talks of $T_2^*$ relaxation when unnatural sources contribute to the transverse relaxation. Such relaxation mainly results from inhomogeneities in the main magnetic field $B_0$, due to intrinsic defects in the magnet itself or susceptibility-induced field distortions produced by the tissue or other materials placed within the field. If one wants to be insensitive to such artificial processes (and regain some signal at the same time), one can use a spin echo sequence.\\

\begin{figure}[h!]
\begin{center}
\includegraphics[width=0.45\textwidth]{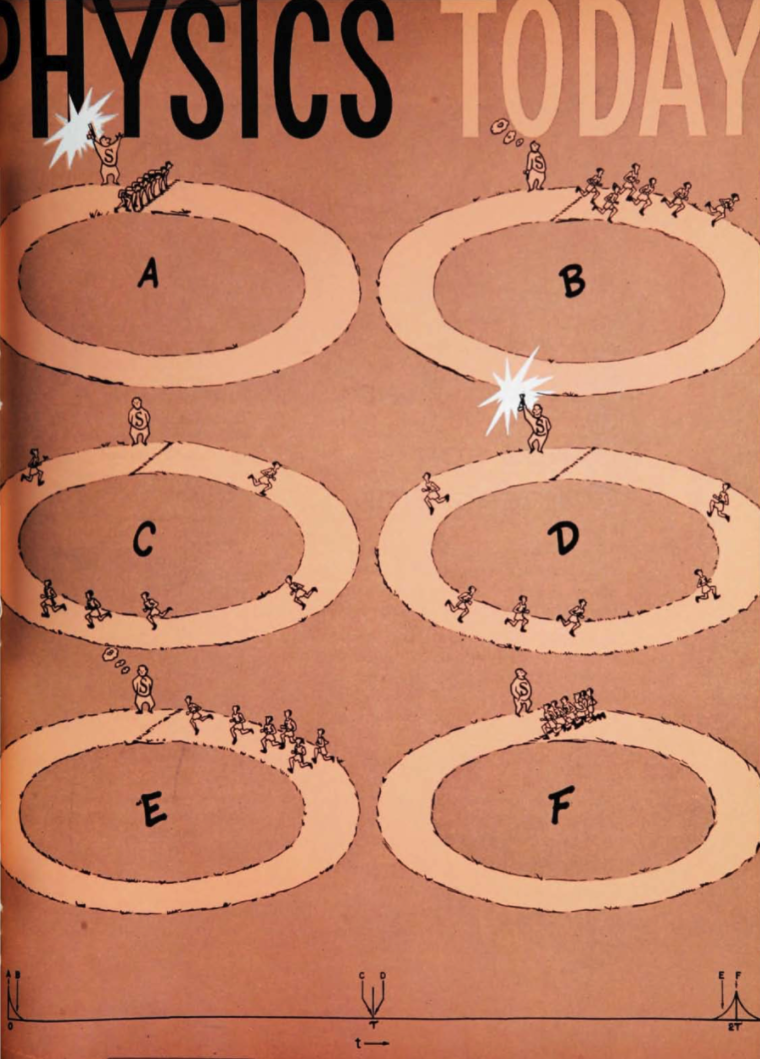}
\caption{Cover of one of the 1953 Physics Today magazines presenting the race analogy for the spin echo sequence.}
\label{Fig_Spin_echo_Physics_Today}
\end{center}
\end{figure}

Proposed in 1950 by Erwin Hahn, the spin echo sequence aims at rephasing the spins dephased by artificial sources of transverse relaxation to obtain a measurable magnetization \cite{Hahn:1950}. This sequence is based on a 90$^\circ$ RF pulse, of duration Eq.~\eqref{Eq_duration_90_RF_pulse}, and a 180$^\circ$ RF pulse, of duration twice of Eq.~\eqref{Eq_duration_90_RF_pulse}. It relies on the idea presented in the cover of Physics Today honoring Hahn's work, Fig.~\ref{Fig_Spin_echo_Physics_Today}. Let us imagine you start a race at time $t=0$ ([A]). Runners usually tend to run at different speeds ([B] and [C]), faster runners overcoming more distance within the same amount of time. After a time $t=t_\text{go back}$, you tell the runners to go back on their race, at the exact same speed they came initially ([D]). The faster runners now have more distance to overcome compared to slower runners ([E]). However, by symmetry, all runners will go to the starting line at the exact same time $t = 2t_\text{go back}$ ([F])! \\

\begin{figure}[h!]
\begin{center}
\includegraphics[width=\textwidth]{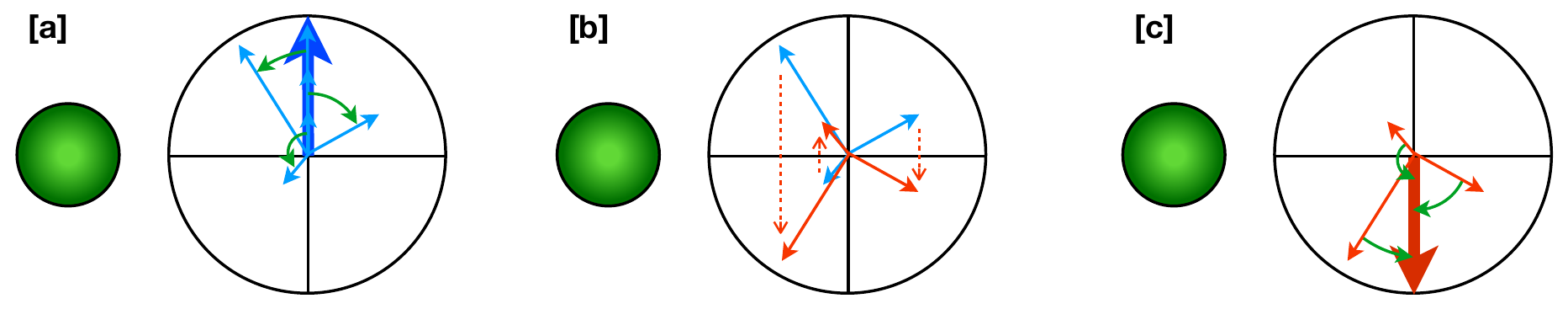}
\caption{\textbf{[a]} Spins usually precess at different rate within the $x$-$y$ plane after the 90$^\circ$ RF pulse. \textbf{[b]} Adding a 180$^\circ$ RF pulse to the sequence acts as a mirror reversing for spins. \textbf{[c]} Still precessing in the exact same way as initially, spins refocus after twice the amount of time between the two RF pulses. The total time spent between [a] and [c] is called the echo time TE.}
\label{Fig_Spin_echo_2}
\end{center}
\end{figure}

The bottom of Fig.~\ref{Fig_Spin_echo_Physics_Today} presents the spin echo sequence as the MR equivalent of this peculiar race. Spins are sent precessing in the $x$-$y$ plane right after the 90$^\circ$ RF pulse ([A], $t=0$). They may precess at different rates because of magnetic disturbances ([B] and [C]). However, applying a 180$^\circ$ RF pulse reverses the situation ([D]) so that spins refocus after the ``echo time" TE. This process is illustrated in more details in Fig.~\ref{Fig_Spin_echo_2}. The overall signal magnitude is proportional to
\begin{equation}
\cbeq{
\mathcal{S} \propto n_\text{p} \, \exp^{-\text{TE}/T_2} \left[ 1- \exp^{-\text{TR}/T_1} \right] 
}\; ,
\end{equation}
where $n_\text{p}$ is the local spin density and TR is called the ``repetition time", the time after which the whole spin echo sequence is repeated. While the $T_2$ term relates to the amount of lost magnetization during the sequence, the $T_1$ term relates to how much magnetization has regrown along the $z$ direction just before applying the next 90$^\circ$ RF pulse.\\

\begin{sidenote}{title = Gradient echoes and $T_2^*$-weighted imaging}
Certain MR measurements rely on gradient echo sequences, where only a 90$^\circ$ RF pulse is applied, followed by a pair of dephasing gradients with opposite polarities (effectively resurrecting the magnetization, albeit $T_2^*$-weighted). Although such gradient echo sequences suffer from inhomogeneity artifacts more than spin echo sequences, this can be put to an advantage to detect local magnetic inhomogeneities caused by hemorrhages or calcifications.
\end{sidenote}

\newpage

\section{The basics of diffusion}

\subsection{Fick's law}

After Robert Brown's first interest in diffusion processes regarding pollen grains on a water surface in the first half of the 19$^\text{th}$ century, Adolf Fick began studying the diffusion of ink in water. Let us imagine a 1D system (associated to the spatial dimension $x$) made up of two separate compartments, one containing water with a high concentration of ink, another containing water with a lower concentration of ink. At time $t=0$, the barrier separating the two compartments is withdrawn, allowing free exchange between them. In general, a current $J(x,t)$ will appear in order to make the concentration of ink $C(x,t)$ uniform across the system (a consequence of maximizing the system's entropy). This current typically writes
\begin{equation}
J = -D\, \frac{\partial C}{\partial x}\, ,
\end{equation}
where $D$ is the diffusion constant of the ink within the water. Conservation of mass imposes that
\begin{equation}
\frac{\partial C}{\partial t} = -\frac{\partial J}{\partial x}\, ,
\end{equation}
which leads to
\begin{equation}
\frac{\partial C}{\partial t} = D\,\frac{\partial^2 C}{\partial x^2}\, .
\end{equation}
If the medium is heterogeneous, \textit{i.e.} $D$ actually depends on $x$, one has instead
\begin{equation}
\frac{\partial C}{\partial t} = \frac{\partial \;}{\partial x} \left[ D(x)\,\frac{\partial C}{\partial x} \right] .
\end{equation}
For a 3D heterogeneous system, this generalizes to
\begin{equation}
\frac{\partial C}{\partial t} = \bm{\nabla}\cdot \left[ D(\mathbf{r})\,\bm{\nabla}C \right] .
\label{Eq_Fick_law}
\end{equation}

\subsection{Probability displacement function}

It's only at the beginning of the 20$^\text{th}$ century that a deeper mathematical framework was developed by Albert Einstein to describe diffusion processes (experimentally confirmed a few years later by Jean Perrin). Starting from Fick's law Eq.~\eqref{Eq_Fick_law}, one can consider that a concentration is very much analogous to a probability and write
\begin{equation}
\frac{\partial \mathcal{P}}{\partial t} = \bm{\nabla}\cdot \left[ D(\mathbf{r})\,\bm{\nabla}\mathcal{P} \right] ,
\label{Eq_Einstein_equation}
\end{equation}
where the probability displacement function $\mathcal{P}(\mathbf{r},t)$ denotes the probability of finding an ink particle at position $\mathbf{r}$ at time $t$ knowing that it started from position $\mathbf{0}$ at time $t=0$. In the case of a homogeneous and isotropic medium, the probability displacement function spatially depends on $r=\Vert\mathbf{r}\Vert$ alone and the previous equation reduces to
\begin{equation}
\frac{\partial \mathcal{P}}{\partial t} (r,t) = D\, \frac{\partial^2\mathcal{P}}{\partial r^2} (r,t)\, ,
\end{equation}
which has a simple Gaussian solution:
\begin{equation}
\cbeq{
\mathcal{P}(r,\tau) = \frac{1}{(4\pi D\tau)^{d/2}}\, \mathrm{exp}\!\left(-\frac{r^2}{4D\tau}\right) 
}
\label{Eq_Gaussian_propagator}
\end{equation}
for a diffusion time $\tau$, with $d$ being the dimension of the system.\\

\begin{figure}[h!]
\begin{center}
\includegraphics[width=0.8\textwidth]{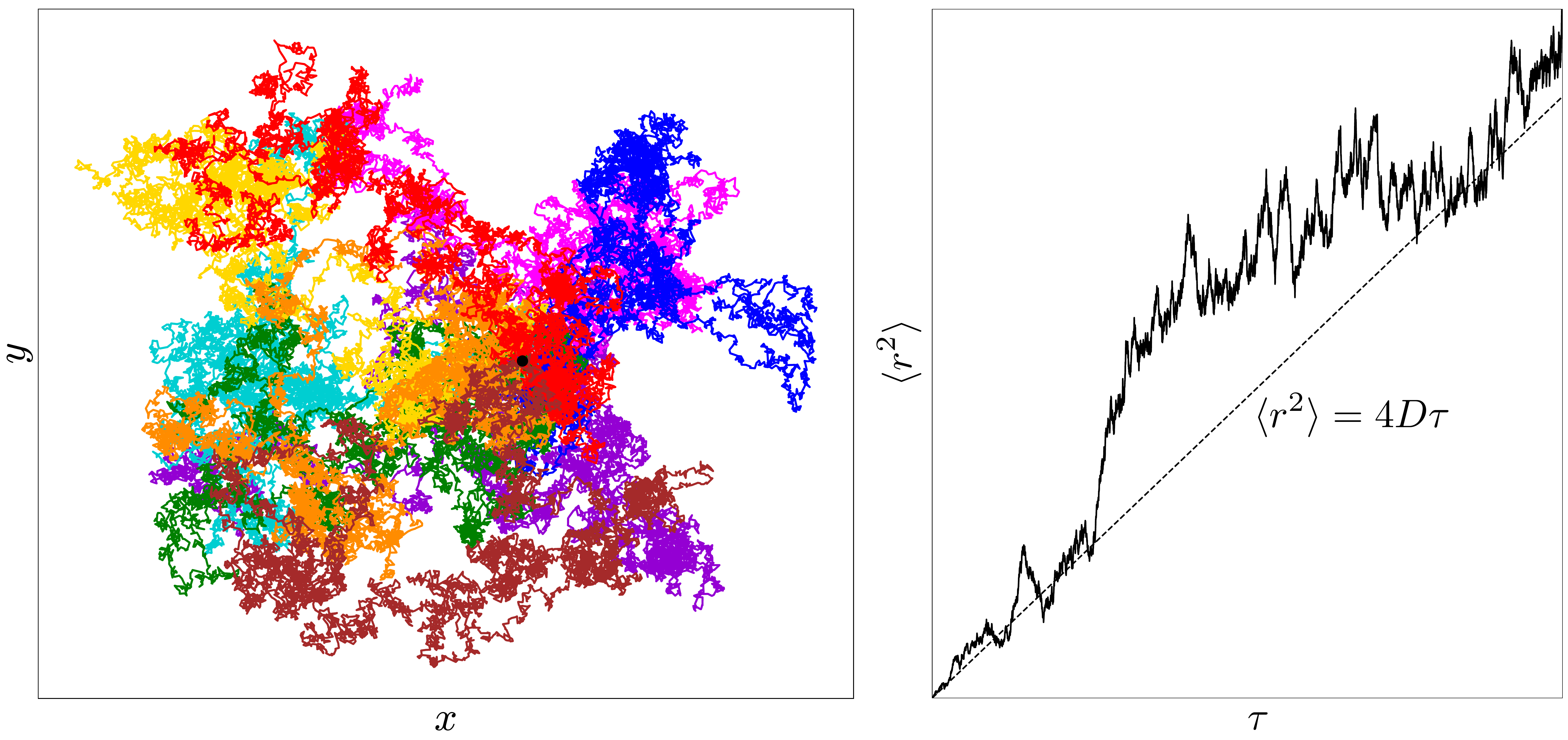}
\caption{\textbf{Left:} Nine two-dimensional random walks, all starting from the black point. \textbf{Right:} Associated mean squared displacement $\langle r^2 \rangle$ (over the nine random walks in the left panel) as a function of diffusion time $\tau$. Dashed line indicates the analytic result Eq.~\eqref{Eq_MSD_d}. Imperfect match to the expected result originates from the fact that only nine paths were used to compute $\langle r^2 \rangle$.}
\label{Fig_brownian}
\end{center}
\end{figure}

From this Gaussian propagator can be defined two important statistics: the mean squared displacement and the velocity autocorrelation function. Let us focus on the 1D case, as it will be relevant in Sec.~\ref{Sec_brownian_continuous}. First, the mean squared displacement is given by
\begin{align}
\langle x^2 \rangle & = \int_{-\infty}^{+\infty}\! \left[x^2\, \mathcal{P}(x,\tau)\right] \mathrm{d}x \nonumber \\
 & = \frac{1}{\sqrt{4\pi D\tau}}\int_{-\infty}^{+\infty}\! x^2\, \mathrm{exp}\!\left(-\frac{x^2}{4D\tau}\right) \mathrm{d}x \nonumber \\
 & = \frac{1}{\sqrt{\pi D\tau}}\int_0^{+\infty}\! x^2\, \mathrm{exp}\!\left(-\frac{x^2}{4D\tau}\right) \mathrm{d}x \nonumber \\
  & = \frac{1}{2}\,\frac{(4D\tau)^{3/2}}{\sqrt{\pi D\tau}}\int_0^{+\infty}\! x\, \mathrm{exp}\!\left(-x^2\right) \underbrace{[2x\,\mathrm{d}x]}_{\mathrm{d}(x^2)} \nonumber \\
 & = \frac{4D\tau}{\sqrt{\pi}} \underbrace{\int_0^{+\infty}\! \sqrt{u}\, \mathrm{exp}\!\left(-u\right) \mathrm{d}u}_{\Gamma(3/2) = \sqrt{\pi}/2} \, ,
\end{align}
so that
\begin{equation}
\cbeq{
\langle x^2 \rangle = 2D\tau
}\; .
\label{Eq_MSD}
\end{equation}
In higher dimensions, since $\langle r^2 \rangle = \sum_i \langle x_i^2\rangle$, one would obtain 
\begin{equation}
\langle r^2 \rangle = 2dD\tau \, ,
\label{Eq_MSD_d}
\end{equation}
as shown in Fig.~\ref{Fig_brownian}. As for the velocity autocorrelation function in 1D, one has
\begin{align}
\underbrace{\langle x^2 \rangle}_{2D\tau} & = \left\langle \left[\int_0^\tau\! v_x(t)\,\mathrm{d}t \right] \times \left[ \int_0^\tau\! v_x(t')\,\mathrm{d}t' \right] \right\rangle \nonumber \\
 & = \int_0^\tau \int_0^\tau \! \langle v_x(t)\,v_x(t')\rangle\, \mathrm{d}t\,\mathrm{d}t' \nonumber \\
 & = \int_0^\tau \int_0^\tau \! \langle v_x(\vert t-t'\vert)\,v_x(0)\rangle\, \mathrm{d}t\,\mathrm{d}t'
 \label{Eq_interm_D_velocity}
\end{align}
because the path average can only depend on the absolute time interval $\vert t-t'\vert$, and not on any specific moment $t_0$. However, in the case of a perfectly free diffusion, velocity correlations are instantly lost at $\vert t-t'\vert > 0$, so that one can usually write
\begin{equation}
\cbeq{
\langle v_x(\vert t-t'\vert)\,v_x(0)\rangle = 2D\, \delta(\vert t-t'\vert)
}\; ,
\label{Eq_auto_velocity_correlation}
\end{equation}
in agreement with Eq.~\eqref{Eq_interm_D_velocity}.\\

\begin{mainpoint}{title = Free (Gaussian) diffusion}
A system is said to be in the free (Gaussian) diffusion regime when its associated probability displacement function is given by the Gaussian function Eq.~\eqref{Eq_Gaussian_propagator} over the probed diffusion time $\tau$. This does not necessarily imply that said system must be isotropic, as the aforementioned Gaussian function can be generalized to an anisotropic medium (see Eq.~\eqref{Eq_Gaussian_propagator_DTI}).
\end{mainpoint}

\subsection{Restricted diffusion and apparent diffusion coefficient}
\label{Sec_restriction}

\begin{figure}[h!]
\begin{center}
\includegraphics[width=\textwidth]{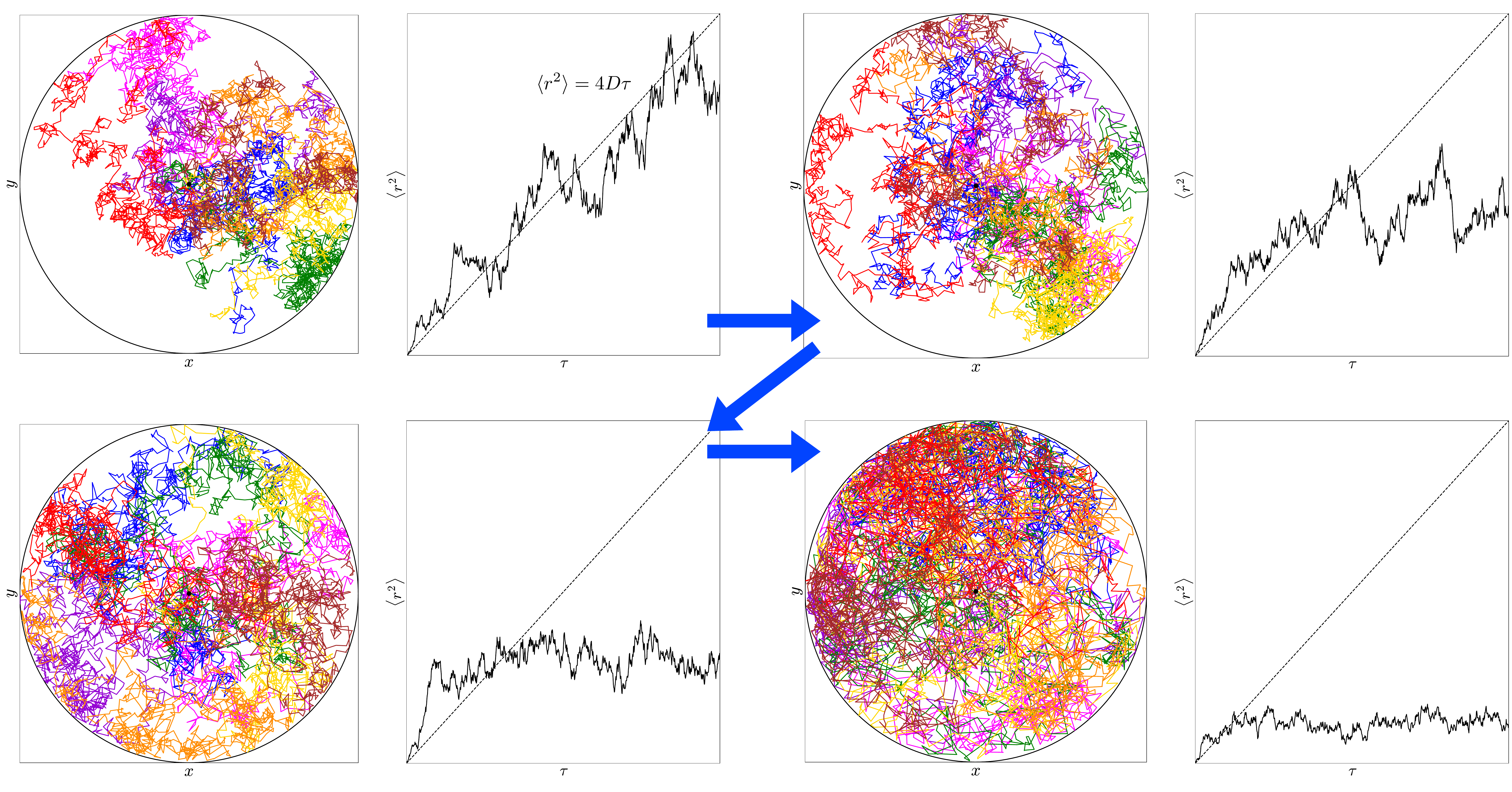}
\caption{Collections of random walks similar to those of Fig.~\ref{Fig_brownian}, with an additional circular barrier to diffusion. Restriction increases along the blue arrows. The more restricted the medium, the faster the mean squared displacement $\langle r^2 \rangle$ saturates below the analytic result Eq.~\eqref{Eq_MSD_d}.}
\label{Fig_brownian_restricted}
\end{center}
\end{figure}

In the presence of restriction in the diffusion medium, the diffusion coefficient becomes time-dependent, as shown in Fig.~\ref{Fig_brownian_restricted}. Indeed, as a result of restriction, the mean squared displacement deviates from the analytic result Eq.~\eqref{Eq_MSD_d} and saturates at long diffusion times. The more restricted the medium, the shorter the timescale $\tau_\text{long}$ after which the long diffusion time regime sets in. If one probes a diffusion time such that $\tau \geq \tau_\text{long}$, an apparent diffusion coefficient $D(\tau)$ will be measured instead of the ``true" diffusion coefficient $D$.\\

\begin{sidenote}{title = Diffusion times and diffusion gradient}
In Sec.~\ref{Sec_MRI_sequences}, we will see that similar diffusion contrasts can be obtained in MRI by varying either the strength of the diffusion gradients or their duration. However, from the standpoint of restriction, these two options are widely unequal. This justifies why comparing two diffusion acquisition schemes should be done at the level of the gradients' strength and duration, and not just at the $b$-value level (see Eq.~\eqref{Eq_b_value}).
\end{sidenote}

\section{Diffusion MRI}
\label{Sec_dMRI}

\subsection{Diffusion-weighted imaging (DWI)}

The DWI derivation below comes from a version of the calculations contained in Ref.~\cite{CampbellThesis:2004}. It intends to start from a discrete brownian motion to describe diffusion and to retrieve, after going to the continuous limit, the typical diffusion-induced signal attenuation of DWI.

\subsubsection{Discrete brownian motion}
\label{Sec_brownian_discrete}

Let us imagine the situation where a given spin-$1/2$ particle diffuses in a medium subjected to a magnetic field oriented along the $z$ axis. By applying a linear field gradient $\mathbf{G}$, the value of the magnetic field changes through space as
\begin{equation}
B(\mathbf{r}) = B_0 + \mathbf{G}\cdot \mathbf{r}\, ,
\label{Eq_B_gradient}
\end{equation}
where $\mathbf{r}$ denotes the position. Now, one considers that the diffusion of the particle is in fact described by a simple discrete brownian motion. From $t=0$ to the path duration $\tau$, the particle makes $N$ steps of equal duration $t_d = \tau/N$. After a given step $i$, the particle's move is represented by the vector $r_d\, \mathbf{u}_i$, where $r_d$ is a constant length and $\mathbf{u}_i$ is a unitary vector that is drawn from the same probability distribution at each step. Assuming the particle starts at position $\mathbf{r} = \mathbf{0}$ at $t=0$, the particle's position after $j$ steps is thus given by
\begin{equation}
\mathbf{r}_j = \mathbf{r}(jt_d) = r_d \sum_{i=1}^j \mathbf{u}_i\, .
\end{equation}
From Eq.~\eqref{Eq_B_gradient}, one finds the magnetic field perceived by the diffusing particle at this $j$-th step,
\begin{equation}
B(jt_d) = B_0 + \mathbf{G}\cdot\mathbf{r}_j = B_0 + r_d \sum_{i=1}^j \mathbf{G}\cdot\mathbf{u}_i\, ,
\end{equation}
and the associated Larmor frequency,
\begin{equation}
\omega_0^\prime (jt_d) = \gamma B(jt_d) = \omega_0 + \underbrace{\gamma r_d \sum_{i=1}^j \mathbf{G}\cdot\mathbf{u}_i}_{\Delta \omega_0(jt_d)} \, .
\end{equation}
Consequently, the field gradient imposes that the particle's dephasing rate changes throughout its motion.\\

The total accrued phase after the $N$ steps (at time $t=\tau$) writes
\begin{equation}
\phi = \sum_{j=1}^N \Delta \omega_0(jt_d) \, t_d = \gamma t_d r_d \sum_{j=1}^N\sum_{i=1}^j \mathbf{G}\cdot\mathbf{u}_i\, .
\label{Eq_accrued_phase_before}
\end{equation}
Let us rearrange the summations of Eq.~\eqref{Eq_accrued_phase_before}. First, one can notice that any $i$ term ($\mathbf{G}\cdot\mathbf{u}_i$) occurs a well-defined number of times in these summations. Indeed, the $i=1$ term occurs $N$ times, the $i=2$ term occurs $N-1$ times, the $i=3$ term occurs $N-2$ times and so on, until the $i=N$ term that occurs only $1$ time. Knowing that the number of occurrences of the $i$ term can be generalized as $\sum_{j=i}^N 1$, one has
\begin{equation}
\phi = \gamma t_d r_d \sum_{i=1}^N\sum_{j=i}^N \mathbf{G}\cdot\mathbf{u}_i\, .
\label{Eq_reminder_discrete}
\end{equation}
Second, the simple change of variable $i = N+1 -p$ gives
\begin{equation}
\phi = \gamma t_d r_d \sum_{p=1}^N\sum_{j=N+1-p}^N \mathbf{G}\cdot\mathbf{u}_{N+1-p}\, ,
\end{equation}
so that the total accrued phase is finally
\begin{equation}
\phi = \gamma t_d r_d \sum_{p=1}^N p\, \mathbf{G}\cdot\mathbf{u}_{N+1-p}\, .
\end{equation}
\smallskip

The total accrued phase can now be seen as a sum over a large number ($N$) of random variables ($\mathbf{G}\cdot\mathbf{u}_i$) all drawn from the same probability distribution. The central limit theorem then ensures that this phase has a Gaussian distribution:
\begin{equation}
\mathcal{P}(\phi) = \frac{1}{\sqrt{2\pi (\langle \phi^2\rangle - \langle \phi\rangle^2)}}\, \exp^{-(\phi - \langle \phi\rangle)^2/2(\langle \phi^2\rangle - \langle \phi\rangle^2)}\, .
\end{equation}
Since there is no accrued phase at the magnet isocenter ($\mathbf{r} = \mathbf{0}$) and diffusion is symmetric (diffusing along a certain direction is as likely as diffusing in the opposite direction), one must have $\langle\phi\rangle = 0$, so that
\begin{equation}
\mathcal{P}(\phi) = \frac{1}{\sqrt{2\pi \langle \phi^2\rangle }}\, \exp^{-\phi^2/2\langle \phi^2\rangle}\, .
\label{Eq_phase_distribution}
\end{equation}
At this point, the overall DWI signal decay can be understood as a consequence of the average phase dispersion (inducing the decay of the net magnetization through time):
\begin{equation}
\mathcal{S} = \mathcal{S}_0 \, \langle\exp^{i\phi}\rangle = \mathcal{S}_0 \int_{-\infty}^{+\infty} \! \exp^{i\phi}\, \mathcal{P}(\phi) \,\mathrm{d}\phi\, .
\end{equation}
From the phase distribution Eq.~\eqref{Eq_phase_distribution}, one finally obtains
\begin{equation}
\mathcal{S} = \mathcal{S}_0 \, \exp^{-\langle \phi^2\rangle/2} \, .
\label{Eq_signal_phase_dispersion}
\end{equation}

\begin{mainpoint}{title = Physical origin of the diffusion-induced signal decay}
The signal decay originates from the diffusion-induced spin dephasing through the finite term $\langle \phi^2\rangle$. In a way, diffusion is analog to $T_2$ relaxation because it also stems from a spin dephasing caused by a varying local $B_0$ field, except that this ``diffusion relaxation" is artificially induced by the applied gradients.
\end{mainpoint}

\subsubsection{Continuous limit - $\mathbold{b}$-value}
\label{Sec_brownian_continuous}

To get closer to realistic things, let us take the continuous limit where the steps of the brownian motion are not discrete anymore, and let us allow the gradient to vary in time. In that case, the accrued phase at the end of the path (time $\tau$) writes
\begin{equation}
\phi(\tau) = \gamma\int_0^\tau \! \mathbf{G}(t) \cdot \mathbf{r}(t) \, \mathrm{d}t \qquad \left( \equiv \gamma t_d \sum_{j=1}^N \mathbf{G}\cdot \mathbf{r}_j \right)\, .
\end{equation}
where the terms in brackets emphasize the link between the continuous case and the discrete one Eq.~\eqref{Eq_reminder_discrete}. By integrating by parts, one obtains
\begin{align}
\phi(\tau) & = \left[ \gamma\left( \int_0^t \! \mathbf{G}(t')\,\mathrm{d}t' \right) \cdot \mathbf{r}(t) \right]_0^\tau - \gamma\int_0^\tau \! \left( \int_0^t\! \mathbf{G}(t')\,\mathrm{d}t' \right) \cdot \mathbf{v}(t)\,\mathrm{d}t \nonumber \\
 & = \left( \gamma\int_0^\tau \! \mathbf{G}(t')\,\mathrm{d}t' \right) \cdot \mathbf{r}(\tau) - \int_0^\tau \! \left( \gamma \int_0^t\! \mathbf{G}(t')\,\mathrm{d}t' \right) \cdot \mathbf{v}(t)\,\mathrm{d}t \, ,
\end{align}
where the velocity $\mathbf{v}(t)$ is, as usual, the time-derivative of the displacement $\mathbf{r}(t)$. Introducing the spin-dephasing vector 
\begin{equation}
\cbeq{
\mathbf{q}(t) = \gamma\int_0^t \! \mathbf{G}(t') \, \mathrm{d}t'
}\; ,
\label{Eq_q_vector}
\end{equation}
one has
\begin{equation}
\phi(\tau) = \mathbf{q}(\tau) \cdot \mathbf{r}(\tau) - \int_0^\tau \! \mathbf{q}(t) \cdot \mathbf{v}(t)\,\mathrm{d}t\, .
\end{equation}
However, the diffusion gradients usually satisfy the echo condition
\begin{equation}
\mathbf{q}(\tau) = \gamma\int_0^\tau \! \mathbf{G}(t) \, \mathrm{d}t = \mathbf{0}
\end{equation}
at the moment of the measurement, to ensure that no artificial dephasing is picked up in the measured signal. Thus, the total accrued phase writes
\begin{equation}
\phi(\tau) = - \int_0^\tau \! \mathbf{q}(t) \cdot \mathbf{v}(t)\,\mathrm{d}t\, .
\end{equation}

Going back to the expression Eq.~\eqref{Eq_signal_phase_dispersion} for the DWI signal attenuation, one still needs to work out $\langle \phi^2\rangle \equiv \langle\phi^2(\tau)\rangle$. Writing
\begin{equation}
\langle\phi^2(\tau)\rangle = \int_0^\tau\int_0^\tau\! q(t)\,q(t')\, \langle v_q(t)\, v_q(t')\rangle\, \mathrm{d}t\,\mathrm{d}t' \, ,
\end{equation}
where $v_q (t)$ denotes the projection of $\mathbf{v}(t)$ along $\mathbf{q}(t)$, and using the fact that the velocity autocorrelation function expresses in the case of a 1D Gaussian diffusion process as Eq.~\eqref{Eq_auto_velocity_correlation}, one has
\begin{equation}
\langle \phi^2(\tau)\rangle = 2D \int_0^\tau q^2(t)\, \mathrm{d}t\, .
\label{Eq_phi_square}
\end{equation}
Defining the $b$-value
\begin{equation}
\cbeq{
b = \int_0^\tau q^2(t)\, \mathrm{d}t
}
\label{Eq_b_value}
\end{equation}
and using the expressions Eqs.~\eqref{Eq_signal_phase_dispersion} and \eqref{Eq_phi_square}, one finally retrieves the typical DWI signal attenuation:
\begin{equation}
\cbeq{
\mathcal{S} = \mathcal{S}_0 \, \exp^{-bD}
}\; .
\label{Eq_diffusion_attenuation}
\end{equation}

\subsubsection{$\mathbold{b}$-value of archetypal MRI sequences}
\label{Sec_MRI_sequences}

One could imagine a very simple sequence based on a 90$^\circ$ RF pulse paired with a slice-select gradient (super short so that no diffusion weighting has to be taken into account) and followed by a time-independent gradient $\mathbf{G}$ (constant magnitude $G$). In that case, the spin dephasing vector and the $b$-value are 
\begin{align}
\mathbf{q}(t) & = \gamma\int_0^t \! \mathbf{G}(t') \, \mathrm{d}t' = \gamma \mathbf{G}t \\
b & = \int_0^\tau q^2(t)\, \mathrm{d}t = \gamma^2G^2 \int_0^\tau\! t^2\,\mathrm{d}t = \frac{\gamma^2G^2\tau^3}{3}\; .
\end{align}
Since the $b$-value increases with diffusion time and causes the diffusion-induced signal attenuation in DWI, one can interpret the $b$-value as the amount of diffusion weighting in an MR sequence. However, increasing the $b$-value by increasing $G$ or by increasing $\tau$ are not equivalent options, as discussed in Sec.~\ref{Sec_restriction}.\\

\begin{figure}[h!]
\begin{center}
\includegraphics[width=0.9\textwidth]{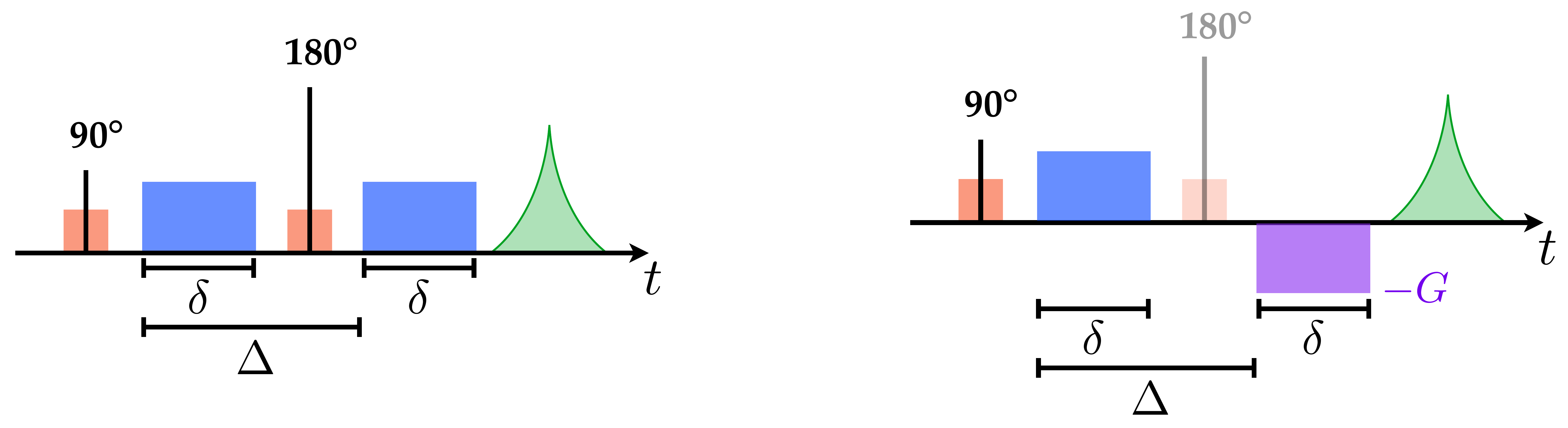}
\caption{\textbf{Left:} Stejskal-Tanner MRI sequence. While the 90$^\circ$ and 180$^\circ$ RF pulses are indicated by black vertical lines, the slice-select gradient and the diffusion gradient are represented in red and blue, respectively. \textbf{Right:} Modified version of the sequence, showing the effective role of the second diffusion gradient after the 180$^\circ$ RF pulse.}
\label{Fig_ST_sequence}
\end{center}
\end{figure}

Now, let us study the celebrated Stejskal-Tanner sequence \cite{Stejskal_Tanner:1965}. Since its development in 1965, it has been the cornerstone of diffusion MRI sequences. It consists in a traditional spin echo sequence where two diffusion gradients pulses of equal duration $\delta$ and amplitude $G$ are inserted, as shown in Fig.~\ref{Fig_ST_sequence}. In that type of sequence, ensuring the spin echo requires that the direct effect of dephasing by the diffusion gradients cancels out at the echo time $t= \mathrm{TE}$. This is not directly apparent in the left-hand side of Fig.~\ref{Fig_ST_sequence}. However, the right-hand side of Fig.~\ref{Fig_ST_sequence} reveals that the 180$^\circ$ RF pulse acts so that the second gradient effectively induces an opposite dephasing to the one induced by the first gradient, actually giving the echo condition
\begin{equation}
\cbeq{
\mathbf{q}(\tau = \mathrm{TE}) = \gamma\int_0^{\mathrm{TE}} \! \mathbf{G}(t)\,\mathrm{d}t = \mathbf{0}
} \; .
\label{Eq_echo_condition}
\end{equation}
The reader is strongly encouraged to play with this kind of sequence on the \href{http://blog.ismrm.org/2017/06/06/dwe-part-2}{ISMRM blog}.\\

\begin{figure}[h!]
\begin{center}
\includegraphics[width=0.4\textwidth]{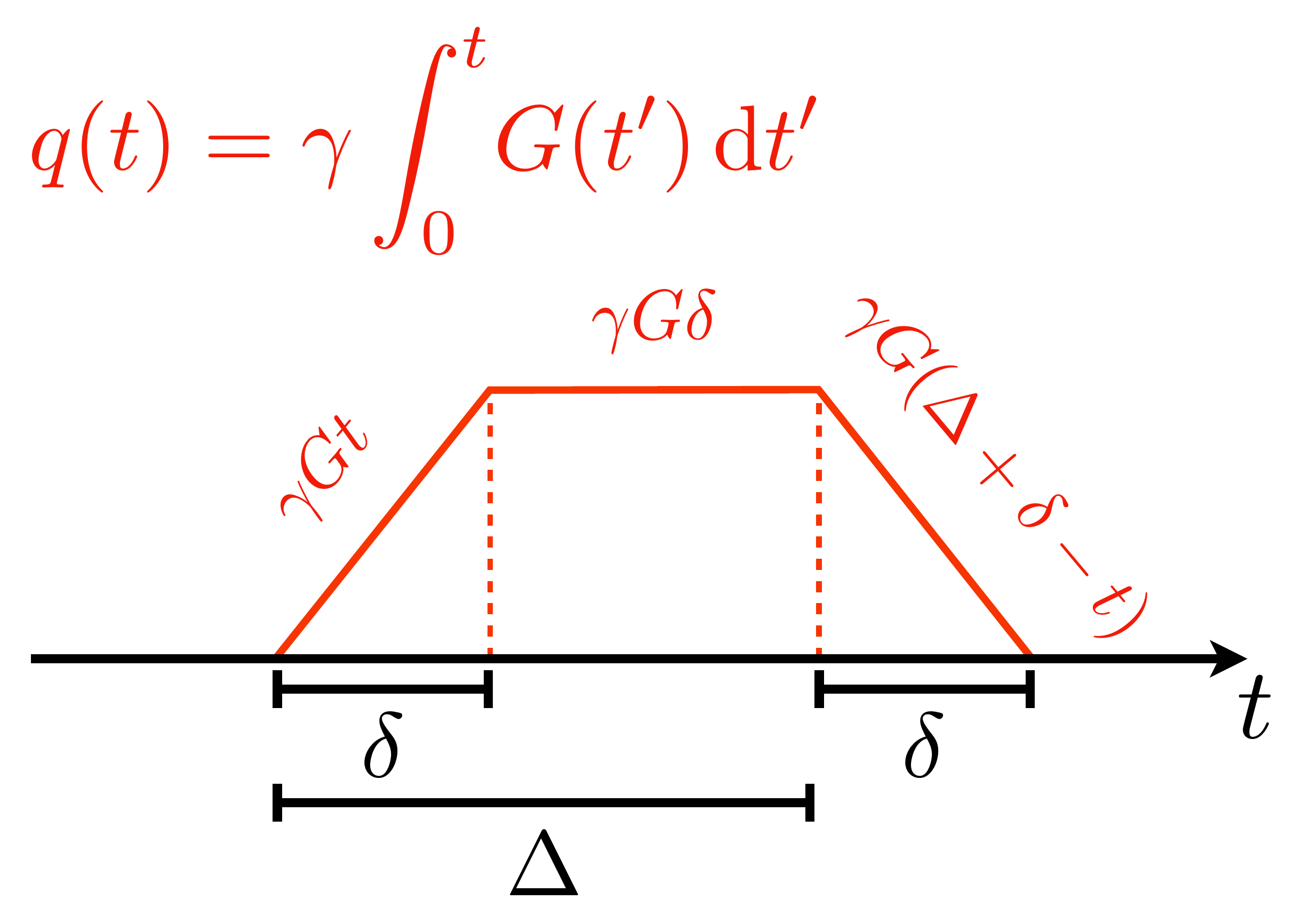}
\caption{Expressions of the spin dephasing vector $q(t)$ during the Stejskal-Tanner sequence.}
\label{Fig_calcul_ST}
\end{center}
\end{figure}

Fig.~\ref{Fig_calcul_ST} enables the calculation of the $b$-value associated to a Stejskal-Tanner sequence. One has
\begin{align}
b & = \int_0^{\Delta + \delta} \! q^2(t)\,\mathrm{d}t \nonumber \\
 & = \int_0^\delta \! q^2(t)\,\mathrm{d}t + \int_\delta^\Delta \! q^2(t)\,\mathrm{d}t + \int_\Delta^{\Delta + \delta} \! q^2(t)\,\mathrm{d}t \nonumber \\
 & = \gamma^2 G^2 \left[\int_0^\delta \! t^2\,\mathrm{d}t + \int_\delta^\Delta \! \delta^2\,\mathrm{d}t + \int_\Delta^{\Delta + \delta} \! (\Delta +\delta - t)^2\,\mathrm{d}t \right] \nonumber \\
 & = \gamma^2 G^2 \delta^2 \left[ \Delta - \frac{\delta}{3} \right] \, ,
\end{align}
where the time between square brackets is usually called the ``effective diffusion time" of the Stejskal-Tanner sequence. \\

\newpage

\newpage

\subsection{Diffusion tensor imaging (DTI)}

\subsubsection{General context - Diffusion tensor}
\label{Sec_Diffusion_tensor_DTI}

\begin{figure}[h!]
\begin{center}
\includegraphics[width=0.8\textwidth]{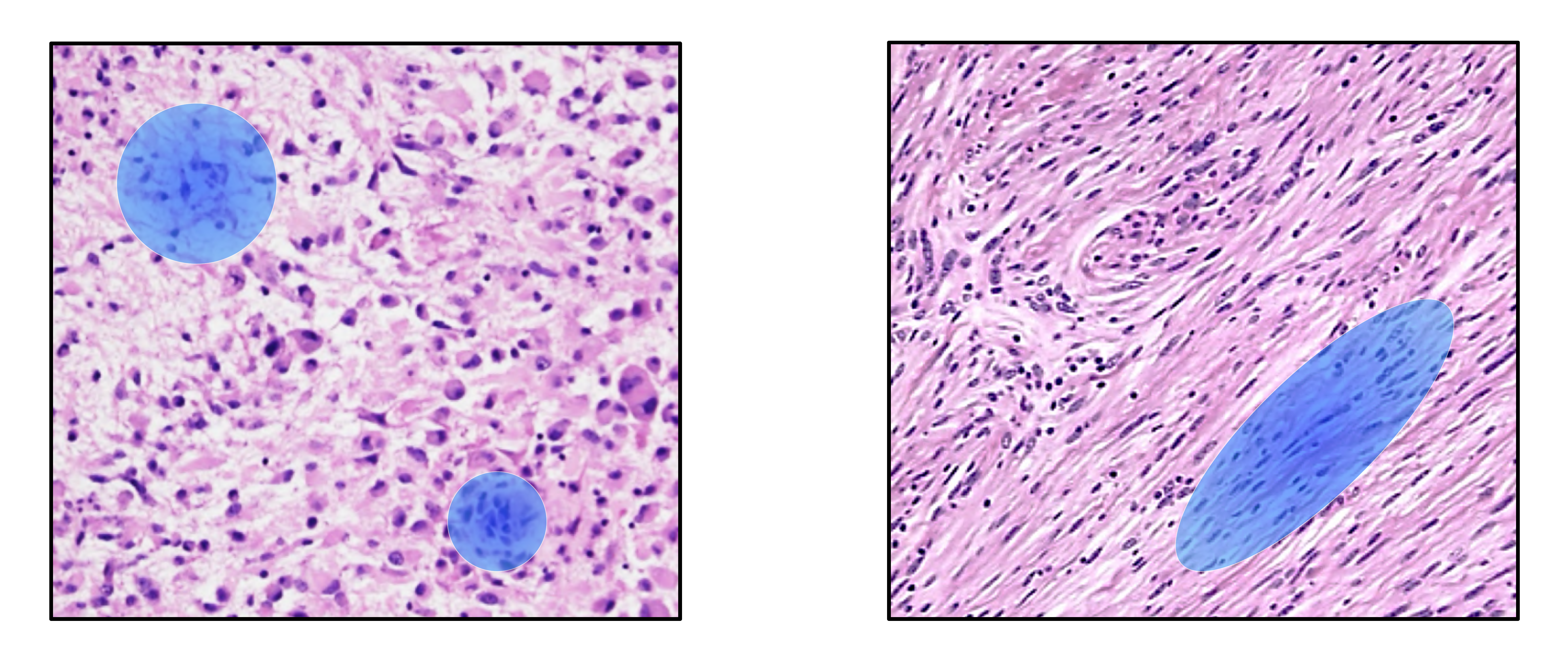}
\caption{\textbf{Left:} Typical glioma structure, where diffusion occurs with varying isotropy at the micrometer scale, hence the description in terms of isotropic diffusion tensors (blue glyphs). \textbf{Right:} Typical meningioma structure, where diffusion occurs anisotropically at the micrometer scale (hence the anisotropic diffusion tensor), but isotropically at the millimeter scale, due to random orientation of the fibrous substructures. Adapted from Ref.~\cite{Szczepankiewicz:2016}.}
\label{Figure_diffusion_biology}
\end{center}
\end{figure}

Developed in 1994, DTI encodes diffusion within a semipositive-definite diffusion tensor that provides a voxel-scale average of intra-voxel diffusion \cite{Basser:1994}
\begin{equation}
\langle\mathbf{D}\rangle =  \begin{pmatrix}
D_{xx} & D_{xy} & D_{xz} \\
D_{yx} & D_{yy} & D_{yz} \\
D_{zx} & D_{zy} & D_{zz}
\end{pmatrix} 
\equiv 
\begin{pmatrix}
D_{xx} & D_{xy} & D_{xz} \\
\cdot & D_{yy} & D_{yz} \\
\cdot & \cdot & D_{zz}
\end{pmatrix}
\label{Eq_diffusion tensor}
\end{equation}
instead of a diffusion constant $D$. The dots in the above formula simply highlight the fact that the diffusion tensor is symmetric (diffusing along a certain direction is as likely as diffusing in the opposite direction). By using a diffusion tensor, DTI enables the use of multiple gradient directions to estimate the intra-voxel diffusion profile. DTI assumes that diffusion is well described by a Gaussian propagator of the form
\begin{equation}
\mathcal{P}(\mathbf{r},\tau) = \frac{1}{\sqrt{(4\pi \tau)^3\, \mathrm{Det}(\langle\mathbf{D}\rangle)}}\, \mathrm{exp}\!\left(-\frac{\mathbf{r}^{\mathrm{T}}\cdot\langle\mathbf{D}\rangle^{-1}\cdot\mathbf{r}}{4\tau}\right) \, ,
\label{Eq_Gaussian_propagator_DTI}
\end{equation}
which resembles the Gaussian propagator obtained by Einstein at the beginning of the 20$^\text{th}$ century (see Eq.~\eqref{Eq_Gaussian_propagator}). Here, the exponent ``$\mathrm{T}$" denotes vector/matrix transposition.  Fig.~\ref{Figure_diffusion_biology} presents two examples of diffusion in the unhealthy human brain (tumors). The reader should already note that a typical voxel volume ($8\;\mathrm{mm}^3$) encapsulates many micrometer-scale cell structures.\\

\begin{figure}[h!]
\begin{center}
\includegraphics[width=0.25\textwidth]{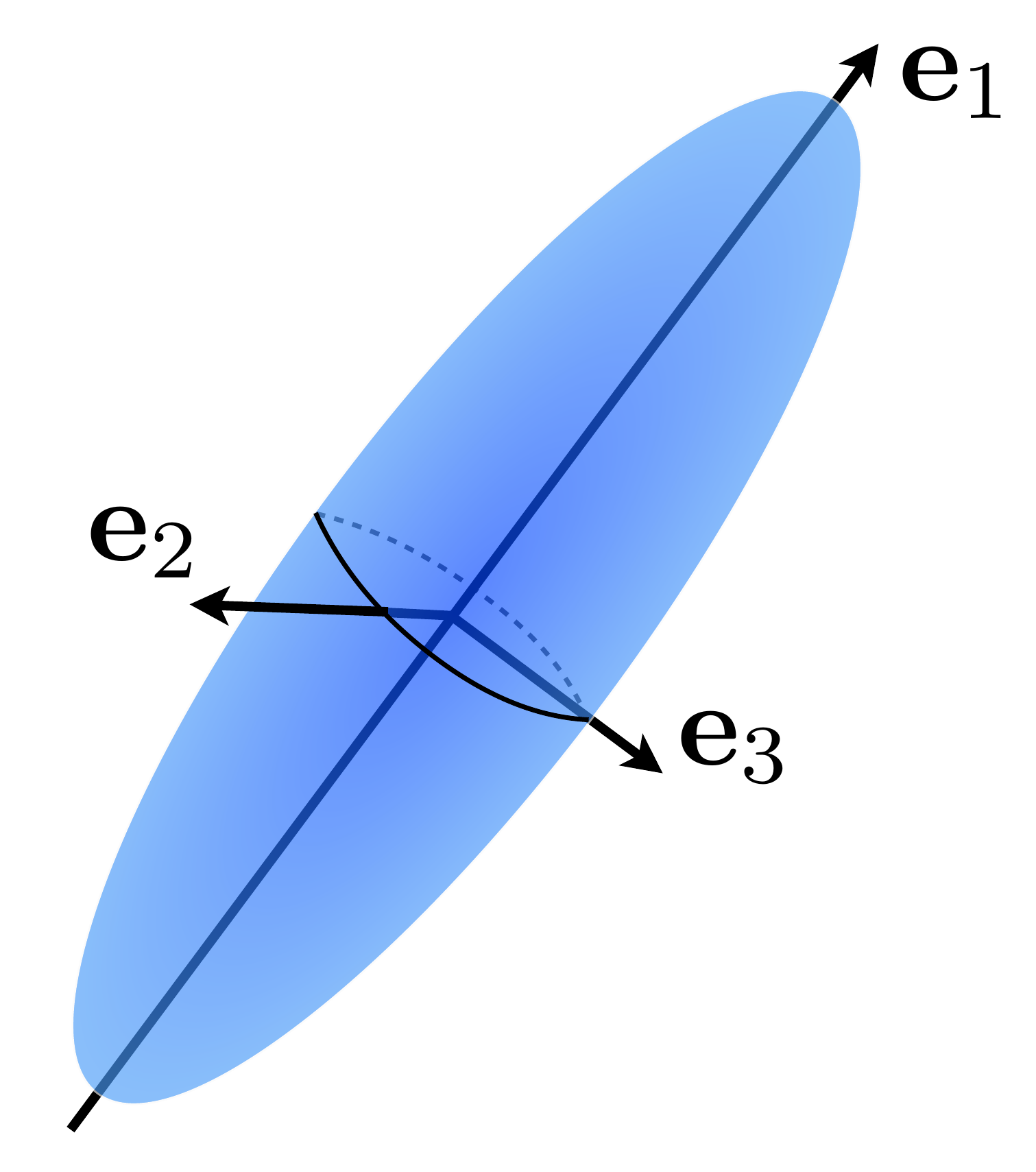}
\caption{Definition of the diffusion tensor's principal axes, given by its eigenvectors $\mathbf{e}_i$. Here, the eigenvalues are such that $\lambda_1 > \lambda_2 = \lambda_3$.}
\label{Figure_tensor}
\end{center}
\end{figure}

As for any semipositive-definite tensor, it is possible to diagonalize the diffusion tensor:
\begin{equation}
\cbeq{
\langle\mathbf{D}\rangle = 
\begin{pmatrix}
D_{xx} & D_{xy} & D_{xz} \\
\cdot & D_{yy} & D_{yz} \\
\cdot & \cdot & D_{zz}
\end{pmatrix}_{\{ \mathbf{u}_x,\mathbf{u}_y,\mathbf{u}_z \}}
\equiv
\begin{pmatrix}
\lambda_1 & 0 & 0 \\
0 & \lambda_2 & 0 \\
0 & 0 & \lambda_3
\end{pmatrix}_{\{ \mathbf{e}_1,\mathbf{e}_2,\mathbf{e}_3 \}}
} \; ,
\end{equation}
which defines its eigenvalues $\lambda_i$ and eigenvectors $\mathbf{e}_i$. While the eigenvectors gives the main axes of the diffusion tensor, the eigenvalues informs on the ``size" of the tensor along these axes, as illustrated in Fig.~\ref{Figure_tensor}. Since the diffusion tensor is symmetric semipositive-definite, we are sure that all its eigenvalues are positive real number and that its eigenvectors are orthogonal to one another. The eigenvalues of the diffusion tensor are used to create a set of metrics describing any diffusion tensor, such as the mean diffusivity (or apparent diffusion coefficient)
\begin{equation}
\mathrm{MD} = \overline{\lambda} = \frac{\lambda_1 + \lambda_2 + \lambda_3}{3} = \frac{\mathrm{Tr}(\langle\mathbf{D}\rangle)}{3}\, ,
\label{Eq_MD}
\end{equation}
expressed in diffusivity unit ($\mathrm{mm}^2/\mathrm{s}$ for instance), or the unitless fractional anisotropy
\begin{equation}
\mathrm{FA} = \sqrt{\frac{3}{2}}\, \sqrt{\frac{\sum_\nu (\lambda_\nu - \overline{\lambda})^2}{\sum_\nu \lambda_\nu^2}}\quad \in [0,1]\, .
\label{Eq_FA}
\end{equation}
\medskip

\begin{mainpoint}{title = When diffusion MRI goes clinical}
While diffusion MRI was propelled as a clinical tool for its unrivalled sensitivity to tissue disruption in cerebral ischemia \cite{Moseley:1990a}, DTI has proven to be a powerful tool because it provides several parameters with seemingly intuitive interpretations. For example, during brain maturation, reduced MD and increased FA in the white matter is interpreted as axon myelination \cite{Lebel:2008}, and the anisotropy serves as a marker of healthy development.
\end{mainpoint}

\subsubsection{DTI's limitations}
\label{Sec_DTI_limitations}

Since DTI only provides a voxel-scale average of the intra-voxel diffusion in Eq.~\eqref{Eq_diffusion tensor}, it shows poor specificity in depicting the precise nature of microstructural tissue changes \cite{Alexander:2001, Alexander:2007, Jones_Cercignani:2010, Jones:2013}, especially in voxels of crossing fascicles \cite{Pierpaoli:1996, Douaud:2011}, which make up 60 to 90\% of the brain \cite{Jeurissen:2013}. This impedes its widespread implementation in a clinical setting. Let us try to understand this from a mathematical standpoint.\\

\begin{figure}[h!]
\begin{center}
\includegraphics[width=0.85\textwidth]{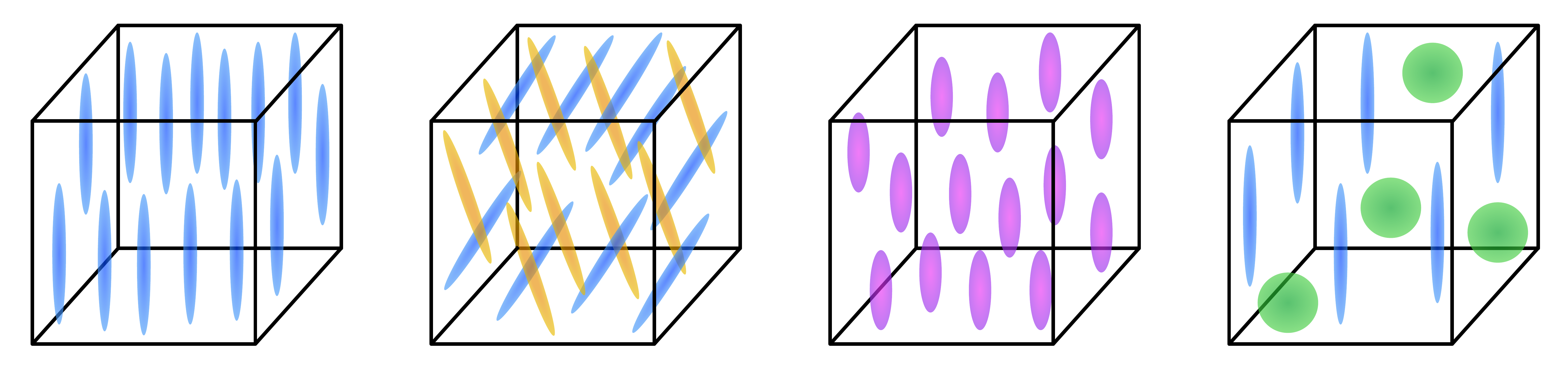}
\caption{Voxels associated to typical brain tissues. \textbf{From left to right:} White matter, crossing fibers, demyelination and white matter inflammation.}
\label{Figure_DTI_Voxels}
\end{center}
\end{figure}

Within DTI, the diffusion-induced signal attenuation Eq.~\eqref{Eq_diffusion_attenuation} can be rewritten as
\begin{equation}
\cbeq{
\mathcal{S}_\text{DTI} = \mathcal{S}_0 \, \exp^{-b\, \mathbf{n}^\mathrm{T} \cdot\langle\mathbf{D}\rangle\cdot \mathbf{n}} 
}\; ,
\label{Eq_DTI_attenuation}
\end{equation}
where $\mathbf{n} = \mathbf{q}/\Vert \mathbf{q} \Vert$ is the unit vector associated to the spin dephasing vector (or $q$-vector) defined in Eq.~\eqref{Eq_q_vector}, which makes $\mathbf{n}^\mathrm{T} \cdot\langle\mathbf{D}\rangle\cdot \mathbf{n}$ the effective average diffusivity along $\mathbf{q}$. However, the heterogeneity of most investigated samples over the voxel scale implies that the measured MR signal is actually the sum of signals arising from a variety of microscopic environments:
\begin{equation}
\cbeq{
\mathcal{S} = \mathcal{S}_0 \int \! \mathcal{P}(\mathbf{D})\, \exp^{-b\, \mathbf{n}^\mathrm{T} \cdot\mathbf{D}\cdot \mathbf{n}} \, \mathrm{d}\mathbf{D}
} \; ,
\label{Eq_signal_tensor_distribution}
\end{equation}
which reads as the Laplace transform of the diffusion tensor distribution $\mathcal{P}(\mathbf{D})$. While computing the diffusion signal from the diffusion tensor distribution is simple (by computing the integral), estimating the diffusion tensor distribution from the diffusion signal (by inverting the above equation) is very difficult. Indeed, the Laplace transform is very well known as an ill-conditioned inverse problem: an infinity of different distributions $\mathcal{P}(\mathbf{D})$ fit the acquired data $\mathcal{S}$ within a reasonable error bar if not enough information (from directions, $b$-values, \textit{etc.}) is provided. Each solution to the Laplace inversion in turn becomes a potential biological scenario to explain the intra-voxel content. DTI constrains the search for the ``right" solution by only considering the intra-voxel average of the diffusion profile, but in doing so it loses its specificity. To highlight that, Fig.~\ref{Figure_DTI_Voxels} presents archetypal voxels associated to typical brain tissues. Despite their striking differences, they are all basically described by the same voxel-averaged diffusion tensor (same MD and same FA) within DTI. Hence, these voxels correspond to identical DTI signals! \\

\begin{sidenote}{title = The ``meat grinder" analogy}
A very visual analogy for any ill-conditioned inverse problem reads as follows: it is easy to grind meat but rather difficult to put the resulting pieces of meat back together afterward. In the same fashion, it is easy to ``grind" the tensor distribution into a signal but rather difficult to put the tensor distribution back together from the signal.
\end{sidenote}

\begin{mainpoint}{title = DTI's lack of specificity}
DTI is inherently unspecific because it only considers the voxel-scale average of the intra-voxel diffusion profile, which can lead to misleading biological interpretations, as shown in Fig.~\ref{Figure_DTI_Voxels}. Without this DTI approximation, actually performing the Laplace inversion Eq.~\eqref{Eq_signal_tensor_distribution} using only linear (unidirectional $q$-vector) gradients is very ill-conditioned for complex voxel populations.
\end{mainpoint}

\subsection{Validity of the Gaussian approximation}
\label{Sec_validity_gaussian}

Most of this section is drawn from Ref.~\cite{SzczepankiewiczThesis:2016}.\\

The diffusion tensor distribution (DTD) model encapsulated in Eq.~\eqref{Eq_signal_tensor_distribution} is actually based on two main assumptions. First, the diffusion in each microscopic environment within the probed voxel is approximately Gaussian. Second, the diffusing particles do not exchange between different microscopic environments during the diffusion encoding. Let us describe the ramifications and validity of these assumptions.

\subsubsection{Non-Gaussian diffusion}

Particles obey Gaussian diffusion only if they propagate in a homogeneous medium that interacts only with itself, such as an infinite body of pure water. Therefore, the diffusion is non-Gaussian \textit{per se} in biological tissues where heterogeneity and obstacles are obvious characteristics \cite{deSwiet_Mitra:1996}, which contradicts the first assumption of the DTD model \cite{Beaulieu:2002}. Let us discuss how this first assumption interacts with three aspects of non-Gaussian diffusion, namely:
\begin{itemize}
\item[\textbf{(1)}] the presence of multiple Gaussian intra-voxel components;
\item[\textbf{(2)}] non-Gaussian phase dispersion;
\item[\textbf{(3)}] time-dependent diffusion.
\end{itemize}
\bigskip

\textbf{(1)} The presence of multiple components with Gaussian diffusion is permitted by the DTD model since it models each component in terms of a diffusion tensor \cite{deSwiet_Mitra:1996, Yablonskiy:2003}. As long as the diffusion in each microscopic environment is approximately Gaussian, the first assumption holds. \\

\textbf{(2)} A non-Gaussian phase distribution may occur, for example, where there is restricted diffusion \cite{Callaghan:1991}, which in turn invalidates the simple exponential relation between the signal and the diffusivity assumed by the DTD model (see the calculations in Secs.~\ref{Sec_brownian_discrete} and \ref{Sec_brownian_continuous}). However, the effects of a non-Gaussian phase distribution are small for moderate attenuation, \textit{i.e.} if the signal is not attenuated below 10\% \cite{Topgaard_Soderman:2003}. This aspect of non-Gaussian diffusion should therefore be negligible in biological tissue at moderate encoding strengths \cite{Nilsson:2010}.\\

\textbf{(3)} Time-dependent diffusivity is caused by an interaction between the geometry of the biophysical content of the diffusing particle and the time during which the diffusion is observed \cite{Stejskal_Tanner:1965, Gore:2010}. For restricted diffusion, the apparent diffusion coefficient (ADC) may therefore depend on the size of the restriction $d$ and the diffusion time $\tau$. Let us denote $D_0$ the intrinsic diffusivity of the diffusion medium. In the regime where $\tau \ll d^2/D_0$, the diffusing particles do not have time to experience the restriction, and the ADC approaches the intrinsic diffusivity $D_0$ \cite{Woessner:1963}. By contrast, when $\tau \gg d^2/D_0$, the restrictions have been probed by most particles and the ADC approaches zero. For these two regimes, the approximation of Gaussian diffusion in each microscopic environment holds. However, in the intermediate regime, when $\tau \sim d^2/D_0$, the ADC will be a function of $\tau$ and $d$, and the diffusion must instead be described by a time-dependent diffusion tensor. A similar dependency exists for hindered diffusion, where the apparent diffusivity transitions from $D_0$ to a lower diffusivity defined by the tortuosity of the environment \cite{Beck_Schultz:1970}. Several studies have demonstrated time-dependent diffusion in neural tissues \cite{Stanisz:1997, Assaf:2000, Does:2003, Assaf:2008, Lundell:2015, Burcaw:2015}, but the effect is probably small for the diffusion times commonly used in conventional experiments \textit{in vivo} \cite{Clark:2001, Ronen:2006, Nilsson:2009, deSantis:2016}. Let us therefore assume that the DTD model is sufficiently accurate to capture the essentials of the diffusion characteristics in tissue, and acknowledge that this assumption must be validated in future studies.


\subsubsection{Exchange}
\label{Sec_exchange}

Diffusing particles may visit multiple microscopic environments during the diffusion time by passing through permeable boundaries that separate the environments. Although exchange is always present up to some degree, effects of exchange can be disregarded within three regimes. The first two encompass the cases where the residence time $t_\text{res.}$ is much longer, or much shorter, than the diffusion time, $t_\text{res.} \ll \tau$ or $t_\text{res.} \gg \tau$, \textit{i.e.} if very few particles have time to exchange or if the time spent in a specific environment is very short \cite{Quirk:2003}. The last regime of interest corresponds to the case where the diffusion characteristics of the exchanging environments are approximately equal, in which case these environments are accurately described by a single diffusion tensor.\\

Effects of exchange have been investigated in the context of diffusion MRI \cite{NilssonThesis:2011, Nilsson:2013b}, and several studies have indicated that the exchange in healthy brain tissue has a negligible effect on the diffusion-weighted signal for conventional diffusion times \cite{Nilsson:2013a, Lampinen:2017}. However, such assumptions may not hold in diseased tissue, where effects of exchange have been demonstrated \cite{Latt:2009}. In a preliminary study of the exchange rate in tumors, it has recently been observed relatively long residence times in the tissue \cite{Lampinen:2017}. Let us therefore assume that exchange has a negligible effect in both healthy tissue and tumoral tissue in the context of this document.\\

\begin{sidenote}{title = More mathematical descriptions of diffusion processes}
May the reader be interested in more mathematical/physical aspects of diffusion, Refs.~\cite{Novikov:2010} and \cite{Kiselev:2017} feature some powerful tools such as the Green's functions and self-energies from high-energy and condensed matter physics. Besides, Refs.~\cite{Novikov:2011} and \cite{Novikov:2014} use the renormalization group and universality classes based on critical exponents (from the same aforementioned fields) to elegantly extract information about the dimension and disorder of biological tissues from time-dependent diffusion.
\end{sidenote}

\section{Symmetric tensors}
\label{Section_sym_tensors}

\subsection{Tensor size and shape}

This section provides several conventions for parameterizing 3$\times$3 symmetric tensors, such as the diffusion tensor $\mathbf{D}$ and the $b$-tensor $\mathbf{B}$. Let us consider a general 3$\times$3 symmetric tensor
\begin{equation}
\bm{\Lambda} = 
\begin{pmatrix}
\lambda_{xx} & \lambda_{xy} & \lambda_{xz} \\
\lambda_{yx} & \lambda_{yy} & \lambda_{yz} \\
\lambda_{zx} & \lambda_{zy} & \lambda_{zz}
\end{pmatrix}
=
\begin{pmatrix}
\lambda_{xx} & \lambda_{xy} & \lambda_{xz} \\
\cdot & \lambda_{yy} & \lambda_{yz} \\
\cdot & \cdot & \lambda_{zz}
\end{pmatrix}\, ,
\end{equation}
where symmetry requires that $\lambda_{ij} = \lambda_{ji}$, yielding 6 independent elements in total. Upon diagonalization, this matrix can express within its principal axis system (basis of eigenvectors) via two different conventions regarding the ordering of its eigenvalues.

\subsubsection{Standard convention}

\begin{figure}[h!]
\begin{center}
\includegraphics[width=\textwidth]{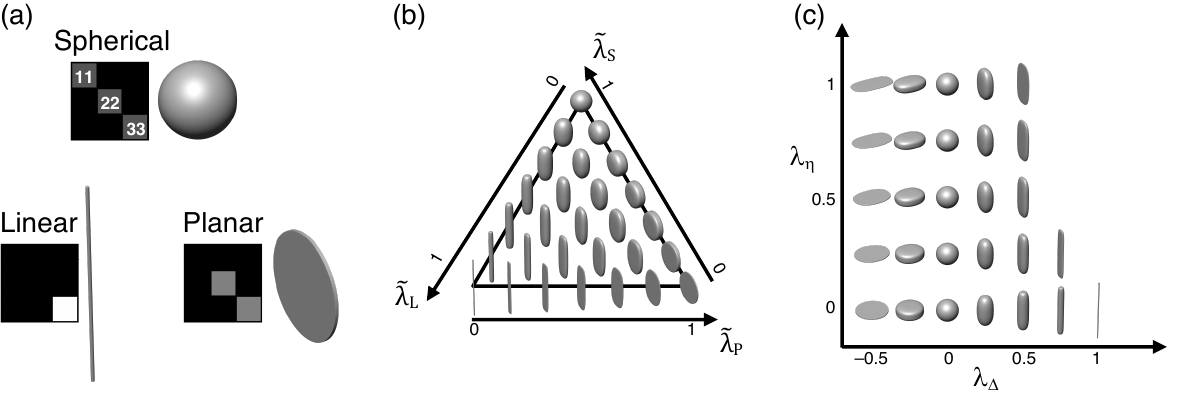}
\caption{Quantitative measures of the shape of 3$\times$3 symmetric tensors. (a) Linear, planar, and spherical tensor shapes represented as grayscale checkerboard plots of the tensor elements in the diagonal basis and superquadratic tensor glyphs with semi-axes corresponding to the tensor eigenvalues. (b) Tensor glyphs in a ternary plot of the normalized linear $\tilde{\lambda}_\text{L}$, planar $\tilde{\lambda}_\text{P}$, and spherical $\tilde{\lambda}_\text{S}$ shape parameters defined in Eq.~\eqref{Eq_normalized_LPS}. (c) Tensor glyphs in a 2D Cartesian plot of the anisotropy $\lambda_\Delta$ and asymmetry $\lambda_\eta$ parameters defined in Eqs.~\eqref{Eq_lambda_Delta} and \eqref{Eq_lambda_eta}. Figure taken from Ref.~\cite{Topgaard_book:2017}.}
\label{Figure_shape_tensors}
\end{center}
\end{figure}

The standard convention corresponds to simply diagonalizing $\bm{\Lambda}$ as
\begin{equation}
\cbeq{
\bm{\Lambda} =
\begin{pmatrix}
\lambda_{11} & 0 & 0 \\
0 & \lambda_{22} & 0 \\
0 & 0 & \lambda_{33}
\end{pmatrix} 
\quad \text{with} \quad
\lambda_{11} \leq \lambda_{22} \leq \lambda_{33}
} \; .
\label{Eq_Lambda_standard_convention}
\end{equation}
As illustrated in Fig.~\ref{Figure_shape_tensors}\textcolor{Red}{.(a)}, the number of non-zero eigenvalues determines the shape of the tensor, represented as a geometrical glyph. For instance, when all eigenvalues are equal, the tensor is spherical, and if only one eigenvalue is non-zero, the tensor is linear. In general, the lengths and directions of the glyph's principal axes are respectively given by the tensor's eigenvalues and corresponding eigenvectors.

\newpage

The size of the tensor $\bm{\Lambda}$ can be quantified in two ways, either by its trace $\mathrm{Tr}(\bm{\Lambda}) = \lambda_{11} + \lambda_{22} + \lambda_{33}$ or by its isotropic average
\begin{equation}
\cbeq{
\lambda_\text{iso} = \frac{\mathrm{Tr}(\bm{\Lambda})}{3} = \frac{\lambda_{11} + \lambda_{22} + \lambda_{33}}{3} 
} \; .
\label{Eq_lambda_iso_trace}
\end{equation}
In this convention, one obtains information about the shape of $\bm{\Lambda}$ by expanding it as
\begin{equation}
\cbeq{
\bm{\Lambda} =
\frac{\lambda_\text{S}}{3}
\begin{pmatrix}
1 & 0 & 0 \\
0 & 1 & 0 \\
0 & 0 & 1
\end{pmatrix}
+
\frac{\lambda_\text{P}}{2}
\begin{pmatrix}
0 & 0 & 0 \\
0 & 1 & 0 \\
0 & 0 & 1
\end{pmatrix}
+
\lambda_\text{L}
\begin{pmatrix}
0 & 0 & 0 \\
0 & 0 & 0 \\
0 & 0 & 1
\end{pmatrix}
} \; ,
\label{Eq_Lambda_LPS}
\end{equation}
where $\lambda_\text{S}$, $\lambda_\text{P}$ and $\lambda_\text{L}$ quantify the amplitudes of the spherical, planar, and linear components, respectively. These coefficients are related to the eigenvalues \textit{via}
\begin{align}
\lambda_\text{S} & = 3 \lambda_{11}\, , \nonumber \\
\lambda_\text{P} & = 2 (\lambda_{22}-\lambda_{11})\, , \label{Eq_link_11_22_33_shape} \\
\lambda_\text{L} & = 3 \lambda_{33}-\lambda_{22}\, . \nonumber
\end{align}
These coefficients satisfy 
\begin{equation}
\lambda_\text{S} + \lambda_\text{P} + \lambda_\text{L} = \lambda_{11} + \lambda_{22} + \lambda_{33} = \mathrm{Tr}(\bm{\Lambda})\, .
\end{equation}
Another convenient way to describe $\bm{\Lambda}$'s shape is to define the normalized parameters \cite{Westin:2002}
\begin{equation}
\cbeq{
\tilde{\lambda}_\alpha = \frac{\lambda_\alpha}{\mathrm{Tr}(\bm{\Lambda})} = \frac{\lambda_\alpha}{3\lambda_\text{iso}} 
} \; ,
\label{Eq_normalized_LPS}
\end{equation}
with $\alpha \equiv \text{L},\text{P},\text{S}$ and
\begin{equation}
\tilde{\lambda}_\text{S} + \tilde{\lambda}_\text{P} + \tilde{\lambda}_\text{L} = 1 \, .
\label{Eq_normalization_LPS}
\end{equation}
The normalization of these shape parameters makes two of them sufficient
to quantify the tensor's shape. From Eq.~\eqref{Eq_link_11_22_33_shape}, it appears clear from the positivity of diffusion eigenvalues that while $\tilde{\lambda}_\text{S}$ has to be positive, $\tilde{\lambda}_\text{P}$ and $\tilde{\lambda}_\text{L}$ can be either positive or negative, as long as the global normalization Eq.~\eqref{Eq_normalization_LPS} is satisfied. The range of available shapes for positive normalized shape parameters of $\bm{\Lambda}$ is shown in Fig.~\ref{Figure_shape_tensors}\textcolor{Red}{.(b)}.

\subsubsection{Haeberlen convention}

In the so-called ``Haeberlen convention" \cite{Haeberlen:1976}, the tensor $\bm{\Lambda}$ is diagonalized as 
\begin{equation}
\cbeq{
\bm{\Lambda} =
\begin{pmatrix}
\lambda_\text{XX} & 0 & 0 \\
0 & \lambda_\text{YY} & 0 \\
0 & 0 & \lambda_\text{ZZ}
\end{pmatrix} 
\quad \text{with} \quad
\vert\lambda_\text{YY}-\lambda_\text{iso} \vert \leq \vert\lambda_\text{XX}-\lambda_\text{iso} \vert \leq \vert\lambda_\text{ZZ}-\lambda_\text{iso} \vert
} \; ,
\label{Eq_Lambda_Haeberlen_convention}
\end{equation}
where the isotropic average $\lambda_\text{iso}$ has the same definition as in the standard convention Eq.~\eqref{Eq_lambda_iso_trace}. This ordering convention assures that $\lambda_\text{ZZ}$ is furthest from the isotropic average while $\lambda_\text{YY}$ is closest. $\bm{\Lambda}$ can then be written in a form that directly reflects its size and shape, albeit different from the standard convention's form, namely
\begin{equation}
\cbeq{
\begin{split}
\bm{\Lambda} =  
\lambda_\text{iso}
\left\{
\begin{pmatrix}
1 & 0 & 0 \\
0 & 1 & 0 \\
0 & 0 & 1
\end{pmatrix}
+
\lambda_\Delta
\left[
\begin{pmatrix}
-1 & 0 & 0 \\
0 & -1 & 0 \\
0 & 0 & 2
\end{pmatrix}
+
\lambda_\eta
\begin{pmatrix}
-1 & 0 & 0 \\
0 & 1 & 0 \\
0 & 0 & 0
\end{pmatrix}
\right]
\right\}
\end{split} 
}\; ,
\label{Eq_Lambda_Delta_eta}
\end{equation}
where 
\begin{align}
\lambda_\text{iso} & = \frac{\lambda_\text{XX} +\lambda _\text{YY} + \lambda_\text{ZZ}}{3} = \frac{\mathrm{Tr}( \bm{\Lambda})}{3}\; , \\
\lambda_\Delta & = \frac{1}{3\lambda_\text{iso}}\left( \lambda_\text{ZZ} - \frac{\lambda_\text{XX} + \lambda_\text{YY}}{2} \right) \in [-0.5,1]  \; , \label{Eq_lambda_Delta}  \\
\lambda_\eta & = \frac{\lambda_\text{YY} - \lambda_\text{XX}}{2\lambda_\text{iso}\lambda_\Delta} \in \left[ \mathrm{max}\left(-\left\vert 1-\frac{1}{\lambda_\Delta} \right\vert, -1 \right),\mathrm{min}\left( \left\vert \frac{1}{\lambda_\Delta}-1 \right\vert,1 \right) \right] \label{Eq_lambda_eta}
\end{align}
are the isotropic diffusivity, the anisotropy parameter and the asymmetry parameter, respectively. Equivalently, one has
\begin{align}
\lambda_\text{XX} & = \lambda_\text{iso}[1-\lambda_\Delta(1+\lambda_\eta)]\, , \\
\lambda_\text{YY} & = \lambda_\text{iso}[1-\lambda_\Delta(1-\lambda_\eta)]\, , \\
\lambda_\text{ZZ} & = \lambda_\text{iso}[1+2\lambda_\Delta] \, .
\end{align}
The bounds for $\lambda_\eta$ in Eq.~\eqref{Eq_lambda_eta} ensure the positivity of the diffusion eigenvalues $\lambda_\text{XX}$ and $\lambda_\text{YY}$ when $\lambda_\Delta >0$. The form of the above equations makes sense: while $\lambda_\text{iso}$ weighs equally on all principal axes of $\bm{\Lambda}$, $\lambda_\Delta$ adds anisotropy by weighing positively on the axial axis (associated to $\lambda_\text{ZZ}$) and negatively on the radial axes, and $\lambda_\eta$ brings asymmetry by unbalancing the weights of radial axes. The relation between tensor shape and the values of $\lambda_\Delta$ and $\lambda_\eta$ is shown for $\lambda_\eta \geq 0$ in Fig.~\ref{Figure_shape_tensors}\textcolor{Red}{.(c)}. \\

In the axisymmetric (axially symmetric) case where $\lambda_\eta = 0$, $\bm{\Lambda}$ writes
\begin{align}
\cbeq{
\bm{\Lambda} =
\begin{pmatrix}
\lambda_\perp & 0 & 0 \\
0 & \lambda_\perp & 0 \\
0 & 0 & \lambda_\parallel
\end{pmatrix}
=
\begin{pmatrix}
\lambda_\text{iso}(1-\lambda_\Delta) & 0 & 0 \\
0 & \lambda_\text{iso}(1-\lambda_\Delta) & 0 \\
0 & 0 & \lambda_\text{iso}(1+2\lambda_\Delta)
\end{pmatrix}
} \; ,
\label{Eq_lambda_para_perp}
\end{align}
where $\lambda_\parallel$ and $\lambda_\perp$ are respectively the axial and radial eigenvalues. Equivalently, one has
\begin{equation}
\cbeq{
\bm{\Lambda} = 
\lambda_\text{iso}
\begin{pmatrix}
1 & 0 & 0 \\
0 & 1 & 0 \\
0 & 0 & 1
\end{pmatrix}
+
\lambda_\text{aniso}
\begin{pmatrix}
-1 & 0 & 0 \\
0 & -1 & 0 \\
0 & 0 & 2
\end{pmatrix}
\quad \text{with}\quad \lambda_\text{aniso} = \lambda_\text{iso}\lambda_\Delta = \frac{\lambda_\parallel - \lambda_\perp}{3}
}\; .
\label{Eq_lambda_iso_aniso}
\end{equation}

\subsubsection{Link between both conventions}
\label{Sec_link_conventions}

Two cases, summarized in Tables~\ref{Table_Case_I} (``Case I", where $\lambda_\text{ZZ}\geq\lambda_\text{iso}$) and \ref{Table_Case_II} (``Case II", where $\lambda_\text{ZZ}\leq\lambda_\text{iso}$), have to be considered to cross the bridge between both previous conventions, as explained in this \href{http://www.ccp14.ac.uk/ccp/web-mirrors/klaus_eichele_software/klaus/nmr/conventions/csa/csa.html}{hyperlink}. Even though Cases I and II, partly illustrated in Figs.~\ref{Figure_shape_tensors}\textcolor{Red}{.(b)} and \ref{Figure_shape_tensors}\textcolor{Red}{.(c)}, are mutually exclusive (except for a spherical tensor), it is simple to describe the shape of any axisymmetric tensor using only one of them. \\

For instance, let us focus on the axisymmetric version of Case I in Table~\ref{Table_Case_I} where $\lambda_\text{ZZ} \geq\lambda_\text{iso}$ and $\lambda_\eta = 0$. The equality $\tilde{\lambda}_\text{P}=0$ seems to imply that no
planar tensor can be described in axisymmetric Case I, even if the planar tensor at $\tilde{\lambda}_\text{P}=1$ in Fig.~\ref{Figure_shape_tensors}\textcolor{Red}{.(b)} clearly is axisymmetric. This is due to the fact that $\tilde{\lambda}_\text{P}$ is directly proportional to the axial asymmetry parameter $\lambda_\eta$ in Table~\ref{Table_Case_I}, only describing planar tensors that contain the axis of symmetry with respect to which $\lambda_\Delta$ and $\lambda_\eta$ are defined ($\text{ZZ}$ in Eq.~\eqref{Eq_Lambda_Delta_eta}, with $\text{33} \equiv \text{ZZ}$ within Case I). Nonetheless, this is not a problem in itself, since only two normalized shape parameters Eq.~\eqref{Eq_normalized_LPS} are sufficient to describe the shape of any symmetric tensor $\bm{\Lambda}$. One can thus describe axisymmetric planar, oblate, spherical, prolate and linear tensors with respect to the same axis of symmetry by only using $\tilde{\lambda}_\text{S} \geq 0$ and $\tilde{\lambda}_\text{L} = 1 − \tilde{\lambda}_\text{S}$ according to Table~\ref{Table_Case_III}. The axisymmetric tensor associated to $\tilde{\lambda}_\text{P}=1$ is obtained upon rotating the axis of symmetry of the axisymmetric planar tensor $(\tilde{\lambda}_\text{S},\tilde{\lambda}_\text{P}) = (1-\lambda_\Delta,\lambda_\Delta) = (3/2, −1/2)$ from Table~\ref{Table_Case_I}.\\

For the rest of this chapter, let us therefore describe the shape of axisymmetric tensors using the extended Case I presented in Table~\ref{Table_Case_III}, where the standard and Haeberlen conventions merge together, as shown in Table~\ref{Table_Case_I}. That way, one is free to use any shape parametrization that best fits any given axisymmetric situation. In the case of non-axisymmetric tensors, one has to remain very careful when transitioning from one convention to the other, using Tables~\ref{Table_Case_I} and \ref{Table_Case_II}.

\newpage
 
\renewcommand{\arraystretch}{1.4}
\begin{table}[h!]
\begin{center}
\begin{tabular}{r|l|l|}
\hline
\multicolumn{1}{|r|}{\cellcolor[gray]{0.9}\textbf{Case I ($\bm{\lambda_\text{ZZ} \geq \lambda_\text{iso}}$)}}
 &  \multicolumn{1}{c|}{Standard convention}
 & \multicolumn{1}{c|}{Haeberlen convention} \\
\hline
\multicolumn{1}{|c|}{\cellcolor[gray]{0.9}Eigenvalues} & \multicolumn{1}{c|}{$(\lambda_{11},\lambda_{22},\lambda_{33})$} & \multicolumn{1}{c|}{$(\lambda_\text{XX},\lambda_\text{YY},\lambda_\text{ZZ})$} \\
\hline
\multicolumn{1}{|r|}{\cellcolor[gray]{0.9}} & $\tilde{\lambda}_\text{S} = 1-\lambda_\Delta(1+\lambda_\eta)$ & $\lambda_\Delta = \tilde{\lambda}_\text{L} + \tilde{\lambda}_\text{P}/4$ \\
\multicolumn{1}{|c|}{\cellcolor[gray]{0.9}Shape parameters} & $\tilde{\lambda}_\text{P} = 4\lambda_\Delta\lambda_\eta/3$ & $\lambda_\eta = 3(1+4\tilde{\lambda}_\text{L}/\tilde{\lambda}_\text{P})^{-1}$ \\
\multicolumn{1}{|c|}{\cellcolor[gray]{0.9}} & $\tilde{\lambda}_\text{L} = \lambda_\Delta(1-\lambda_\eta/3)$ & $\lambda_\Delta\lambda_\eta = 3\tilde{\lambda}_\text{P}/4$ \\
\hline
\multicolumn{1}{|c|}{\cellcolor[gray]{0.9}Axisymmetric case} & $\tilde{\lambda}_\text{S} = 1-\lambda_\Delta$ & $\lambda_\Delta = \tilde{\lambda}_\text{L} = 1- \tilde{\lambda}_\text{S}$ \\
\multicolumn{1}{|c|}{\cellcolor[gray]{0.9}$(\tilde{\lambda}_\text{P} = 0,\lambda_\eta = 0)$} & $\tilde{\lambda}_\text{L} = \lambda_\Delta$ &  \\
\hline
\end{tabular}
\end{center}
\caption{Link between the standard convention Eq.~\eqref{Eq_Lambda_standard_convention} and the Haeberlen convention Eq.~\eqref{Eq_Lambda_Haeberlen_convention} in the case where $\lambda_\text{ZZ} \geq \lambda_\text{iso}$ in the Haeberlen convention. While $\tilde{\lambda}_\text{S}$, $\tilde{\lambda}_\text{P}$ and $\tilde{\lambda}_\text{L}$ are
the normalized spherical, planar and linear parameters of Eq.~\eqref{Eq_normalized_LPS} that satisfy the normalization Eq.~\eqref{Eq_normalization_LPS}, $\lambda_\Delta$ and $\lambda_\eta$ are the anisotropy and asymmetry parameters of Eqs.~\eqref{Eq_lambda_Delta} and \eqref{Eq_lambda_eta}, defined with respect to the $\text{33} \equiv \text{ZZ}$ eigenvector.}
\label{Table_Case_I}
\end{table} 

\vspace*{\fill}
 
\begin{table}[h!]
\begin{center}
\begin{tabular}{r|l|l|}
\hline
\multicolumn{1}{|r|}{\cellcolor[gray]{0.9}\textbf{Case II ($\bm{\lambda_\text{ZZ} \leq \lambda_\text{iso}}$)}}
 &  \multicolumn{1}{c|}{Standard convention}
 & \multicolumn{1}{c|}{Haeberlen convention} \\
\hline
\multicolumn{1}{|c|}{\cellcolor[gray]{0.9}Eigenvalues} & \multicolumn{1}{c|}{$(\lambda_{11},\lambda_{22},\lambda_{33})$} & \multicolumn{1}{c|}{$(\lambda_\text{ZZ},\lambda_\text{YY},\lambda_\text{XX})$} \\
\hline
\multicolumn{1}{|r|}{\cellcolor[gray]{0.9}} & $\tilde{\lambda}_\text{S} = 1+2\lambda_\Delta(1+\lambda_\eta)$ & $\lambda_\Delta = -(\tilde{\lambda}_\text{L} + \tilde{\lambda}_\text{P})/2$ \\
\multicolumn{1}{|c|}{\cellcolor[gray]{0.9}Shape parameters} & $\tilde{\lambda}_\text{P} = -2\lambda_\Delta(1-\lambda_\eta/3)$ & $\lambda_\eta = 6(1+\tilde{\lambda}_\text{P}/\tilde{\lambda}_\text{L})^{-1}$ \\
\multicolumn{1}{|c|}{\cellcolor[gray]{0.9}} & $\tilde{\lambda}_\text{L} = -2\lambda_\Delta\lambda_\eta/3$ & $\lambda_\Delta\lambda_\eta = -3\tilde{\lambda}_\text{L}$ \\
\hline
\multicolumn{1}{|c|}{\cellcolor[gray]{0.9}Axisymmetric case} & $\tilde{\lambda}_\text{S} = 1+2\lambda_\Delta$ & $\lambda_\Delta = -\tilde{\lambda}_\text{P}/2 = -(1- \tilde{\lambda}_\text{S})/2$ \\
\multicolumn{1}{|c|}{\cellcolor[gray]{0.9}$(\tilde{\lambda}_\text{L} = 0,\lambda_\eta = 0)$} & $\tilde{\lambda}_\text{P} = -2\lambda_\Delta$ & \\
\hline
\end{tabular}
\end{center}
\caption{Equivalent of Table~\ref{Table_Case_I} in the case where $\lambda_\text{ZZ} \leq \lambda_\text{iso}$ in the Haeberlen convention. Here, $\lambda_\Delta$ and $\lambda_\eta$ are defined with respect to the $\text{11} \equiv \text{ZZ}$ eigenvector.}
\label{Table_Case_II}
\end{table} 

\vspace*{\fill}

\begin{table}[h!]
\begin{center}
\begin{tabular}{|c|c|c|}
\hline
\cellcolor[gray]{0.9}\textbf{Value of} $\mathbold{\lambda_\Delta}$ & \cellcolor[gray]{0.9}\textbf{Value of} $\mathbold{(\tilde{\lambda}_\text{S}, \tilde{\lambda}_\text{L})}$ & \cellcolor[gray]{0.9} $\bm{\Lambda}$\textbf{'s shape} \\
\hline
$-0.5$ & $(3/2, -1/2)$ & Planar \\
\hline
$0$ & $(1, 0)$ & Spherical \\
\hline
$1$ & $(0, 1)$ & Linear\\
\hline
\end{tabular}
\end{center}
\caption{Different shape parametrizations of axisymmetric tensors around the $\text{33} \equiv \text{ZZ}$ eigenvector, from planar via spherical to linear tensors. Oblate and prolate tensors have intermediate values between those indicated above.}
\label{Table_Case_III}
\end{table}

\newpage

\subsection{Alternative metrics of tensor anisotropy}

\begin{figure}[h!]
\begin{center}
\includegraphics[width=\textwidth]{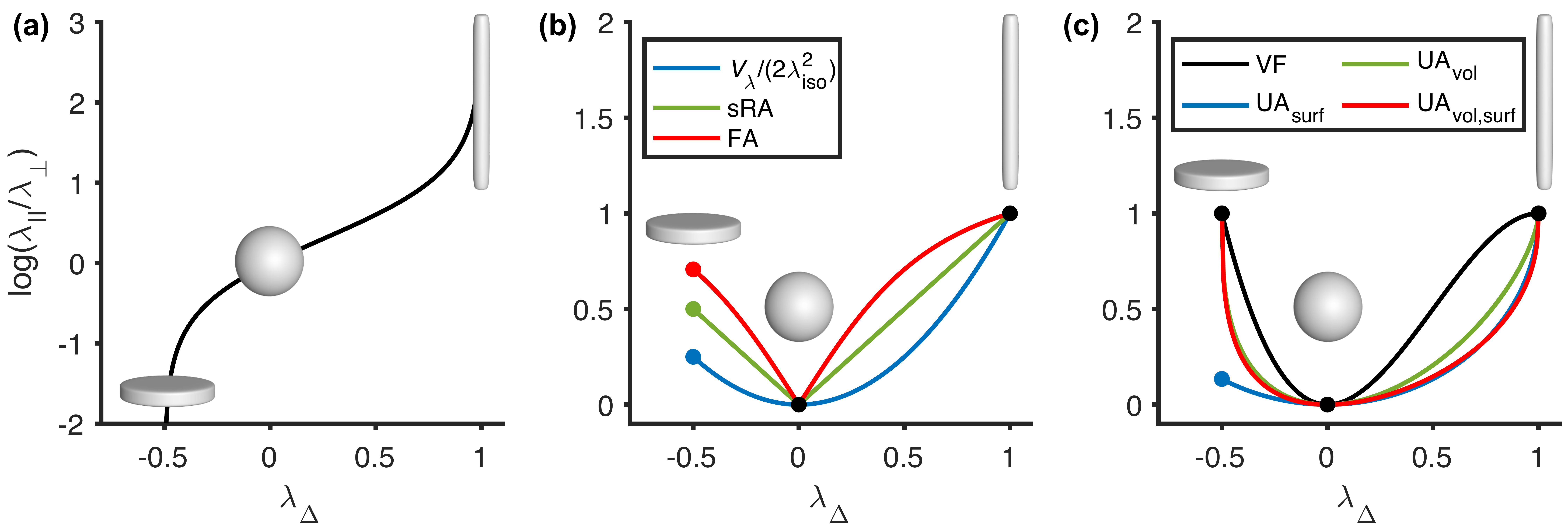}
\caption{Alternative measures of tensor anisotropy vs. the anisotropy parameter $\lambda_\Delta$ for axisymmetric tensors. (a) Logarithm of the ratio between the parallel and perpendicular eigenvalues $\lambda_\parallel$ and $\lambda_\perp$ defined in Eq.~\eqref{Eq_lambda_para_perp}. (b) Normalized variance of eigenvalues $V_\lambda/(2\lambda_\text{iso}^2)$, with $V_\lambda$ defined in Eq.~\eqref{Eq_variance_eigenvalues}, scaled relative anisotropy (sRA) Eq.~\eqref{Eq_sRA}, and fractional anisotropy (FA) Eq.~\eqref{Eq_FA_2}. (c) Volume fraction (VF) Eq.~\eqref{Eq_VF} and ultimate anisotropy indices (UA$_{\cdots}$) Eq.~\eqref{Eq_UA}. The tensor glyphs highlight the parameter values for planar, spherical, and linear tensor shapes.}
\label{Figure_metrics_anisotropy}
\end{center}
\end{figure}

From the previous section, the anisotropy parameter $\lambda_\Delta\in [-0.5, 1]$ defined in Eq.~\eqref{Eq_lambda_Delta} and, for axisymmetric tensors, the ratio between the parallel and perpendicular eigenvalues $\lambda_\parallel/\lambda_\perp\in[0, +\infty[$ in Eq.~\eqref{Eq_lambda_para_perp} appear as direct metrics to quantify tensor anisotropy. However, alternative metrics exist. For instance, a mathematically well-defined quantity to measure the shape of $\bm{\Lambda}$ is the variance of its eigenvalues
\begin{equation}
V_\lambda = \frac{(\lambda_{11} -\lambda_\text{iso})^2+(\lambda_{22} -\lambda_\text{iso})^2+(\lambda_{33} -\lambda_\text{iso})^2}{3}\, ,
\end{equation}
which writes in terms of $\lambda_\Delta$ and $\lambda_\eta$ as
\begin{equation}
\cbeq{
V_\lambda = 2(\lambda_\text{iso}\lambda_\Delta)^2\, \frac{\lambda_\eta^2+3}{3} = 2\lambda_\text{aniso}^2\, \frac{\lambda_\eta^2+3}{3}
}\; ,
\label{Eq_variance_eigenvalues}
\end{equation}
with $\lambda_\text{aniso}=\lambda_\text{iso}\lambda_\Delta$ from Eq.~\eqref{Eq_lambda_iso_aniso}. Many anisotropy indices ranging from 0 (isotropic) to 1 (purely anisotropic) can also be defined, showing different sensitivities to signal-to-noise ratio (SNR) \cite{Kingsley_Monahan:2005}. A non-exhaustive list includes the scaled relative anisotropy (sRA) \cite{Conturo:1996}
\begin{equation}
\text{sRA} = \frac{1}{\lambda_\text{iso}}\sqrt{\frac{(\lambda_{11} -\lambda_\text{iso})^2+(\lambda_{22} -\lambda_\text{iso})^2+(\lambda_{33} -\lambda_\text{iso})^2}{6}}\; ,
\label{Eq_sRA}
\end{equation}
the fractional anisotropy (FA) \cite{Basser_Pierpaoli:1996}
\begin{equation}
\text{FA} = \sqrt{\frac{3}{2}}\,\sqrt{\frac{(\lambda_{11} -\lambda_\text{iso})^2+(\lambda_{22} -\lambda_\text{iso})^2+(\lambda_{33} -\lambda_\text{iso})^2}{\lambda_{11}^2+\lambda_{22}^2+\lambda_{33}^2}}\; ,
\label{Eq_FA_2}
\end{equation}
the volume fraction (VF) \cite{Alexander:2000}
\begin{equation}
\text{VF} = 1-\frac{\lambda_{11}\lambda_{22}\lambda_{33}}{\lambda_\text{iso}^3}\; ,
\label{Eq_VF}
\end{equation}
and the ultimate anisotropy indices (UA$_{\cdots}$) \cite{Ulug_vanZijl:1999}
\begin{align}
\text{UA}_\text{surf} & = 1-\frac{1}{\lambda_\text{iso}^2}\sqrt{\frac{\lambda_{11}\lambda_{22}+\lambda_{11}\lambda_{33}+\lambda_{22}\lambda_{33}}{3}} \nonumber \\
\text{UA}_\text{vol} & = 1-\frac{(\lambda_\text{11}\lambda_{22}\lambda_{33})^{1/3}}{\lambda_\text{iso}} \label{Eq_UA} \\
\text{UA}_\text{vol,surf} & =1-\frac{(\lambda_\text{11}\lambda_{22}\lambda_{33})^{1/3}}{\sqrt{(\lambda_{11}\lambda_{22}+\lambda_{11}\lambda_{33}+\lambda_{22}\lambda_{33})/3}} \nonumber \; .
\end{align}
All these metrics are related to the variance Eq.~\eqref{Eq_variance_eigenvalues}. For instance, the FA writes as
\begin{equation}
\text{FA} = \sqrt{\frac{3}{2}}\left( 1+\frac{\lambda_\text{iso}^2}{V_\lambda} \right)^{-1/2} \quad , \quad V_\lambda = 2\lambda_\text{iso}^2\left( \frac{3}{\text{FA}^2}-2 \right)^{-1} \; .
\label{Eq_FA_variance_eigenvalues}
\end{equation}
It is also worth noting that the above expressions do not depend on choosing between Tables~\ref{Table_Case_I} and \ref{Table_Case_II} to link the standard and Haeberlen conventions, since they all are invariant upon exchange of $\lambda_{11}$ and $\lambda_{33}$. The dependence on $\lambda_\Delta$ of these metrics is illustrated in Fig.~\ref{Figure_metrics_anisotropy}.\\

\subsection{Eigenvalue average and variance}

When it comes to describing the statistics of diffusion eigenvalues, let us introduce more general notations \cite{SzczepankiewiczThesis:2016} that will naturally bridge the gap between macroscopic and microscopic anisotropy metrics later on in this chapter. We denote
\begin{equation}
\mathbb{E}_\lambda [\bm{\Lambda}] = \frac{1}{3}\sum_{i=1,2,3} \lambda_{ii} = \frac{1}{3}\sum_{i=\mathrm{X},\mathrm{Y},\mathrm{Z}} \lambda_{ii} \equiv \lambda_\text{iso}
\label{Eq_average_eigenvalue}
\end{equation}
and
\begin{equation}
\mathbb{V}_\lambda [\bm{\Lambda}] = \frac{1}{3}\sum_{i=1,2,3} (\lambda_{ii}-\mathbb{E}_\lambda [\bm{\Lambda}])^2 = \frac{1}{3}\sum_{i=\mathrm{X},\mathrm{Y},\mathrm{Z}} (\lambda_{ii}-\mathbb{E}_\lambda [\bm{\Lambda}])^2 \equiv V_\lambda
\label{Eq_variance_eigenvalue}
\end{equation}
the expectation and variance of $\bm{\Lambda}$'s eigenvalues, respectively.

\newpage

\section{Tensor-valued encoding}

\subsection{Going multidimensional}

Part of this section is drawn from Ref.~\cite{Topgaard:2017}.

\subsubsection{Inspiration from multidimensional solid-state NMR}
\label{Sec_solid_NMR_inspiration}

We mentioned in Sec.~\ref{Sec_DTI_limitations} that performing the Laplace inversion Eq.~\eqref{Eq_signal_tensor_distribution} using only linear (unidirectional $q$-vector) gradients is very ill-conditioned for complex voxel populations. While still working with nuclear magnetic resonance (NMR), let us distance ourselves from diffusion MRI (dMRI) to talk about physical chemistry. Indeed, the reader will soon realize that part of the answer to that inversion problem lies in a dMRI analog of physical chemistry NMR called tensor-valued encoding (or $b$-tensor encoding).\\

In NMR spectroscopy, one wants to measure a quantity called the chemical shift, \textit{i.e.} the resonant frequency of a nucleus's spin in a magnetic field $\mathbf{B}_0$ relative to a standard in the same applied magnetic field $\mathbf{B}_0$. To clear things up, the reader has to know that some atomic nuclei possess a non-zero magnetic moment (nuclear spin) that gives rise to different energy levels and resonance frequencies in an applied magnetic field. However, the actual magnetic field experienced by a nucleus is affected by the surrounding electrons, since any magnetic perturbation will be screened by these electrons (a natural consequence of Lenz law in magnetic induction). Since the electron distribution around a given nucleus (\textit{e.g.} $^1$H, $^{13}$C, $^{15}$N) usually varies according to the environment surrounding the atom (binding partners, bond lengths, angles between bonds, \textit{etc}.), so does the local magnetic field at the nucleus. The complexity and anisotropy of the electronic environment of the nucleus is thus reflected in the nuclear spin energy levels and resonance frequencies. The variations of NMR frequencies of the same kind of nucleus, due to variations its electronic surrounding, is called the chemical shift $\sigma$. The size of the chemical shift is given with respect to a reference frequency $\nu_\text{ref}$ (usually associated to a molecule with a barely distorted electron distribution): $\sigma = (\nu_\text{sample} - \nu_\text{ref})/\nu_\text{ref}$. Since the chemical shift depends on the orientation of the applied magnetic field $\mathbf{B}_0$, it is often encoded in a semipositive-definite symmetric tensor $\mathbold{\sigma}$.\\

To understand how solid-state NMR could come to the rescue of diffusion MRI, one has to take a closer look at Ref.~\cite{Andrew:1959}, where the investigated sample is allowed to effectively rotate around a fixed magnetic field $\mathbf{B}_0$. By choosing the fixed angle of rotation
\begin{equation}
\theta_\text{m} = \arccos\!\left( \frac{1}{\sqrt{3}} \right) = \arctan\!\left(\sqrt{2}\right) \simeq 54.7^\circ \, ,
\label{Eq_magic_angle}
\end{equation}
called the ``magic angle", the measured signal only describes the isotropic part of the chemical shift. Therefore, it is possible to enhance the specificity of physical chemistry NMR through a clever rotation of the sample in a fixed magnetic field! This could prove useful, should it be transferable to diffusion MRI. \\

\begin{sidenote}{title = The magic angle}
The magic angle $\theta_\text{m}$ of Eq.~\eqref{Eq_magic_angle} is the root of the second-order Legendre polynomial $P_2(\cos\theta) = [3\cos^2\theta-1]/2$, meaning 
\begin{equation}
P_2(\cos\theta_\text{m}) = \frac{3\cos^2\theta_\text{m} - 1}{2} = 0\, .
\label{Eq_Legendre}
\end{equation}
Hence, any interaction whose angular dependency is given by this polynomial, such as dipole-dipole interaction Eq.~\eqref{Eq_dipole_dipole_interaction}, vanishes at the magic angle! This is why certain biological structures with ordered collagen oriented at the magic angle, such as tendons and ligaments, may appear hyperintense in some MR sequences (no dipole-dipole interaction means less $T_1$ and $T_2$ relaxations, so more signal): one speaks of ``magic angle artifact". 
\end{sidenote}\bigskip

When analyzing the orientational dependence of the interactions giving birth to chemical shift, it is more common to use the rotating frame of reference where the sample is at rest. Within that frame, one can consider the unit vector $\mathbf{n}(t)$, defined \textit{via} 
\begin{equation}
\mathbf{B}_0(t) = B_0(t)\, \mathbf{n}(t)\, ,
\end{equation}
as changing orientation through time (describing a cone at fixed magic angle in Ref.~\cite{Andrew:1959}). It is then possible to write an analog of Eq.~\eqref{Eq_DTI_attenuation},
\begin{equation}
\mathcal{S}(t) = \mathcal{S}_0\, \mathrm{exp}\!\left(-i\gamma \int_0^t B_0(t')\;\mathbf{n}^{\mathrm{T}}(t')\cdot \mathbold{\sigma}\cdot \mathbf{n}(t')\, \mathrm{d}t'\right)\, ,
\label{Eq_signal_solid_state_NMR_1}
\end{equation}
and an analog of Eq.~\eqref{Eq_signal_tensor_distribution},
\begin{equation}
\mathcal{S}(t) = \mathcal{S}_0 \int \! \mathcal{P}(\mathbold{\sigma})\, \mathrm{exp}\!\left(-i\gamma \int_0^t B_0(t')\;\mathbf{n}^{\mathrm{T}}(t')\cdot \mathbold{\sigma}\cdot \mathbf{n}(t')\, \mathrm{d}t'\right) \mathrm{d}\mathbold{\sigma}\, .
\label{Eq_signal_solid_state_NMR_2}
\end{equation}
Everything related to this technique, broadly called ``multidimensional solid-state NMR", can be found in Ref.~\cite{MD-dMRI_solid-state_book:1994}.

\subsubsection{Transfer to diffusion MRI}

\begin{figure}[h!]
\begin{center}
\includegraphics[width=0.5\textwidth]{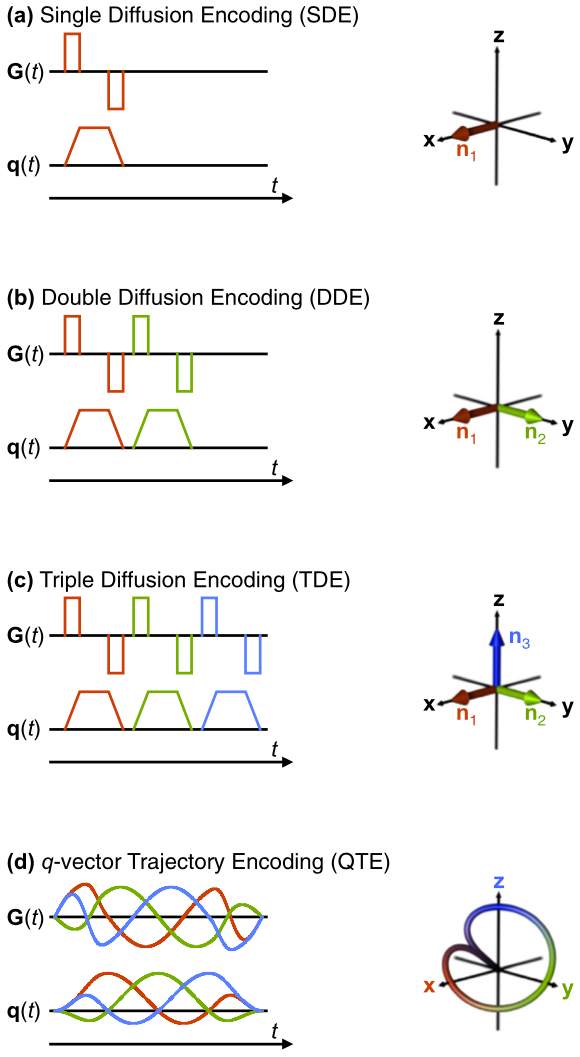}
\caption{Different examples of $q$-trajectories. While the three top panels are trivial (the $q$-vector only points out and back in along the main axes of the frame of reference), the bottom panel presents a non-trivial $q$-trajectory. Figure taken from Ref.~\cite{Topgaard:2017}.}
\label{Figure_Topgaard_trajectory}
\end{center}
\end{figure}

Many analogs between physical chemistry NMR and diffusion MRI have already been highlighted. For instance, the fact that both chemical shift and diffusion are encoded in semipositive-definite symmetric tensors. To complete this picture, one analogy remains to be addressed: the probing vector. Within physical chemistry NMR, the magnetic field $\mathbf{B}_0$ probes chemical shift. In the case of diffusion MRI, it is rather the spin-dephasing vector $\mathbf{q}$ (hence the gradient) that probes diffusion. Indeed, the applied magnetic field $\mathbf{B}_0$ only serves the purpose of setting a natural frequency that can be matched by an RF pulse to satisfy the resonance condition (see Sec.~\ref{Sec_resonance}). 
Therefore, transferring multidimensional solid-state NMR to diffusion MRI naturally means rotating the $q$-vector through time with respect to the magnetic field $\mathbf{B}_0$ and the brain of the patient, which remain fixed in space (it is very difficult to rotate a person's brain in a scanner). The sequences based on a rotating spin-dephasing vector $\mathbf{q}(t) = q(t)\, \mathbf{n}(t)$ are called ``$q$-trajectories". An example of non-trivial $q$-trajectory is provided in the bottom panel of Fig.~\ref{Figure_Topgaard_trajectory}.

\subsubsection{Encoding $\mathbold{b}$-tensor and Frobenius inner product}

From a more mathematical perspective, the Bloch-Torrey equation \cite{Torrey:1956} (generalization of the Bloch equation Eq.~\eqref{Eq_Bloch_equation} to diffusion) for the density of transverse magnetization $m_{xy}(\mathbf{r},t)$ in the rotating frame of reference writes \cite{Price_book:2009}
\begin{equation}
\frac{\mathrm{d}m_{xy}}{\mathrm{d}t}(\mathbf{r},t) = [-i\gamma\, \mathbf{G}(t)\cdot \mathbf{r}]\, m_{xy}(\mathbf{r},t) + \bm{\nabla}\cdot \mathbf{D}\cdot\bm{\nabla} m_{xy}(\mathbf{r},t)\, ,
\end{equation}
where the first term describes the change in $\mathbf{B}_0$ brought by the diffusion gradient (in the same way that a slice-select gradient) and the second term is the equivalent of Eq.~\eqref{Eq_Fick_law} considering that magnetization is the analog of an orientational concentration of spin. This equation has the solution \cite{Price_book:2009}
\begin{align}
m_{xy}(\mathbf{r},t) & = m_0\; \mathrm{exp}\!\left(-i\mathbf{q}(t)\cdot \mathbf{r} - \int_0^t\!\mathbf{q}^\text{T}(t')\cdot \mathbf{D}\cdot\mathbf{q}(t')\, \mathrm{d}t' \right) \nonumber \\
 & = m_0\; \mathrm{exp}\!\left(-i\mathbf{q}(t)\cdot \mathbf{r} - \int_0^t\!q^2(t')\,\mathbf{n}^\text{T}(t')\cdot \mathbf{D}\cdot\mathbf{n}(t')\, \mathrm{d}t' \right)\, .
\end{align}
Typically, diffusion NMR measurements acquire the signal at a time $\tau$ where the echo condition $\mathbf{q}(\tau)=\mathbf{0}$ is fulfilled, thus nulling the previous imaginary term. Finally, integrating $m_{xy}(\mathbf{r},\tau)$ over the investigated sample volume yields the signal $\mathcal{S}$ according to
\begin{equation}
\mathcal{S} = \mathcal{S}_0 \int \! \mathcal{P}(\mathbf{D})\; \mathrm{exp}\!\left( - \int_0^\tau\!q^2(t)\,\mathbf{n}^\text{T}(t)\cdot \mathbf{D}\cdot\mathbf{n}(t)\, \mathrm{d}t \right) \mathrm{d}\mathbf{D} \, ,
\label{Eq_signal_dMRI_2}
\end{equation}
thereby obtaining dMRI equivalents of the solid-state NMR Eqs.~\eqref{Eq_signal_solid_state_NMR_1} and \eqref{Eq_signal_solid_state_NMR_2}. Introducing the $b$-tensor
\begin{equation}
\cbeq{
\mathbf{B} = \int_0^\tau q^2(t)\, \mathbf{n}(t) \cdot\mathbf{n}^{\mathrm{T}}(t)\,\mathrm{d}t
} 
\label{Eq_b_tensor_definition}
\end{equation}
and the Frobenius inner product (or generalized scalar product)
\begin{equation}
\cbeq{
\mathbf{B}:\mathbf{D} = \sum_{ij} B_{ij} D_{ij} = \int_0^\tau q^2(t)\;\mathbf{n}^{\mathrm{T}}(t)\cdot \mathbf{D}\cdot \mathbf{n}(t)\, \mathrm{d}t
}\; ,
\label{Eq_Frobenius_inner_product_integral}
\end{equation}
the signal Eq.~\eqref{Eq_signal_dMRI_2} reduces to
\begin{equation}
\cbeq{
\mathcal{S} = \mathcal{S}_0 \int \! \mathcal{P}(\mathbf{D})\, \exp^{- \mathbf{B}:\mathbf{D}}\, \mathrm{d}\mathbf{D} 
}\; .
\label{Eq_signal_tensor_distribution_b_tensor}
\end{equation}

\vspace*{\fill}

\newpage

\begin{sidenote}{title = $b$-tensor encoding or multidimensional diffusion MRI?}
It is tempting to directly call $b$-tensor encoding ``multidimensional diffusion MRI" (MD-dMRI), in comparison with multidimensional NMR. However, the MD-dMRI concept is larger, as it encompasses variable exchange weighting with filter exchange imaging (FEXI) \cite{Lampinen:2017, Lasic:2011}, flow-compensation for improved intra-voxel incoherent motion (IVIM) \cite{LeBihan:1986}, \textit{etc}.
\end{sidenote}

\begin{mainpoint}{title = $b$-tensor encoding and Laplace inversion}
The reader should be careful: $b$-tensor encoding does not actually solve the ill-conditioned inverse problem mentioned at Eq.~\eqref{Eq_signal_tensor_distribution}. It merely takes advantage of a rotating $q$-vector to separate and correlate the isotropic and anisotropic parts of the diffusion tensor. In doing so, it enables acquiring more diverse pieces of diffusion information compared to linear encoding alone (corresponding to $\mathbf{B} = b\,\mathbf{n}\cdot \mathbf{n}^\text{T}$). This beautiful idea of mimicking solid-state NMR to benefit the field of diffusion MRI has emerged during the first half of the 2010's \cite{Eriksson:2013, Lasic:2014, Eriksson:2015, Westin:2016}.
\end{mainpoint}

\subsection{Diffusion weighting}

\subsubsection{Typical parametrizations of the diffusion and encoding tensors}

The diffusion tensor is usually expressed in its eigenbasis following the Haeberlen convention Eq.~\eqref{Eq_Lambda_Haeberlen_convention}:
\begin{equation}
\cbeq{
\mathbf{D} =  
D_\text{iso}
\left\{
\begin{pmatrix}
1 & 0 & 0 \\
0 & 1 & 0 \\
0 & 0 & 1
\end{pmatrix}
+
D_\Delta
\left[
\begin{pmatrix}
-1 & 0 & 0 \\
0 & -1 & 0 \\
0 & 0 & 2
\end{pmatrix}
+
D_\eta
\begin{pmatrix}
-1 & 0 & 0 \\
0 & 1 & 0 \\
0 & 0 & 0
\end{pmatrix}
\right]
\right\}
}\; ,
\label{Eq_diffusion_tensor_eigenbasis}
\end{equation}
where $D_\text{iso}$ is the isotropic diffusivity, $D_\Delta \in [-0.5,1]$ is the diffusion anisotropy parameter and $D_\eta$ is the diffusion asymmetry. As for the encoding tensor $\mathbf{B}$, on the one hand, it can be expressed using the Haeberlen convention Eq.~\eqref{Eq_Lambda_Haeberlen_convention}
\begin{equation}
\cbeq{
\mathbf{B} =
\frac{b}{3}
\left\{
\begin{pmatrix}
1 & 0 & 0 \\
0 & 1 & 0 \\
0 & 0 & 1
\end{pmatrix}
+
b_\Delta
\left[
\begin{pmatrix}
-1 & 0 & 0 \\
0 & -1 & 0 \\
0 & 0 & 2
\end{pmatrix}
+
b_\eta
\begin{pmatrix}
-1 & 0 & 0 \\
0 & 1 & 0 \\
0 & 0 & 0
\end{pmatrix}
\right]
\right\}
} \; ,
\label{Eq_b_tensor_eigenbasis_1}
\end{equation}
where 
\begin{equation}
\cbeq{
b = \mathrm{Tr} (\mathbf{B}) \equiv 3\, b_\text{iso}
}\; ,
\end{equation}
is the usual $b$-value, $b_\Delta \in [-0.5,1]$ is the encoding anisotropy parameter, and $b_\eta$ is the encoding asymmetry. On the other hand, $\mathbf{B}$ can be expressed in the standard convention Eq.~\eqref{Eq_Lambda_standard_convention}
\begin{equation}
\cbeq{
\mathbf{B} =
\frac{b_\text{S}}{3}
\begin{pmatrix}
1 & 0 & 0 \\
0 & 1 & 0 \\
0 & 0 & 1
\end{pmatrix}
+
\frac{b_\text{P}}{2}
\begin{pmatrix}
0 & 0 & 0 \\
0 & 1 & 0 \\
0 & 0 & 1
\end{pmatrix}
+
b_\text{L}
\begin{pmatrix}
0 & 0 & 0 \\
0 & 0 & 0 \\
0 & 0 & 1
\end{pmatrix}
} \; ,
\label{Eq_b_tensor_eigenbasis_2}
\end{equation}
where $b_\text{S}$, $b_\text{P}$, $b_\text{L}$ are the spherical, planar and linear components of the $b$-tensor, respectively. \\

But how exactly does choosing a certain shape for the $b$-tensor affect in any way the probed diffusion signal?

\subsubsection{Link between encoding tensor's shape a probed diffusion pattern}
\label{Sec_b-tensor_shape_probed_diffusion} 
 
One knows from Eq.~\eqref{Eq_signal_tensor_distribution_b_tensor} that the diffusion encoding depends on the Frobenius inner product Eq.~\eqref{Eq_Frobenius_inner_product_integral}
\begin{equation}
\mathbf{B}:\mathbf{D} = \sum_{ij} B_{ij} D_{ij}\, .
\end{equation}
However, the diffusion tensor Eq.~\eqref{Eq_diffusion_tensor_eigenbasis} and the $b$-tensor Eq.~\eqref{Eq_b_tensor_eigenbasis_1} are not expressed in the same basis (but rather their respective eigenbases), which makes the calculation of the generalized tensor product more complex than it seems to be. A simple way of dealing with this problem is to take the $b$-tensor's eigenbasis (``eigenframe") as a reference and to work out the expression of the diffusion tensor in that basis.\\

%

This gives the following diffusion tensor, expressed in the $b$-tensor's eigenbasis:
\begin{equation}
\mathbf{D}^{(\mathbf{B})} (\alpha, \beta, \gamma) = \begin{pmatrix}
D_\text{XX}^{(\mathbf{B})} & D_\text{XY}^{(\mathbf{B})} & D_\text{XZ}^{(\mathbf{B})} \\
\cdot & D_\text{YY}^{(\mathbf{B})} & D_\text{YZ}^{(\mathbf{B})} \\
\cdot & \cdot & D_\text{ZZ}^{(\mathbf{B})}
\end{pmatrix} = \mathbf{R}_\text{Euler}(\alpha, \beta, \gamma) \cdot \mathbf{D} \cdot\mathbf{R}_\text{Euler}^\mathrm{T}(\alpha, \beta, \gamma) \, ,
 \label{Eq_Change_basis_D}
\end{equation}
where
\begin{align}
\mathbf{R}_\text{Euler}(\alpha, \beta, \gamma) & = \mathbf{R}_\text{Z}(\gamma) \cdot\mathbf{R}_\text{Y}(\beta)\cdot\mathbf{R}_\text{Z}(\alpha) \nonumber \\
& = 
\begin{pmatrix}
\cos\alpha\cos\beta\cos\gamma - \sin\alpha\sin\gamma & - \sin\alpha\cos\beta\cos\gamma - \cos\alpha\sin\gamma & \sin\beta\cos\gamma \\
\cos\alpha \cos\beta\sin\gamma  + \sin\alpha\cos\gamma & - \sin\alpha \cos\beta\sin\gamma + \cos\alpha\cos\gamma & \sin\beta\sin\gamma \\
-\cos\alpha \sin\beta & \sin\alpha\sin\beta & \cos\beta
\end{pmatrix}
\end{align}
is the Euler rotation matrix, depending on the rotation matrices
\begin{align}
\mathbf{R}_\text{Z}(\alpha) & = 
\begin{pmatrix}
\cos\alpha & -\sin\alpha & 0 \\
\sin\alpha & \cos\alpha & 0 \\
0 & 0 & 1
\end{pmatrix} \, , \label{Eq_Euler_matrix_Z}  \\
\mathbf{R}_\text{Y}(\beta) & = 
\begin{pmatrix}
\cos\beta & 0 & \sin\beta \\
0 & 1 & 0 \\
-\sin\beta & 0 & \cos\beta
\end{pmatrix} \, , \label{Eq_Euler_matrix_Y} \\
\mathbf{R}_\text{Z}(\gamma) & = 
\begin{pmatrix}
\cos\gamma & -\sin\gamma & 0 \\
\sin\gamma & \cos\gamma & 0 \\
0 & 0 & 1
\end{pmatrix} \, ,
\end{align}
for all three Euler angles $\alpha$, $\beta$ and $\gamma$ separating the $b$-tensor's eigenbasis from the diffusion tensor's eigenbasis. One can notice that the symmetry of the diffusion tensor is preserved upon rotation, because rotations are unitary transformations. The same invariance applies to the trace of the diffusion tensor:
\begin{equation}
\mathrm{Tr}( \mathbf{D}^{(\mathbf{B})} )= D^{(\mathbf{B})}_\text{XX} + D^{(\mathbf{B})}_\text{YY} + D^{(\mathbf{B})}_\text{ZZ} = D_\text{XX} + D_\text{YY} + D_\text{ZZ} = \mathrm{Tr}( \mathbf{D} ) = 3D_\text{iso} \, . 
\end{equation}
Once the components of the rotated diffusion tensor obtained, one can use the expression of the above trace and the $b$-tensor's form derived through expanding Eq.~\eqref{Eq_b_tensor_eigenbasis_1},
\begin{equation}
\mathbf{B} =  
\begin{pmatrix}
b[1-b_\Delta(1+b_\eta)]/3  & 0 & 0 \\
0 & b[1-b_\Delta(1-b_\eta)]/3 & 0 \\
0 & 0 & b(1+2b_\Delta)/3
\end{pmatrix}\; ,
\label{Eq_b_tensor_eigenbasis_3}
\end{equation}
to work out the generalized scalar product:
\begin{align}
\mathbf{B} : \mathbf{D} & = \sum_{ij} B_{ij}\, D_{ij} \nonumber \\
 & = b_\text{XX}D^{(\mathbf{B})}_\text{XX}  + b_\text{YY}D^{(\mathbf{B})}_\text{YY} + b_\text{ZZ}D^{(\mathbf{B})}_\text{ZZ} \nonumber \\
 & = \frac{b}{3} \left[ [1-b_\Delta (1+ b_\eta)] D^{(\mathbf{B})}_\text{XX}  + [1-b_\Delta (1- b_\eta)] D^{(\mathbf{B})}_\text{YY} + (1 + 2b_\Delta) D^{(\mathbf{B})}_\text{ZZ} \right] \nonumber \\
 & = b \left[ D_\text{iso} + b_\Delta \left(D^{(\mathbf{B})}_\text{ZZ} - D_\text{iso}\right) + \frac{b_\Delta b_\eta}{3} \left(D^{(\mathbf{B})}_\text{YY} - D^{(\mathbf{B})}_\text{XX} \right) \right] \, .
\label{Eq_Start_calculations}
\end{align}
Going back to Eq.~\eqref{Eq_Change_basis_D}, one finds
\begin{align}
D^{(\mathbf{B})}_\text{XX}(\alpha,\beta,\gamma) & = \left( \cos\alpha \cos\beta\cos\gamma - \sin\alpha\sin\gamma \right)^2 D_\text{XX} + \left( \sin\alpha \cos\beta\cos\gamma + \cos\alpha\sin\gamma \right)^2 D_\text{YY} \nonumber \\
 & \qquad + \sin^2\beta\cos^2\gamma \, D_\text{ZZ} \, , \\
D^{(\mathbf{B})}_\text{YY}(\alpha,\beta,\gamma) & = \left( \cos\alpha \cos\beta\sin\gamma + \sin\alpha\cos\gamma \right)^2 D_\text{XX} + \left( \sin\alpha \cos\beta\sin\gamma - \cos\alpha\cos\gamma \right)^2 D_\text{YY} \nonumber \\
 & \qquad + \sin^2\beta\sin^2\gamma \, D_\text{ZZ} \, , \\
D^{(\mathbf{B})}_\text{ZZ}(\alpha,\beta) & = \cos^2\alpha \sin^2\beta \, D_\text{XX} + \sin^2\alpha \sin^2\beta \, D_\text{YY} + \cos^2\beta \, D_\text{ZZ} \, , 
\label{Eq_expression_D_ZZ_b}
\end{align}
where $D_\text{XX}$, $D_\text{YY}$ and $D_\text{ZZ}$ are the eigenvalues of the diffusion tensor from Eq.~\eqref{Eq_diffusion_tensor_eigenbasis}. Combining these results with the expressions of the diffusion eigenvalues as a function of $D_\text{iso}$, $D_\Delta$ and $D_\eta$ turns Eq.~\eqref{Eq_Start_calculations} into
\begin{align}
\frac{\mathbf{B} : \mathbf{D}}{b \, D_\text{iso}} & = 1 + b_\Delta D_\Delta \left[2P_2(\cos\beta) - D_\eta \sin^2\beta \cos(2\alpha) - b_\eta \sin^2\beta \cos(2\gamma)\right] \nonumber \\
 & \qquad\qquad + \frac{2b_\Delta b_\eta D_\Delta D_\eta}{3} \, \sin(2\alpha)\sin(2\gamma)\cos\beta \, ,
\end{align}
where the second-order Legendre polynomial $P_2(\cos\beta) = [3\cos^2\beta - 1]/2$ already appeared within the framework of solid-state NMR in Eq.~\eqref{Eq_Legendre}. \\

Three special cases using fewer angular variables can now be distinguished. If one considers an axisymmetric diffusion tensor and a general encoding tensor ($D_\eta = 0$):
\begin{equation}
\cbeq{
\left[\mathbf{B} : \mathbf{D}\right]_{D_\eta=0} = b \, D_\text{iso} \left( 1 + b_\Delta D_\Delta \left[2P_2(\cos\beta) - b_\eta \sin^2\beta \cos(2\gamma)\right] \right)
} \; .
\label{Eq_generalized_scalar_product_axisym_D}
\end{equation}
If one considers a general diffusion tensor and an axisymmetric encoding tensor ($b_\eta = 0$):
\begin{equation}
\cbeq{
\left[\mathbf{B} : \mathbf{D}\right]_{b_\eta=0} = b \, D_\text{iso} \left( 1 + b_\Delta D_\Delta \left[2P_2(\cos\beta) - D_\eta \sin^2\beta \cos(2\alpha)\right] \right) 
} \; .
\label{Eq_generalized_scalar_product_axisym_b}
\end{equation}
Finally, if one considers axisymmetric diffusion and encoding tensors ($D_\eta = 0$ and $b_\eta = 0$):
\begin{equation}
\cbeq{
\left[\mathbf{B} : \mathbf{D}\right]_{D_\eta=0, b_\eta=0} = b \, D_\text{iso} \left[ 1 + 2b_\Delta D_\Delta P_2(\cos\beta)\right] 
}\; .
\label{Eq_generalized_scalar_product_axisym_b_D}
\end{equation}
It seems clear from the previous equations that while spherical b-tensors isolate information about diffusion isotropy ($D_\text{iso}$), linear and planar b-tensors both encode for diffusion isotropy, anisotropy ($D_\Delta$), asymmetry ($D_\eta$) and orientation, albeit in different ways.

\subsubsection{Qualitative understanding in terms of projections}

The Frobenius inner product can be easily interpreted as a generalized scalar product in the axisymmetric diffusion Eq.~\eqref{Eq_generalized_scalar_product_axisym_D} and axisymmetric encoding Eq.~\eqref{Eq_generalized_scalar_product_axisym_b} cases. Indeed, it can then always be rewritten as a generalized scalar product using an ``effective diffusivity" $D^\text{eff}(\theta,\phi)$ \textit{via}
\begin{equation}
\cbeq{
\mathbf{B} : \mathbf{D} = b\, D^\text{eff}(\theta,\phi) = b\left[D^\text{eff}_1 \sin^2\theta \cos^2\phi + D^\text{eff}_2 \sin^2\theta \sin^2\phi  + D^\text{eff}_3 \cos^2\theta\right] 
} \; ,
\label{Eq_generalized_scalar_product_effective_diffusivity}
\end{equation}
where, for the axisymmetric diffusion case Eq.~\eqref{Eq_generalized_scalar_product_axisym_D}, $\theta \equiv \beta$, $\phi\equiv \alpha$, and
\begin{equation}
\begin{split}
D^\text{eff}_1 & = D_\text{iso}[1- D_\Delta b_\Delta (1+b_\eta)]\, , \\
D^\text{eff}_2 & = D_\text{iso}[1- D_\Delta b_\Delta (1-b_\eta)]\, ,  \\
D^\text{eff}_3 & = D_\text{iso}(1 + 2D_\Delta b_\Delta)\, ,
\end{split}
\label{Eq_component_Deff_axisymmetric_D}
\end{equation}
and for the axisymmetric encoding case Eq.~\eqref{Eq_generalized_scalar_product_axisym_b}, $\theta \equiv \beta$, $\phi\equiv \gamma$, and
\begin{equation}
\begin{split}
D^\text{eff}_1 & = D_\text{iso}[1- D_\Delta b_\Delta (1+D_\eta)]\, , \\
D^\text{eff}_2 & = D_\text{iso}[1- D_\Delta b_\Delta (1-D_\eta)] \, ,  \\
D^\text{eff}_3 & = D_\text{iso}(1 + 2D_\Delta b_\Delta)\, .
\end{split}
\label{Eq_component_Deff_axisymmetric_b}
\end{equation}
One recognizes the squares of the cartesian coordinates expressed with spherical variables (also called ``directional cosines") in Eq.~\eqref{Eq_generalized_scalar_product_effective_diffusivity}, simply because two rotation matrices are required in order to rotate tensors (against one rotation matrix to rotate vectors in an usual scalar product). In these terms, one can say that the diffusion tensor (diffusion pattern) probed by tensor-valued encoding results from the projection of a more general diffusion tensor onto the shape of the $b$-tensor. It is of crucial importance for the reader to realize that not only do the associated projections depend on the diffusion tensor's parameters, $D_\text{iso}$, $D_\Delta$ and $D_\eta$, but they also depend on the $b$-tensor's parameters $b_\Delta$ and $b_\eta$!\\

\begin{mainpoint}{title = The logic behind $b$-tensor encoding}
Eq.~\eqref{Eq_generalized_scalar_product_effective_diffusivity} shows that the shape chosen for the $b$-tensor actually selects through projections which part of the diffusion tensor is really probed in the MR signal: the effective diffusivity $D^\text{eff}(\alpha,\beta)$. Therefore, the diffusion MR data can now be more specific! However, Sec.~\ref{Sec_Tensor_metrics} will show that things are not as simple in practice when diffusion is probed in a heterogeneous medium.
\end{mainpoint}

\subsection{How to design simple gradient waveforms}

Part of this section is drawn from Ref.~\cite{Topgaard:2017}. It focuses on very simple sequences that are far from being optimized in any way, which is discussed at the end of the section. However, the mathematical building of such sequences gives some insight into the different roles played by the norm and the orientation of the spin-dephasing vector $\mathbf{q}(t)$.

\subsubsection{Building up the gradient waveform}

\begin{figure}[h!]
\begin{center}
\includegraphics[width=0.7\textwidth]{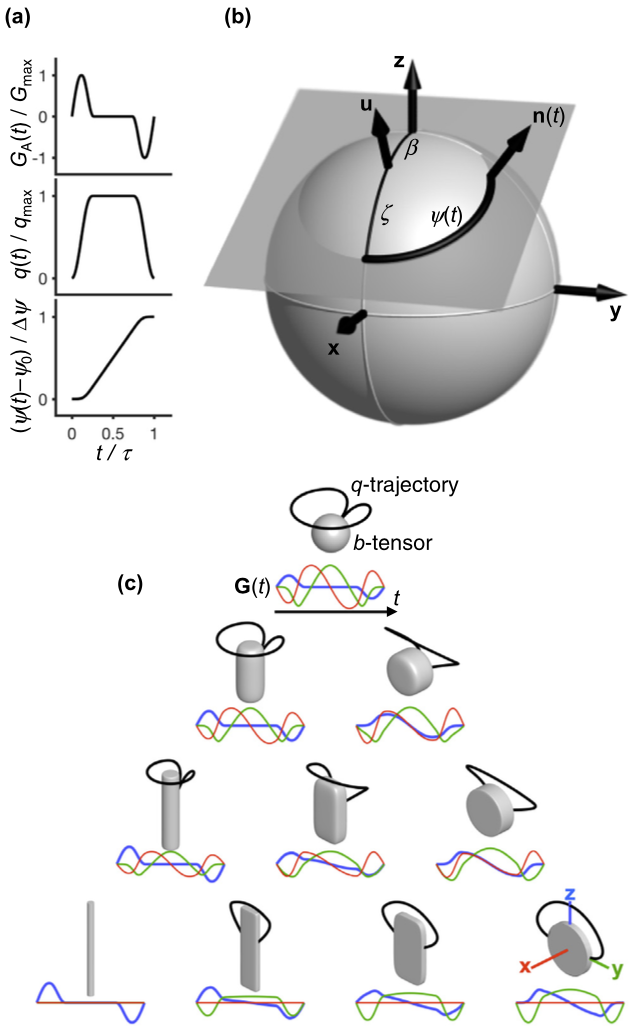}
\caption{\textbf{a)} Design of an axial gradient waveform $G_\text{A}(t)$ satisfying the echo condition Eq.~\eqref{Eq_echo_condition} for the $q$-vector, ensuring the correct $b$-value ($b$-tensor size) and parametrizing the angle of rotation $\psi(t)$ during the $q$-trajectory \textit{via} Eq.~\eqref{Eq_angle_q_trajectory}. \textbf{b)} Framework characterizing the rotation of the unit vector $\mathbf{n}(t)$ around the axis of rotation $\mathbf{u}$. \textbf{c)} Representations of different $q$-trajectories with their corresponding $b$-tensors and gradient waveforms. Figure taken from Ref.~\cite{Topgaard:2017}.}
\label{Figure_Topgaard_trajectory_detail}
\end{center}
\end{figure}

Remaining within the framework of axisymmetric $b$-tensor encoding, \textit{i.e.} $b_\eta = 0$, one needs to create a gradient waveform that respects the echo condition Eq.~\eqref{Eq_echo_condition} while ensuring the correct size and shape for the $b$-tensor. To solve this problem, it is important to factorize the $q$-vector into the usual ``norm-orientation" form: $\mathbf{q}(t) = q(t)\, \mathbf{n}(t)$, where $\mathbf{n}(t)$ is a unit vector. Now, obtaining the right echo condition and $b$-tensor size, \textit{i.e.} the right $b$-value, only depends on the norm $q(t)$. Indeed, one can choose any axial (one-directional or one-dimensional) traditional gradient waveform $G_\text{A}(t)$ (such as the Stejskal-Tanner one, see Sec.~\ref{Sec_MRI_sequences}) to build the norm $q(t)$, since this kind of waveform already satisfies the echo condition Eq.~\eqref{Eq_echo_condition} and can be molded so that it imposes the right $b$-value to the overall sequence. Therefore, one has
\begin{equation}
q(t) = \gamma \int_0^t \! G_\text{A} (t') \,\mathrm{d}t' \qquad \text{and} \qquad b = \int_0^\tau \! q^2(t) \, \mathrm{d}t \, .
\label{Eq_axial_waveform}
\end{equation}
This step is shown in the two top panels of Fig.~\ref{Figure_Topgaard_trajectory_detail}\textcolor{Red}{a}.\\

Now, obtaining the right $b$-tensor shape solely depends on making the unit vector $\mathbf{n}(t)$ rotate through time along the correct trajectory, in analogy with multidimensional solid-state NMR experiments (see Sec.~\ref{Sec_solid_NMR_inspiration}). Mimicking the latter experiments, let us make the unit vector $\mathbf{n}(t)$ follow the intersection between a plane and a unit sphere, as shown in Fig.~\ref{Figure_Topgaard_trajectory_detail}\textcolor{Red}{b}. This plane is perpendicular to the axis of rotation $\mathbf{u}$, so that the $q$-vector always remains at an angle $\zeta \in [0,\pi/2]$ from this axis. While the axis of rotation is at an angle $\beta$ from the lab frame $z$-axis, the plane is located at a distance $\cos\zeta$ from the origin of the lab frame of reference. To finish setting up the stage for the next calculations, one defines the angle of rotation of the unit vector $\mathbf{n}(t)$ around the axis of rotation $\mathbf{u}$ as
\begin{equation}
\psi(t) = \psi_0 + \frac{\Delta \psi}{b}\int_0^t\! q^2(t')\,\mathrm{d}t' \, ,
\label{Eq_angle_q_trajectory}
\end{equation}
where $\psi_0 = \psi(t=0)$ is the initial angle of rotation and $\Delta \psi = \psi(\tau) - \psi_0$ is the total angle overcome during the $q$-trajectory, so that the angle of rotation $\psi(t)$ is in fact parametrized by the axial gradient waveform $G_\text{A}(t)$ \textit{via} Eq.~\eqref{Eq_axial_waveform}, as shown in the bottom panel of Fig.~\ref{Figure_Topgaard_trajectory_detail}\textcolor{Red}{a}.

\newpage

\vspace*{\fill}

\vspace*{\fill}

\newpage

Going back to Fig.~\ref{Figure_Topgaard_trajectory_detail}\textcolor{Red}{b}, one writes the $q$-vector in the basis $\{\mathbf{u}\}$ linked to the axis of rotation (the third component is along this axis) as 
\begin{equation}
\mathbf{q}(t) = q(t) \, \mathbf{n}(t) = \left[\gamma \int_0^t \! G_\text{A} (t') \,\mathrm{d}t' \right] 
\begin{pmatrix}
\cos(\psi(t)) \sin\zeta \\
\sin(\psi(t)) \sin\zeta \\
\cos\zeta
\end{pmatrix}_{\!\!\{\mathbf{u}\}}
\label{Eq_expression_n_q_u}
\end{equation}
using Eq.~\eqref{Eq_axial_waveform}. Given the fact that the real gradient generated by the $q$-trajectory should satisfy $\mathbf{q}(t) = \gamma \int_0^t \! \mathbf{G} (t') \,\mathrm{d}t'$, it is obtained as
\begin{equation}
\mathbf{G}(t) 
= 
\frac{1}{\gamma}\, \frac{\mathrm{d}\mathbf{q}}{\mathrm{d}t} 
= 
G_\text{A}(t) \begin{pmatrix}
\cos(\psi(t)) \sin\zeta \\
\sin(\psi(t)) \sin\zeta \\
\cos\zeta
\end{pmatrix}_{\!\!\{\mathbf{u}\}}
+ \frac{q(t)}{\gamma} \begin{pmatrix}
\displaystyle -\frac{\mathrm{d}\psi}{\mathrm{d}t}\sin(\psi(t)) \sin\zeta \\[7pt]
\displaystyle \frac{\mathrm{d}\psi}{\mathrm{d}t}\cos(\psi(t)) \sin\zeta \\[7pt]
\cos\zeta
\end{pmatrix}_{\!\!\{\mathbf{u}\}} \, .
\end{equation}
One can then use Eq.~\eqref{Eq_angle_q_trajectory} to work out the time derivative of the rotation angle $\psi(t)$:
\begin{equation}
\frac{\mathrm{d}\psi}{\mathrm{d}t} = \frac{\Delta \psi}{b}\, q^2(t)\, ,
\label{Eq_link_angle_q}
\end{equation}
so that
\begin{equation}
\mathbf{G}(t) =
\begin{pmatrix}
\displaystyle G_\text{A}(t) \cos(\psi(t)) \sin\zeta - \frac{\Delta\psi}{\gamma b}\, q^3(t) \sin(\psi(t))\sin \zeta \\[7pt]
\displaystyle G_\text{A}(t) \sin(\psi(t)) \sin\zeta + \frac{\Delta\psi}{\gamma b}\, q^3(t) \cos(\psi(t))\sin \zeta \\[7pt]
G_\text{A}(t) \cos\zeta
\end{pmatrix}_{\!\!\{\mathbf{u}\}}
\, .
\end{equation}
Introducing the complex radial gradient
\begin{equation}
G_\text{R} (t) = \left[ G_\text{A}(t) + i\, \frac{\Delta \psi}{\gamma b}\, q^3(t) \right] \exp^{i\psi(t)} = \left[ G_\text{A}(t) + i\, \frac{\Delta \psi}{\gamma b}\, q^3(t) \right] \left[ \cos(\psi(t)) + i \sin(\psi(t))\right] \, ,
\end{equation}
one finally writes the total gradient in a short-hand notation:
\begin{equation}
\mathbf{G}(t) =
\begin{pmatrix}
\displaystyle \mathrm{Re} \left[ G_\text{R}(t)\right] \sin\zeta \\
\displaystyle \mathrm{Im} \left[ G_\text{R}(t)\right] \sin\zeta \\
G_\text{A}(t) \cos\zeta
\end{pmatrix}_{\!\!\{\mathbf{u}\}}
\, .
\end{equation}
The reader may notice that the name ``radial gradient" comes from the fact that this gradient only appears in the radial components of the total gradient.\\

In the axisymmetric case, only a rotation of the gradient waveform axis $\mathbf{u}$ by the Euler angle $\beta$ is relevant, since the $q$-trajectory is axisymmetric and is consequently invariant under $\alpha$ and $\gamma$ rotations. The rotated gradient waveform, expressed in the lab frame of reference $\{ \text{lab}\}$, therefore writes
\begin{equation}
\mathbf{G}_\text{rot.}(t) = \mathbf{R}_\text{Y}(\beta) \, \mathbf{G}(t) = 
\begin{pmatrix}
\mathrm{Re} \left[ G_\text{R}(t)\right] \sin\zeta \cos\beta + G_\text{A}(t) \cos\zeta \sin\beta \\
\mathrm{Im} \left[ G_\text{R}(t)\right] \sin\zeta \\
G_\text{A}(t) \cos\zeta \cos\beta - \mathrm{Re} \left[ G_\text{R}(t)\right] \sin\zeta \sin\beta
\end{pmatrix}_{\!\!\{\text{lab}\}}
\, ,
\end{equation}
where the rotation matrix $\mathbf{R}_\text{Y}(\beta)$ is defined in Eq.~\eqref{Eq_Euler_matrix_Y}.

\subsubsection{Corresponding $\mathbold{b}$-tensor's eigenvalues}

One can know work out the expression of the $b$-tensor from Eq.~\eqref{Eq_b_tensor_definition}. Indeed, using Eq.~\eqref{Eq_expression_n_q_u} and Eq.~\eqref{Eq_link_angle_q} to write 
\begin{align}
\mathbf{n}(t) \cdot\mathbf{n}^{\mathrm{T}}(t) & =
\begin{pmatrix}
\cos^2(\psi(t)) \sin^2\zeta & \displaystyle \frac{\sin(2\psi(t)) \sin^2\zeta}{2}  & \displaystyle \frac{\cos(\psi(t))\sin(2\zeta)}{2}   \\[7pt]
\displaystyle \frac{\sin(2\psi(t)) \sin^2\zeta}{2} & \sin^2(\psi(t)) \sin^2\zeta & \displaystyle \frac{\sin(\psi(t))\sin(2\zeta)}{2}  \\[7pt]
\displaystyle \frac{\cos(\psi(t))\sin(2\zeta)}{2} & \displaystyle \frac{\sin(\psi(t))\sin(2\zeta)}{2} & \cos^2\zeta
\end{pmatrix}_{\!\!\{\mathbf{u}\}} \!\!\!\equiv \mathbf{n}(\psi(t)) \cdot\mathbf{n}^{\mathrm{T}}(\psi(t)) 
\label{Eq_matrix_n_nT}
\end{align}
and
\begin{equation}
q^2(t) = \frac{b}{\Delta\psi} \, \frac{\mathrm{d}\psi}{\mathrm{d}t}\, ,
\end{equation}
the $b$-tensor is given by
\begin{equation}
\mathbf{B} = \frac{b}{\Delta\psi} \int_0^\tau \mathbf{n}(\psi(t)) \cdot\mathbf{n}^{\mathrm{T}}(\psi(t))\, \frac{\mathrm{d}\psi}{\mathrm{d}t} \,\mathrm{d}t\, .
\end{equation}
By operating the change of variable $t \longrightarrow \psi(t) \equiv \psi$, one then obtains
\begin{align}
\mathbf{B} & = \frac{b}{\Delta\psi} \int_{\psi_0}^{\psi_0 + \Delta \psi} \mathbf{n}(\psi) \cdot\mathbf{n}^{\mathrm{T}}(\psi)\, \mathrm{d}\psi \nonumber \\
& = \frac{b}{\Delta\psi} \int_{\psi_0}^{\psi_0 + \Delta \psi} 
\begin{pmatrix}
\cos^2\psi \sin^2\zeta & \displaystyle \frac{\sin(2\psi) \sin^2\zeta}{2}  & \displaystyle \frac{\cos\psi\sin(2\zeta)}{2}   \\[7pt]
\displaystyle \frac{\sin(2\psi) \sin^2\zeta}{2} & \sin^2\psi \sin^2\zeta & \displaystyle \frac{\sin\psi\sin(2\zeta)}{2}  \\[7pt]
\displaystyle \frac{\cos\psi\sin(2\zeta)}{2} & \displaystyle \frac{\sin\psi\sin(2\zeta)}{2} & \cos^2\zeta
\end{pmatrix}_{\!\!\{\mathbf{u}\}}  
\!\!\! \mathrm{d}\psi \, .
\label{Eq_matrix_n_nT2}
\end{align}
\smallskip

Now, let us consider the case of a full rotation around the axis $\mathbf{u}$, \textit{i.e.} $\Delta \psi \equiv 2\pi\; \mathrm{rad}$, which is perfectly illustrated in Fig.~\ref{Figure_Topgaard_trajectory_detail}\textcolor{Red}{c}. Once integrated over $\psi$, all the non diagonal terms in Eq.~\eqref{Eq_matrix_n_nT2} then cancel, since the different non diagonal cosines and sines are integrated over a multiple of their period. As for the diagonal terms, one can use
\begin{equation}
\int_{\psi_0}^{\psi_0 + 2\pi} \! \cos^2\psi \, \mathrm{d}\psi = \frac{1}{2} \int_{\psi_0}^{\psi_0 + 2\pi} \! [1+ \cos(2\psi)] \, \mathrm{d}\psi = \pi
\end{equation}
and
\begin{equation}
\int_{\psi_0}^{\psi_0 + 2\pi} \! \sin^2\psi \, \mathrm{d}\psi = \frac{1}{2} \int_{\psi_0}^{\psi_0 + 2\pi} \! [1 - \cos(2\psi)] \, \mathrm{d}\psi = \pi
\end{equation}
to find that
\begin{equation}
\mathbf{B} = b 
\begin{pmatrix}
\displaystyle \frac{\sin^2\zeta}{2} & 0 & 0  \\[7pt]
0 & \displaystyle  \frac{\sin^2\zeta}{2} & 0  \\[7pt]
0 & 0 & \cos^2\zeta
\end{pmatrix}_{\!\!\{\mathbf{u}\}}
\, .
\end{equation}
\bigskip

At this point, it may be relevant to wonder within which eigenbasis one is actually working: Eq.~\eqref{Eq_b_tensor_eigenbasis_1} or Eq.~\eqref{Eq_b_tensor_eigenbasis_2}? To answer that, let us take two limit cases. First, let us take $\zeta$ equal to the magic angle $\zeta_\text{m}$. From Eq.~\eqref{Eq_Legendre}, one has $P_2 (\cos\zeta_\text{m}) = [3\cos^2\zeta_\text{m} - 1]/2 = 0$, so that $\cos^2\zeta_\text{m} = 1/3$, which implies that $\sin^2\zeta_\text{m} = 2/3$ and
\begin{equation}
\mathbf{B}(\zeta = \zeta_\text{m}) = \frac{b}{3} 
\begin{pmatrix}
1 & 0 & 0  \\
0 & 1 & 0  \\
0 & 0 & 1
\end{pmatrix}_{\!\!\{\mathbf{u}\}} 
= b_\text{iso}
\begin{pmatrix}
1 & 0 & 0  \\
0 & 1 & 0  \\
0 & 0 & 1
\end{pmatrix}_{\!\!\{\mathbf{u}\}}
\, .
\end{equation}
One retrieves the solid-state NMR experiment described in Sec.~\ref{Sec_solid_NMR_inspiration}! Second, let us take $\zeta = 0$. From Fig.~\ref{Figure_Topgaard_trajectory_detail}\textcolor{Red}{b}, the $\zeta = 0$ $q$-trajectory is clearly associated to a linear $b$-tensor. In that case, one has
\begin{equation}
\mathbf{B}(\zeta = 0) = b 
\begin{pmatrix}
0 & 0 & 0  \\
0 & 0 & 0  \\
0 & 0 & 1
\end{pmatrix}_{\!\!\{\mathbf{u}\}} 
\, .
\end{equation}
While the first case corresponds to a purely spherical $b$-tensor ($b_\text{S} = b$, $b_\text{P} = 0$ and $b_\text{L} = 0$), the second case corresponds to a purely linear $b$-tensor ($b_\text{S} = 0$, $b_\text{P} = 0$ and $b_\text{L} = b$). Planar encoding is obtained for $\zeta = \pi/2$. Therefore, only the basis used in Eq.~\eqref{Eq_b_tensor_eigenbasis_2} fits these observations. Then, one easily obtains
\begin{equation}
(b_\text{S}, b_\text{P}, b_\text{L}) = \bigg( b\left[ 1 - P_2 (\cos\zeta) \right] , \; 0,\;  b\, P_2 (\cos\zeta) \bigg)
\qquad \text{and} \qquad
(b_\Delta, b_\eta) = \bigg( P_2(\cos\zeta) ,\; 0 \bigg) 
\end{equation}
from Eq.~\eqref{Eq_b_tensor_eigenbasis_2} and Sec.~\ref{Sec_link_conventions}.\\

\vspace*{\fill}

\begin{sidenote}{title = First attempts at $b$-tensor encoding}
Non-trivial $b$-tensor encoding has already been investigated in 1990 by the use of double diffusion encoding in order to yield planar tensor encoding \cite{Cory:1990}, and in 1995 by the use of a spherically encoded sequence \cite{Wong:1995}. However, nothing as generalized as the present approach had been done at the time.
\end{sidenote}

\medskip

\begin{mainpoint}{title = More optimized $q$-space trajectories}
The above sequences are not optimized because they rely on conal $q$-trajectories. Very optimized sequences would remain as long as possible either on the edges of the cube of maximum gradient amplitude ($L_1$ norm) or on the surface of a sphere of large gradient amplitude ($L_2$ norm) to ensure non time-consuming diffusion weighting. While the former is technically more optimized, the latter allows for rotation of gradient waveforms (for planar encoding for example). Such optimized $q$-trajectories can be found in Ref.~\cite{Sjolund:2015}. \\

Another aspect to take into account is ``Maxwell compensation". Using multiple strong gradients at the same time enhances cross terms in $\Vert \mathbf{G}\Vert^2$ developing progressively from the isocenter in the scanner, leading to misestimation of diffusivities away from the isocenter. These terms, called ``Maxwell terms", can be nulled or minimized when designing the sequence, as discussed in Ref.~\cite{Szczepankiewicz_Maxwell:2019} \\

A last important aspect of sequence optimization lies in the frequency content of the waveforms. Let us take the example of a spherical acquisition. If the three underlying gradients of this sequence do not share the same frequency content, they actually probe different diffusion times, hence are anisotropic with respect to time-dependent diffusion. Solving this is called ``spectral tuning" or ``spectral matching". The reader is invited to take a look at Refs.~\cite{Lundell_ISMRM:2017, Lasic:2017} for more details about this, especially on how to use ``detuned" sequences to probe time-dependent diffusion.
\end{mainpoint}

\section{Ensembles of diffusion tensors}
\label{Sec_Tensor_metrics}

Most of this section is drawn from Refs.~\cite{SzczepankiewiczThesis:2016} and \cite{Reymbaut_book:2019}.

\subsection{Diffusion tensor distributions}

As discussed in Sec.~\ref{Sec_DTI_limitations}, most investigated samples/tissues are heterogeneous over the spatial resolution of diffusion NMR/MRI measurements, which implies that the measured diffusion signal is a combination of signals arising from a variety of microstructural environments. This means that drawing from a voxel-scale diffusion tensor to describe a single effective diffusion environment (\textit{i.e.} DTI) could easily lead to incoherent microstructural interpretations when it comes to studying microscopic diffusion anisotropy. Indeed, a collection of randomly oriented anisotropic structures would yield a globally isotropic voxel-scale diffusion tensor, missing all microscopic anisotropy. In order to describe a collection of diffusion environments, a new mathematical object is required: the diffusion tensor distribution (DTD).\\

For simplicity, let us focus on axisymmetric diffusion tensors, parametrized $(D_\text{iso},D_\Delta)$ or $(D_\parallel,D_\perp)$ following Eq.~\eqref{Eq_lambda_para_perp}, that possess only four independent elements (including angular dependency). This reduction can be understood in terms of irrelevant degrees of freedom. Indeed, with axisymmetry, the asymmetry parameter becomes irrelevant and only two angles are required to describe orientation. The full four-dimensional DTD can now be written as the joint probability distribution $\mathcal{P}(\mathbf{D})\equiv\mathcal{P}(D_\text{iso},D_\Delta, \theta,\phi)$, normalized as
\begin{equation}
\cbeq{
\int\! \mathcal{P}(\mathbf{D})\, \mathrm{d}\mathbf{D} \equiv \int_0^{+\infty}\int_{-0.5}^1\int_0^{2\pi}\int_0^\pi \!\mathcal{P}(D_\text{iso},D_\Delta, \theta,\phi) \, \sin\theta\, \mathrm{d}\theta\,\mathrm{d}\phi\,\mathrm{d}D_\Delta\, \mathrm{d}D_\text{iso} = 1
} \; . 
\end{equation}
Note that we will denote all probability distributions simply as ``$\mathcal{P}$" and will let the arguments clarify the type of distribution.

\subsection{Size, shape and orientation distributions}

Lower-dimensional projections of the full DTD can also be studied. For instance, integration over orientations leads to the two-dimensional size-shape distribution
\begin{equation}
\mathcal{P}(D_\text{iso}, D_\Delta) = \int_0^{2\pi}\int_0^\pi\! \mathcal{P}(D_\text{iso}, D_\Delta)\, \sin\theta\, \mathrm{d}\theta\,\mathrm{d}\phi\, ,
\end{equation}
which, upon integration over the anisotropy dimension, gives the one-dimensional size distribution
\begin{equation}
\mathcal{P}(D_\text{iso}) = \int_{-0.5}^1\! \mathcal{P}(D_\text{iso},D_\Delta)\, \mathrm{d}D_\Delta\, ,
\end{equation}
or, upon integration over the isotropy dimension, gives the one-dimensional shape distribution
\begin{equation}
\mathcal{P}(D_\Delta) = \int_{0}^{+\infty}\! \mathcal{P}(D_\text{iso},D_\Delta)\, \mathrm{d}D_\text{iso}\, .
\end{equation}
To quantify the angular part of $\mathcal{P}(D_\text{iso}, D_\Delta,\theta,\phi)$, one can instead compute the orientation distribution function (ODF)
\begin{equation}
\mathcal{P}(\theta,\phi) = \int_0^{+\infty} \int_{-0.5}^1\!\mathcal{P}(D_\text{iso}, D_\Delta,\theta,\phi)\, \mathrm{d}D_\Delta\, \mathrm{d}D_\text{iso}\, .
\label{Eq_ODF}
\end{equation}

\subsection{Ensemble averages and variances}
\label{Eq_ensemble_averages_variances}

In practice, even projections such as $\mathcal{P}(D_\text{iso}, D_\Delta)$ and $\mathcal{P}(D_\text{iso})$ can be challenging to estimate. Therefore, it is sometimes useful to focus on various means and variances of these distributions. Denoting $\langle \cdot \rangle$ and $\mathrm{Var}(\cdot)$ the macroscopic ensemble (\textit{i.e.} voxel-scale) average and variance, respectively, and using the eigenvalue average and variance Eqs.~\eqref{Eq_average_eigenvalue} and \eqref{Eq_variance_eigenvalue}, such quantities include the mean diffusivity
\begin{equation}
\cbeq{
\mathrm{MD} = \langle D_\text{iso}\rangle = \langle \mathbb{E}_\lambda[\mathbf{D}] \rangle = \mathbb{E}_\lambda[\langle \mathbf{D}\rangle] = \int_0^{+\infty}\! D_\text{iso}\,\mathcal{P}(D_\text{iso})\, \mathrm{d}D_\text{iso}
}\; ,
\label{Eq_MD_DTD}
\end{equation}
the variance of isotropic diffusivities
\begin{equation}
\cbeq{
V_\text{iso} = \mathrm{Var}(\mathbb{E}_\lambda[\mathbf{D}]) = \int_0^{+\infty}\! (D_\text{iso}-\langle D_\text{iso} \rangle)^2 \mathcal{P}(D_\text{iso})\, \mathrm{d}D_\text{iso} = \langle D_\text{iso}^2\rangle - \langle D_\text{iso}\rangle^2
}\; ,
\label{Eq_Viso_DTD}
\end{equation}
the average variance of diffusion tensor eigenvalues
\begin{align}
\cbeq{
\langle V_\lambda \rangle = \langle \mathbb{V}_\lambda [\mathbf{D}] \rangle = \int_0^{+\infty} \int_{-0.5}^1\! V_\lambda(D_\text{iso},D_\Delta)\, \mathcal{P}(D_\text{iso},D_\Delta)\, \mathrm{d}D_\Delta\, \mathrm{d}D_\text{iso} = 2\langle D_\text{aniso}^2\rangle
}\; ,
 \label{Eq_average_variance_Daniso}
\end{align}
following Eq.~\eqref{Eq_variance_eigenvalues}, and the variance of variances of diffusion tensor eigenvalues
\begin{equation}
\cbeq{
V_\Delta = \mathrm{Var}(\mathbb{V}_\lambda[\mathbf{D}]) = 4\,\mathrm{Var}(D_\text{aniso}^2) = 4\left[ \langle D_\text{aniso}^4 \rangle - \langle D_\text{aniso}^2\rangle^2 \right]
} \; . 
\end{equation}
While $\langle D_\text{iso}\rangle$ and $V_\text{iso}$ relate to the mean and spread of tensor sizes, $\langle V_\lambda\rangle$ and $V_\Delta$ capture microscopic diffusion anisotropy and the variance in microscopic diffusion tensor shapes. As for orientational averages, they write similarly as
\begin{equation}
\langle f(\theta,\phi) \rangle = \int_0^{2\pi}\int_0^\pi f(\theta,\phi)\, \mathcal{P}(\theta,\phi)\, \sin\theta\, \mathrm{d}\theta\, \mathrm{d}\phi\, ,
\end{equation}
where $f(\theta,\phi)$ denotes any angular function.

\subsection{Simple ensemble-averaged diffusion tensor - Saupe order tensor}

Let us consider a voxel consisting of identically shaped axisymmetric diffusion tensors whose orientations may obey any ODF. Starting from the diagonal parametrization Eq.~\eqref{Eq_lambda_para_perp} for $\mathbf{D}$, rotation  according to $\mathbf{D}(\theta,\phi)=\mathbf{R}(\theta,\phi)\cdot \mathbf{D}\cdot \mathbf{R}^\text{T}(\theta,\phi)$, with $\mathbf{R}(\theta,\phi) = \mathbf{R}_\text{Z}(\phi)\cdot\mathbf{R}_\text{Y}(\theta)$ (see Eqs.~\eqref{Eq_Euler_matrix_Z} and \eqref{Eq_Euler_matrix_Y}), yields the general diffusion tensor
\begin{equation}
\mathbf{D}(\theta,\phi) = D_\text{iso}
\left[ 
\mathbf{I}_3 + D_\Delta
\begin{pmatrix}
3u_x^2-1 & 3u_x u_y & 3u_x u_z \\
3u_x u_y & 3u_y^2-1 & 3u_y u_z \\
3u_x u_z & 3u_y u_z & 3u_z^2-1
\end{pmatrix}
\right] ,
\end{equation}
with the directional cosines $u_x = \sin\theta\cos\phi$, $u_y = \sin\theta\sin\phi$ and $u_z = \cos\theta$. The ensemble average of this last equation gives
\begin{equation}
\cbeq{
\langle \mathbf{D}\rangle = D_\text{iso}(\mathbf{I}_3 + 2D_\Delta \mathbf{S}) 
} \; ,
\label{Eq_link_averaged_D_Saupe}
\end{equation}
where $\mathbf{I}_3$ is the 3$\times$3 identity matrix, $\langle \mathbf{D}\rangle$ is the ensemble-averaged diffusion tensor and $\mathbf{S}$ is the Saupe order tensor.\\

\begin{sidenote}{title = Saupe order tensor}
The Saupe order tensor $\mathbf{S}$, that usually quantifies orientational order in liquid crystals \cite{Saupe:1963}, is given by 
\begin{equation}
\mathbf{S} = \frac{1}{2} \left\langle
\begin{pmatrix}
3u_x^2-1 & 3u_x u_y & 3u_x u_z \\
3u_x u_y & 3u_y^2-1 & 3u_y u_z \\
3u_x u_z & 3u_y u_z & 3u_z^2-1
\end{pmatrix}
\right\rangle \, .
\end{equation}
More generally, its elements write $S_{ij} = [3\langle u_i u_j \rangle - \delta_{ij}]/2$. This 3$\times$3 tensor is symmetric and traceless, implying that it contains five independent parameters and that it cannot be directly visualized as a glyph (since it has no size). This tensor can still be diagonalized in its eigenbasis, yielding
\begin{equation}
\mathbf{S} = 
\begin{pmatrix}
S_\text{XX} & 0 & 0 \\
0 & S_\text{YY} & 0 \\
0 & 0 & S_\text{ZZ} 
\end{pmatrix}
\end{equation}
with the elements ordered according to Eq.~\eqref{Eq_Lambda_Haeberlen_convention}. Since $\mathbf{S}$ is traceless, only two quantities, $S_\text{ZZ}$ and $S_\text{XX}-S_\text{YY}$, are needed to define its shape. The employed ordering convention imposes that $\vert S_\text{ZZ}\vert > \vert S_\text{XX} - S_\text{YY} \vert$. This justifies the fact that $S_\text{ZZ}$ is usually called the principal order parameter, ranging from $−1/2$ to $+1$, as shown later in Eq.~\eqref{Eq_link_averaged_D_Saupe}. While perfect alignment in a single direction corresponds to $S_\text{ZZ}=1$, random orientations in a plane perpendicular to the ZZ eigenvector gives $S_\text{ZZ} = −1/2$. The value $S_\text{ZZ}=0$ could indicate an isotropic distribution of orientations, as well as more exotic orientation distributions, \textit{e.g.} three orthogonal directions with equal probability. In order to obtain a glyph representation of $\mathbf{S}$, one can use the shifted Saupe tensor
\begin{equation}
\mathbf{S}^\prime = \frac{1}{3}\, (\mathbf{I}_3 + 2\mathbf{S})\, .
\end{equation}
The eigenvalues of this symmetric and unit-trace tensor are all positive, covering the range from $0$ to $1$. Its shape can be interpreted as a coarse version of the ODF Eq.~\eqref{Eq_ODF}.
\end{sidenote}
\medskip

Going back to Eq.~\eqref{Eq_link_averaged_D_Saupe}, this remarkably simple relation between microscopic anisotropy $D_\Delta$, orientational order $\mathbf{S}$ (or $\mathbf{S}^\prime$), and macroscopic anisotropy $\langle \mathbf{D}\rangle$ indicates that the eigenvectors of $\langle \mathbf{D}\rangle$ and $\mathbf{S}^{(\prime)}$ are identical, and that the shape of $\langle \mathbf{D}\rangle$ can be interpreted as the shape of $\mathbf{S}^\prime$ modulated by the value of $D_\Delta$. Noticing that the ZZ element $\lambda_\text{ZZ}$ of $\langle \mathbf{D}\rangle$ in its eigenbasis writes $\lambda_\text{ZZ}=D_\text{iso}(1+2D_\Delta S_\text{ZZ})$, one can define the anisotropy parameter $\langle D_\Delta \rangle$ of the ensemble-averaged diffusion tensor by analogy with Eq.~\eqref{Eq_lambda_para_perp}:
\begin{equation}
\cbeq{
\langle D_\Delta \rangle = D_\Delta S_\text{ZZ}
}\; ,
\label{Eq_average_Delta_Szz}
\end{equation}
offering an even simpler relationship between microscopic anisotropy $D_\Delta$, orientational order $S_\text{ZZ}$ and macroscopic anisotropy $\langle D_\Delta \rangle$.\\

\begin{mainpoint}{title = Macroscopic metrics VS microscopic metrics}
The fractional anisotropy (FA) Eq.~\eqref{Eq_FA_variance_eigenvalues} can be rewritten as
\begin{equation}
\mathrm{FA} = \sqrt{\frac{3}{2}} \left( 1 + \frac{\mathbb{E}_\lambda^2[\langle \mathbf{D}\rangle]}{\mathbb{V}_\lambda[\langle \mathbf{D}\rangle]} \right)^{-1/2}\, .
\label{Eq_FA_mathbb}
\end{equation}
The value of this formalism is to underline the mathematical difference between macroscopic and microscopic metrics in terms of diffusion eigenvalue statistics and ensemble statistics. Indeed, comparing the microscopic metrics of Sec.~\ref{Eq_ensemble_averages_variances} with the macroscopic FA, it is clear that while macroscopic metrics require to evaluate eigenvalue statistics on the ensemble- averaged diffusion tensor, microscopic metrics rely on taking ensemble statistics of various eigenvalue statistics computed over the microscopic tensors.
\end{mainpoint}

\section{Extracting microscopic diffusion anisotropy from the signal}
\label{Sec_new_metrics}

This section, mostly taken from Ref.~\cite{Reymbaut_book:2019}, presents different methods designed to disentangle the effects of cell sizes, shapes, and orientations directly on the outcome of the measurement, without any biophysical modeling. It starts with techniques based on the orientationally averaged (powder-averaged) signal, analogous to the signal measured in diffusion powders, and finishes with the tensor covariance approximation, that does not require any signal averaging.

\subsection{Powder-averaging}

\subsubsection{Signal measured in diffusion ``powders"}

The special case of an isotropic distribution of orientations is typically obtained in X-ray diffraction and solid-state NMR by crushing the investigated sample into a fine powder. Let us write the diffusion signal using our updated notations:
\begin{equation}
\frac{\mathcal{S}}{\mathcal{S}_0} = \int_0^{+\infty}\int_{-0.5}^1\int_0^{2\pi}\int_0^\pi \! \mathrm{exp}(-\mathbf{B}:\mathbf{D})\, \mathcal{P}(D_\text{iso},D_\Delta, \theta,\phi) \, \sin\theta\, \mathrm{d}\theta\,\mathrm{d}\phi\,\mathrm{d}D_\Delta\, \mathrm{d}D_\text{iso} \, ,
\end{equation}
where the Frobenius inner product Eq.~\eqref{Eq_generalized_scalar_product_axisym_D} depends on the four integration dimensions. If the investigated sample constitutes a ``powder" of diffusion tensors with identical size and shape, this expression reduces to
\begin{equation}
\frac{\mathcal{S}}{\mathcal{S}_0} = \int_0^{+\infty}\int_{-0.5}^1\int_0^{2\pi}\int_0^\pi \! \mathrm{exp}(-\mathbf{B}:\mathbf{D})\, \mathcal{P}(\theta,\phi) \, \sin\theta\, \mathrm{d}\theta\,\mathrm{d}\phi \, ,
\end{equation}
Assuming that the angles $\theta$ and $\phi$ refer to the orientation of the diffusion tensor in the $b$-tensor's eigenbasis, one can use the expression Eq.~\eqref{Eq_generalized_scalar_product_effective_diffusivity} of the Frobenius inner product as a generalized scalar product defining an effective diffusivity $D^\text{eff}$ as a function of the diffusivities $D^\text{eff}_1$, $D^\text{eff}_2$ and $D^\text{eff}_3$ Eq.~\eqref{Eq_component_Deff_axisymmetric_D}. Drawing from Ref.~\cite{Schmidt-Rohr_Spiess_Book:1994}, one can replace the above integral over orientations with an integral over effective diffusion coefficients $D^\text{eff} \equiv D$, yielding the one-dimensional Laplace transform
\begin{equation}
\cbeq{
\frac{\mathcal{S}}{\mathcal{S}_0} = \int_0^{+\infty} \! \mathrm{exp}(-bD)\, \mathcal{P}(D) \, \mathrm{d}D
}
\label{Eq_Laplace_1D}
\end{equation}
for the apparent diffusion coefficient (ADC) distribution $\mathcal{P}(D)$. Notice that $\mathcal{P}(D)$ now contains information about both the shapes of the b- and diffusion tensors. By analogy between the ADC distribution $\mathcal{P}(D)$ and the lineshape function originating from chemical shift anisotropy \cite{Bloembergen:1953}, one obtains 
\begin{equation}
\mathcal{P}(D)=
\begin{cases}
\displaystyle \frac{1}{\pi\sqrt{(D^\text{eff}_3-D)(D^\text{eff}_2-D^\text{eff}_1)}}\, K\!\left[ \frac{(D^\text{eff}_3-D^\text{eff}_2)(D-D^\text{eff}_1)}{(D^\text{eff}_3-D)(D^\text{eff}_2-D^\text{eff}_1)} \right] & \text{for $D^\text{eff}_1<D<D^\text{eff}_2$} \\
\displaystyle \frac{1}{\pi\sqrt{(D^\text{eff}_3-D^\text{eff}_2)(D-D^\text{eff}_1)}}\, K\!\left[ \frac{(D^\text{eff}_3-D)(D^\text{eff}_2-D^\text{eff}_1)}{(D^\text{eff}_3-D^\text{eff}_2)(D-D^\text{eff}_1)} \right] & \text{for $D^\text{eff}_2<D<D^\text{eff}_3$} \\
0 & \text{otherwise,}
\end{cases}
\label{Eq_bigass_P_ADC}
\end{equation}
where $K(\cdot)$ is the complete elliptic integral of the first kind. Here, the ordering convention of the effective diffusivities has to be reversed if $b_\Delta D_\Delta < 0$. A more convenient expression for $\mathcal{P}(D)$ can be derived in the case of axisymmetric encoding tensors, where $D^\text{eff}_1=D^\text{eff}_2=D_\text{iso}(1-b_\Delta D_\Delta)$ and $D^\text{eff}_3 = D_\text{iso}(1+2b_\Delta D_\Delta)$. Indeed, using the special value $K(0)=\pi/2$, one has 
\begin{equation}
\mathcal{P}(D) = \frac{1}{2\sqrt{3D_\text{iso}D_\Delta b_\Delta[D-\mathrm{D}_\text{iso}(1-b_\Delta D_\Delta)]}}
\end{equation}
for $\mathrm{min}[D_\text{iso}(1-b_\Delta D_\Delta),D_\text{iso}(1+2 b_\Delta D_\Delta)]<D<\mathrm{max}[D_\text{iso}(1-b_\Delta D_\Delta),D_\text{iso}(1+2 b_\Delta D_\Delta)]$, and $\mathcal{P}(D)=0$ otherwise. Injecting this distribution in Eq.~\eqref{Eq_Laplace_1D} yields \cite{Eriksson:2015}
\begin{equation}
\cbeq{
\frac{\mathcal{S}}{\mathcal{S}_0} = 
 \frac{\sqrt{\pi}}{2} \, \frac{\mathrm{erf} \!\left(\sqrt{3\,bb_\Delta D_\text{iso}D_\Delta} \right)}{\sqrt{3\,bb_\Delta D_\text{iso}D_\Delta}}\; \mathrm{exp}\left(-b\, D_\text{iso}(1-b_\Delta D_\Delta)\right)
 }\; ,
 \label{Eq_kernel_powder_averaged}
\end{equation}
with the error function
\begin{equation}
\mathrm{erf} \; : \; x \longmapsto \frac{2}{\sqrt{\pi}} \int_0^x\! \exp^{-t^2}\, \mathrm{d}t \, .
\end{equation}
Ref.~\cite{Herberthson:2019} provides exact expressions for the signal acquired \textit{via} general gradient waveforms from a powder of generally shaped diffusion tensors.

\subsubsection{Powder-averaged signal}

The above expressions have been obtained assuming an isotropic distribution of diffusion tensors with identical size and shape. In the case where an ensemble of diffusion tensors exhibits favored orientations, any effect of anisotropy can be removed by acquiring the signal for many orientations of the $b$-tensor and considering instead the orientationally averaged signal
\begin{equation}
\cbeq{
\frac{\overline{\mathcal{S}}}{\mathcal{S}_0}(b,b_\Delta, b_\eta) = \frac{1}{4\pi} \iint_{\mathrm{S}(1)} \frac{\mathcal{S}}{\mathcal{S}_0}(b,b_\Delta, b_\eta, \Theta, \Phi) \, \sin\Theta\, \mathrm{d}\Theta\, \mathrm{d}\Phi 
}\; ,
\label{Eq_powder-average}
\end{equation}
where $\mathrm{S}(1)$ denotes the unit sphere and $(\Theta, \Phi)$ gives the orientation of each acquired $b$-tensor. However, enough orientations must be acquired in order for the powder-averaged signal to be truly rotationally invariant \cite{Szczepankiewicz_ISMRM:2016}.\\

In analogy with corresponding approaches for calculating powder lineshapes in solid-state NMR spectroscopy \cite{Eden:2003}, such acquisition schemes have been dubbed ``powder-averaged" \cite{Lasic:2014}. Assuming that the studied volume consists of diffusion tensors of identical size and shape, the resulting powder-averaged signal Eq.~\eqref{Eq_powder-average} expresses like the aforementioned signal Eq.~\eqref{Eq_kernel_powder_averaged} arising from powder samples. Therefore, Ref.~\cite{Herberthson:2019} provides exact expressions for powder- averaged signals in various acquisition and diffusion contexts.

\subsection{Cumulant expansion of the powder-averaged signal}

\subsubsection{Collection of identically shaped diffusion tensors}

Originally developed in the field of dynamical light scattering in the 1970s \cite{Frisken:2001,Koppel:1972}, one of the most straightforward way to extract information from a powder-averaged signal is to write its cumulant expansion. Let us first consider the powder-averaged signal $\overline{\mathcal{S}}$ of Eq.~\eqref{Eq_powder-average} and assume a volume consisting of diffusion tensors of identical size and shape, so that the expression Eq.~\eqref{Eq_Laplace_1D} of the diffusion signal as a 1D Laplace transform holds. If $\mathcal{P}(D)$ is normalized to unity, one can write
\begin{align}
\frac{\overline{\mathcal{S}}}{\mathcal{S}_0} & = \mathrm{exp}(-b\langle D \rangle) \int_0^{+\infty}\! \mathrm{exp}(-b(D-\langle D \rangle))\, \mathcal{P}(D)\,\mathrm{d}D \nonumber \\
 & = \mathrm{exp}(-b\langle D \rangle) \int_0^{+\infty}\! \left[ \sum_{m=0}^{+\infty} \frac{[-b(D-\langle D \rangle)]^m}{m!} \right] \mathcal{P}(D)\,\mathrm{d}D \nonumber \\
 & = \mathrm{exp}(-b\langle D \rangle)\sum_{m=0}^{+\infty} \frac{\mu_m}{m!}\, (-b)^m \nonumber \\
 & = \mathrm{exp}(-b\langle D \rangle) \left[ 1 + \sum_{m=2}^{+\infty} \frac{\mu_m}{m!}\, (-b)^m \right]\, ,
 \label{Eq_interm_cumulant}
\end{align}
where we introduce the mean apparent diffusivity and the central moments
\begin{align}
\langle D\rangle & = \int_0^{+\infty} \! D \, \mathcal{P}(D) \,\mathrm{d}D \\
\mu_m & = \int_0^{+\infty} \! (D-\langle D \rangle)^m \, \mathcal{P}(D) \,\mathrm{d}D \, ,
\label{Eq_central_moments}
\end{align}
with $\mu_0 = 1$ and $\mu_1 = 0$. Taking the logarithm of Eq.~\eqref{Eq_interm_cumulant} and its low $b$-value limit, one obtains the cumulant expansion
\begin{equation}
\cbeq{
\ln\! \left( \frac{\overline{\mathcal{S}}}{\mathcal{S}_0} \right) \underset{b\to 0}{\simeq} -\langle D \rangle b + \frac{\mu_2}{2}\, b^2 - \cdots
}\; ,
\label{Eq_cumulant_expansion}
\end{equation}
where the low $b$-value limit corresponds to using $\ln(1+X) \simeq X$ when $X\to 0$. While $\langle D \rangle$ describes the initial slope of $\ln(\overline{\mathcal{S}}/\mathcal{S}_0)$ as a function of $b$, $\mu_2$ reports on the initial deviation from monoexponential decay.\\

\begin{figure}[h!]
\begin{center}
\includegraphics[width=\textwidth]{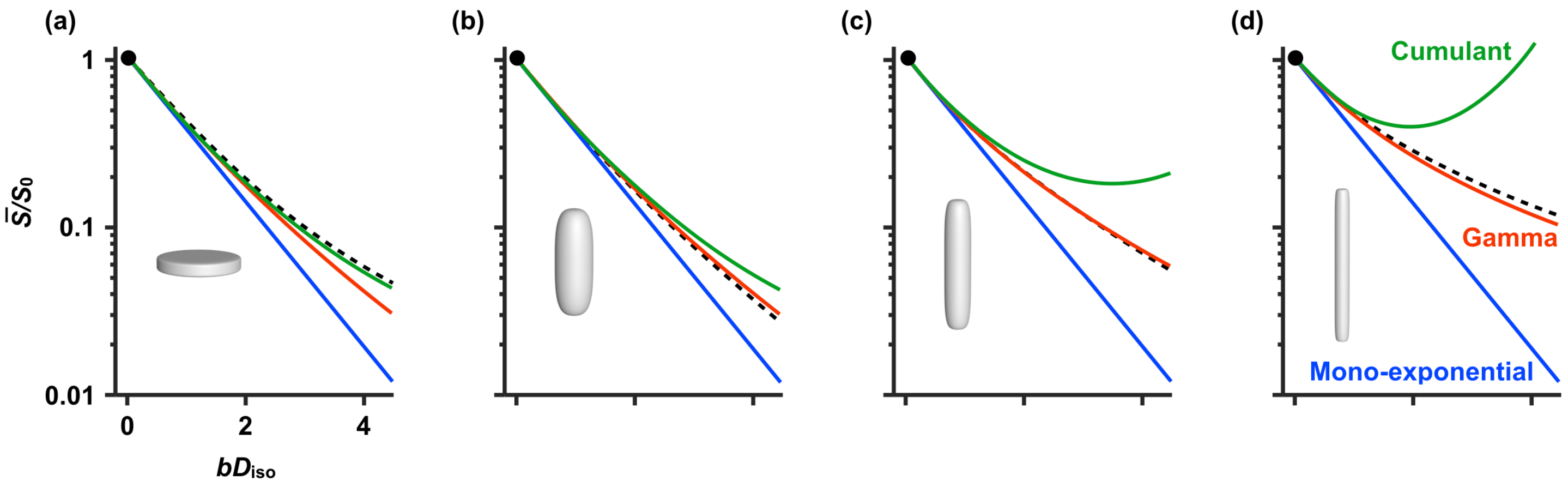}
\caption{Validity of the monoexponential (blue), two-term cumulant (green), and Gamma distribution (red) approximations (see Sec.~\ref{Sec_Gamma_distribution}) to the exact (dashed) powder-averaged signal attenuation given by Eq.~\eqref{Eq_kernel_powder_averaged} for linear diffusion encoding ($b_\Delta = 1$) applied to powders of axisymmetric diffusion tensors with normalized anisotropy $D_\Delta = -0.4$ (a), $0.4$ (b), $0.6$ (c), and $0.8$ (d). The glyphs represent the microscopic diffusion tensors. The cumulant and gamma curves are calculated with Eqs.~\eqref{Eq_cumulant_expansion} and \eqref{Eq_signal_Gamma_mu2}, respectively, using the true value of the second moment $\mu_2$ according to Eq.~\eqref{Eq_second_moment_f_variance}. Adapted from Refs.~\cite{Topgaard_book:2017} and \cite{Reymbaut_book:2019}.}
\label{Figure_Gamma_cumulant}
\end{center}
\end{figure}

While the average apparent diffusivity $\langle D \rangle$ does not contain any information about macroscopic nor microscopic anisotropy, the second moment $\mu_2$ does. Indeed, for the distribution Eq.~\eqref{Eq_bigass_P_ADC}, one shows that \cite{VanderHart_Gutowsky:1968}
\begin{align}
\langle D \rangle & = \frac{D^\text{eff}_1 + D^\text{eff}_2 + D^\text{eff}_3}{3} = D_\text{iso} \\ 
\mu_2 & = \frac{4}{45}\, [(D^\text{eff}_1-D^\text{eff}_3)^2 + (D^\text{eff}_2-D^\text{eff}_1)(D^\text{eff}_2-D^\text{eff}_3)]\, .
\label{Eq_mu2_D1_D2_D3}
\end{align}
Upon expansion, the second central moment is straightforwardly related to the variance of diffusion eigenvalues Eq.~\eqref{Eq_variance_eigenvalues} \textit{via}
\begin{equation}
\cbeq{
\mu_2 = \frac{2f}{5}\, V_\lambda = \frac{4f}{5}\, D_\text{aniso}^2
}\; ,
\label{Eq_second_moment_f_variance}
\end{equation}
where $f \in [0,1]$ is the $b$-tensor-dependent scaling factor
\begin{equation}
\cbeq{
f = b_\Delta^2\, \frac{b_\eta^2 + 3}{3} =
\begin{cases}
0 & \text{for spherical encoding} \\
1/4 & \text{for planar encoding} \\
1 & \text{for linear encoding} 
\end{cases}
}\; .
\label{Eq_scaling_f}
\end{equation}
Although very direct, the cumulant expansion cannot explain the signal at high $b$-value, as shown in Eq.~\eqref{Eq_cumulant_expansion} and Fig.~\ref{Figure_Gamma_cumulant}.

\subsubsection{Collection of diffusion tensors with various sizes and shapes}

\begin{figure}[h!]
\begin{center}
\includegraphics[width=\textwidth]{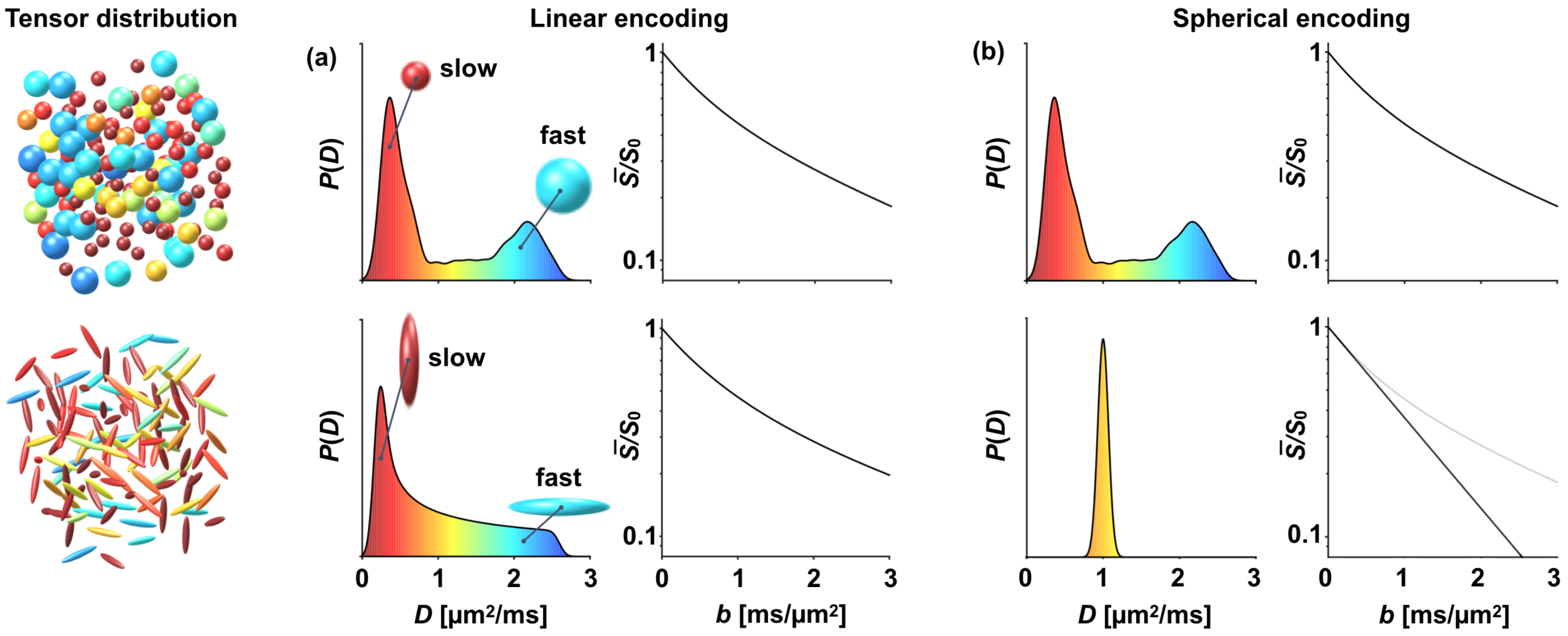}
\caption{Detection of microscopic diffusion anisotropy by comparing linear (a) and spherical (b) diffusion encodings for two drastically different diffusion tensor distributions: a distribution of spherical tensors with various sizes, and a distribution of randomly oriented identically shaped prolate tensors. The apparent diffusion coefficient distribution $\mathcal{P}(D)$ and powder-averaged signal decay $\overline{\mathcal{S}}/\mathcal{S}_0$ is shown for each encoding. The color-coding matches each contribution to $\mathcal{P}(D)$ with its corresponding population of microscopic diffusion tensors. The apparent diffusion coefficient distributions associated to linear encoding were designed with comparable means and variances. Figure obtained from M. Nilsson.}
\label{Figure_Encoding_S_b}
\end{center}
\end{figure}

When the investigated volume contains diffusion tensors of various sizes and shapes, the variance in the ADC distribution $\mathcal{P}(D)$ probed by powder-averaged signals originates from two sources: isotropic heterogeneity and microscopic anisotropy itself. This is illustrated in Fig.~\ref{Figure_Encoding_S_b} that displays the ADC distribution, and linear and spherical signals associated to two archetypal volumes: a distribution of spherical tensors with various sizes (isotropic heterogeneity), and a distribution of randomly oriented identically shaped prolate tensors (microscopic anisotropy without orientational order).\\

While isotropic heterogeneity is encoded equally by the deviation from monoexponentiality of the linear, planar and spherical signals, microscopic anisotropy appears only in the linear and planar signals. This is due to the fact that the cumulant expansion Eq.~\eqref{Eq_cumulant_expansion} of these signals actually encodes for
\begin{align}
\langle D \rangle & = \langle D_\text{iso}\rangle \\
\mu_2^\text{tot} & = V_\text{iso} + \frac{2f}{5}\,\langle V_\lambda \rangle\, ,
\end{align}
where $\langle D_\text{iso}\rangle$ is the mean diffusivity Eq.~\eqref{Eq_MD_DTD}, $V_\text{iso}$ is the variance of isotropic diffusivities Eq.~\eqref{Eq_Viso_DTD}, and $\langle V_\lambda \rangle$ is the mean variance of diffusion tensor eigenvalues Eq.~\eqref{Eq_average_variance_Daniso}, as shown in the context of chemical shift anisotropy for NMR spectroscopy of solids \cite{VanderHart_Gutowsky:1968}. Here, $f$ remains given by Eq.~\eqref{Eq_scaling_f}, which explains why linear encoding ($f = 1$) is sensitive to both isotropic heterogeneity ($V_\text{iso}$) and microscopic anisotropy ($\langle V_\lambda \rangle$), contrary to spherical encoding ($f = 0$) that does not capture microscopic anisotropy but instead reports exclusively on
\begin{equation}
\cbeq{
\mu_2^\text{sph} = \mu_2^\text{tot}(f=0) = V_\text{iso}
}\; .
\end{equation}
The amount of microscopic diffusion anisotropy is encoded in the difference 
\begin{equation}
\cbeq{
\mu_2^\text{lin} - \mu_2^\text{sph} = \mu_2^\text{tot}(f=1) - \mu_2^\text{tot}(f=0) = \frac{2f}{5}\,\langle V_\lambda \rangle = \frac{4f}{5}\,\langle D_\text{aniso}^2 \rangle
}\; .
\label{Eq_diff_second_moments}
\end{equation}
Note that the different moments are sometimes scaled and normalized as the ``mean kurtoses" \cite{Szczepankiewicz:2015, Szczepankiewicz:2016, Szczepankiewicz_DIVIDE:2019} $\mathrm{MK}_\text{iso} = 3\mu_2^\text{sph}/\langle D_\text{iso}\rangle^2$ and $\mathrm{MK}_\text{aniso} = 3(\mu_2^\text{lin}-\mu_2^\text{sph})/\langle D_\text{iso}\rangle^2$ in order to match the naming conventions of the diffusional kurtosis field \cite{Jensen:2005}. A measure of these second moments using a Gamma distribution fitting (see Sec.~\ref{Sec_Gamma_distribution}) in porous media is presented in Ref.~\cite{Lasic:2014}.

\subsection{Microscopic fractional anisotropy and orientational order parameter}

Since the fractional anisotropy FA Eq.~\eqref{Eq_FA_variance_eigenvalues} is the \textit{de facto} standard for reporting voxel-averaged diffusion anisotropy in conventional DTI \cite{Basser:1994}, it would be convenient to recast $\langle D \rangle$, $\mu_2^\text{lin}$ and $\mu_2^\text{sph}$ into a microscopic equivalent of the FA. This can easily be done by defining the microscopic fractional anisotropy ($\mu\mathrm{FA}$) as
\begin{equation}
\mu\mathrm{FA} = \sqrt{\frac{3}{2}} \left( 1 + \frac{\langle D \rangle^2}{\langle V_\lambda \rangle} \right)^{-1/2} = \sqrt{\frac{3}{2}} \left( 1 + \frac{2}{5}\,\frac{\langle D \rangle^2}{\mu_2^\text{lin}-\mu_2^\text{sph}} \right)^{-1/2} \in [0,1] \; .
\label{Eq_microFA}
\end{equation}
However, another version of the $\mu$FA, that takes into account the effect of isotropic variance is found in Ref.~\cite{Szczepankiewicz:2016}:
\begin{equation}
\cbeq{
\mu\mathrm{FA} = \sqrt{\frac{3}{2}}\,\sqrt{\frac{\langle \mathbb{V}_\lambda[\mathbf{D}]\rangle}{\langle\mathbb{E}_\lambda[ \mathbf{D}]^2 \rangle+\langle \mathbb{V}_\lambda[\mathbf{D}]\rangle}} = \sqrt{\frac{3}{2}} \left( 1 + \frac{2}{5}\,\frac{\langle D \rangle^2 + \mu_2^\text{sph}}{\mu_2^\text{lin}-\mu_2^\text{sph}} \right)^{-1/2} \in [0,1]
}\; .
\label{Eq_microFA_with_Viso}
\end{equation}
Note that, compared to Eq.~\eqref{Eq_FA_mathbb}, the order of averaging across eigenvalues is no longer arbitrary since the expected value is squared, \textit{i.e.} $\langle\mathbb{E}_\lambda[ \mathbf{D}]^2 \rangle \neq \mathbb{E}_\lambda[ \langle\mathbf{D}\rangle]^2$ if $V_\text{iso}\neq 0$.\\

The values of FA and $\mu\mathrm{FA}$ are equal in the ideal case of a single diffusion tensor component with no orientation dispersion. For instance, diffusion between parallel planes is characterized by $\mu\mathrm{FA}=\mathrm{FA}=\sqrt{1/2}$ and diffusion within parallel narrow tubes presents $\mu\mathrm{FA}=\mathrm{FA}=1$. It is crucial to realize that only the combined acquisition of linear and spherical signals (or at least two distinct diffusion encoding shapes) enables the estimation of the $\mu\mathrm{FA}$. Indeed, $\mu\mathrm{FA}$ misestimations have recently been reported in the use of spherical mean technique for powder-averaged linearly encoded signals \cite{Henriques:2019}.\\

\begin{mainpoint}{title = Distinguishing between prolate and oblate diffusion tensors}
Even though the $\mu$FA can detect and quantify microscopic anisotropy, it cannot in itself tease apart oblate ($D_\Delta < 0$) and prolate ($D_\Delta > 0$) diffusion tensors. Indeed, according to Eqs.~\eqref{Eq_variance_eigenvalues} and \eqref{Eq_second_moment_f_variance}, the values of $V_\lambda$ and $\mu_2$ are proportional to $D_\Delta^2$. However, if it can be assumed that only oblate or prolate microscopic diffusion tensors contribute to the signal, then the sign of $D_\Delta$ can be robustly determined by fitting Eq.~\eqref{Eq_kernel_powder_averaged} to acquired data \cite{Eriksson:2015}. After estimating the microscopic parameters $D_\text{iso}$ and $D_\Delta$, as well as the voxel-averaged diffusion tensor $\langle\mathbf{D}\rangle$ by conventional DTI analysis, the Saupe order tensor $\mathbf{S}$ can be calculated through inversion of Eq.~\eqref{Eq_link_averaged_D_Saupe}.
\end{mainpoint}
\medskip

Let us now define a measure of orientational order by considering a volume containing axisymmetric diffusion tensors of varying orientations, characterized by the axial and radial diffusivities $D_\parallel$ and $D_\perp$ defined in Eq.~\eqref{Eq_lambda_para_perp}. Assuming that the distribution of diffusion orientations is also axisymmetric around a certain ensemble symmetry axis, each diffusion tensor contributes to the effective diffusivity along the symmetry axis as
\begin{equation}
D(\theta) = D_\parallel\cos^2\theta + D_\perp\sin^2\theta = \frac{D_\parallel + 2D_\perp}{3} + \frac{2}{3}\, (D_\parallel-D_\perp)\, P_2(\cos\theta)\, ,
\end{equation}
where $\theta$ is the angle between the diffusion tensor's principal axis and the ensemble symmetry axis. Denoting $\langle D\rangle = (D_\parallel + 2D_\perp)/3$, the the macroscopic axial and radial diffusivity are given by the ensemble averages
\begin{align}
\langle D_\parallel \rangle & = \langle D \rangle + \frac{2}{3}(D_\parallel-D_\perp) \langle P_2(\cos\theta)\rangle \\
\langle D_\perp \rangle & = \langle D \rangle +\frac{2}{3}(D_\parallel-D_\perp) \left\langle P_2\left(\cos\left(\theta+\frac{\pi}{2}\right)\right)\right\rangle = \langle D \rangle - \frac{2}{3}(D_\parallel-D_\perp) \langle P_2(\cos\theta)\rangle\, ,
\end{align}
so that one can define the orientational order parameter (OP) \cite{Saupe:1963,Hong:1991} as
\begin{equation}
\mathrm{OP} = \langle P_2(\cos\theta)\rangle = \frac{\langle D_\parallel \rangle - \langle D_\perp\rangle}{D_\parallel-D_\perp} = \frac{\langle D_\Delta\rangle}{D_\Delta} = S_\text{ZZ}
\label{Eq_OP}
\end{equation}
using Eqs.~\eqref{Eq_lambda_para_perp}, \eqref{Eq_link_averaged_D_Saupe} and \eqref{Eq_average_Delta_Szz}. However, this expression is not easy to measure from the signal. While Eq.~\eqref{Eq_mu2_D1_D2_D3} gives the second moment yielded by a single diffusion tensor,
\begin{equation}
\mu_2 = \frac{4}{45}\, (D_\parallel - D_\perp)^2\, ,
\label{Eq_mu2_direct}
\end{equation}
the same formula can be applied to the ensemble averaged diffusion tensor $\langle \mathbf{D}\rangle$, of axial and radial diffusivities $\langle D_\parallel
\rangle$ and $\langle D_\perp \rangle$, to obtain the second moment yielded by
\begin{equation}
\mu_2^\text{FA} =\frac{4}{45}\, (\langle D_\parallel\rangle - \langle D_\perp\rangle)^2 \, ,
\label{Eq_mu2_FA}
\end{equation}
where the superscript ``FA" indicates that this second moment directly relates to the macroscopic FA Eq.~\eqref{Eq_FA_variance_eigenvalues} for $\langle \mathbf{D}\rangle$ \textit{via}
\begin{equation}
\mathrm{FA} = \sqrt{\frac{3}{2}} \left( 1 + \frac{2}{5}\, \frac{\langle D \rangle^2}{\mu_2^\text{FA}} \right)^{-1/2}\, .
\end{equation}
Eqs.~\eqref{Eq_mu2_direct} and \eqref{Eq_mu2_FA} can now be combines to match the order parameter Eq.~\eqref{Eq_OP}:
\begin{equation}
\mathrm{OP} = \sqrt{\frac{\mu_2^\text{FA}}{\mu_2}}\, .
\end{equation}
However, this approach does not take into account the effect of isotropic heterogeneity. As shown previously for the $\mu$FA Eq.~\eqref{Eq_microFA}, this can be achieved by simply replacing $\mu_2$ by $\mu_2^\text{lin}-\mu_2^\text{sph}$ Eq.~\eqref{Eq_diff_second_moments} in the previous result, suggesting the definition
\begin{equation}
\cbeq{
\mathrm{OP} = \sqrt{\frac{\mu_2^\text{FA}}{\mu_2^\text{lin}-\mu_2^\text{sph}}} \in [0,1]
}\; .
\end{equation}

Other measures of microscopic anisotropy have been designed. For instance, a key alternative to $\mu$FA, the microscopic anisotropy (MA) index, lies in the work of Refs.~\cite{Lawrenz:2010, Lawrenz:2011,Lawrenz:2013,Lawrenz:2015,Lawrenz:2016,Lawrenz:2019} This index is defined through a fourth-order tensor approach linked to the Taylor expansion of the spin density distribution’s Fourier transform within microenvironments probed with double diffusion encoding. It is fourth-order in spin-dephasing vector’s norm $q$, hence second-order in $b$-value. Therefore, the MA index is another measure of deviation from monoexponentiality in the signal, such as $\mu$FA. One should also mention the fractional compartment eccentricity (FE) of Ref.~\cite{Jespersen:2013} that also compares variances in diffusion eigenvalues at the macroscopic and microscopic scales.

\subsection{Imposing a functional form for the ADC distribution}

\subsubsection{A limited freedom of choice}

Even a simple Laplace transform such as Eq.~\eqref{Eq_Laplace_1D} is hard to inverse into the probability distribution $\mathcal{P}(D)$. Even though the cumulant approach Eq.~\eqref{Eq_cumulant_expansion} offers a very direct way to obtain the main cumulants of this distribution, it suffers from an inherent overshooting at high $b$-values \cite{Lasic:2014}, as shown in Fig.~\ref{Figure_Gamma_cumulant}. Besides, the convergence of the cumulant expansion is very slow in the case of randomly oriented anisotropic diffusion components, even though the corresponding signal can be expressed in a simple analytic form \cite{Topgaard:2017, Eriksson:2013, deAlmeidaMartins_Topgaard:2016}. \\

Conversely, the problem of analyzing the diffusion-weighted signal can instead be considered from the perspective of finding a suitable approximation to the probability distribution $\mathcal{P}(\mathbf{D})$ or its first two moments, \textit{i.e.} the so-called ``statistical approach". A convenient functional form to approximate $\mathcal{P}(\mathbf{D})$ should have five characteristics:
\begin{itemize}
\item[$\bullet$] presenting no negative diffusivities;
\item[$\bullet$] presenting no negative probabilities;
\item[$\bullet$] possessing a simple analytic Laplace transform;
\item[$\bullet$] being described by few shape parameters;
\item[$\bullet$] being able to capture a wide range of diffusion distributions.
\end{itemize}

\begin{sidenote}{title = The statistical approach is agnostic and fast}
A very nice way of looking at the statistical approach comes from F. Szczepankiewicz. This kind of approach is ``agnostic", in the sense that no assumptions nor leap of faith are required, and ``fast", in the sense that it extrapolates in the right direction, without recovering the full DTD, which is non needed if one can still find the main metrics! However, state-of-the-art inversion methods now often rely on mixing strong statistical aspects with little, yet hopefully solid, biophysical insight. Although biophysical approaches will not be fully described in this document, a nice review can be found in Ref.~\cite{NilssonThesis:2011}.
\end{sidenote}
\bigskip

\begin{figure}[h!]
\begin{center}
\includegraphics[width=\textwidth]{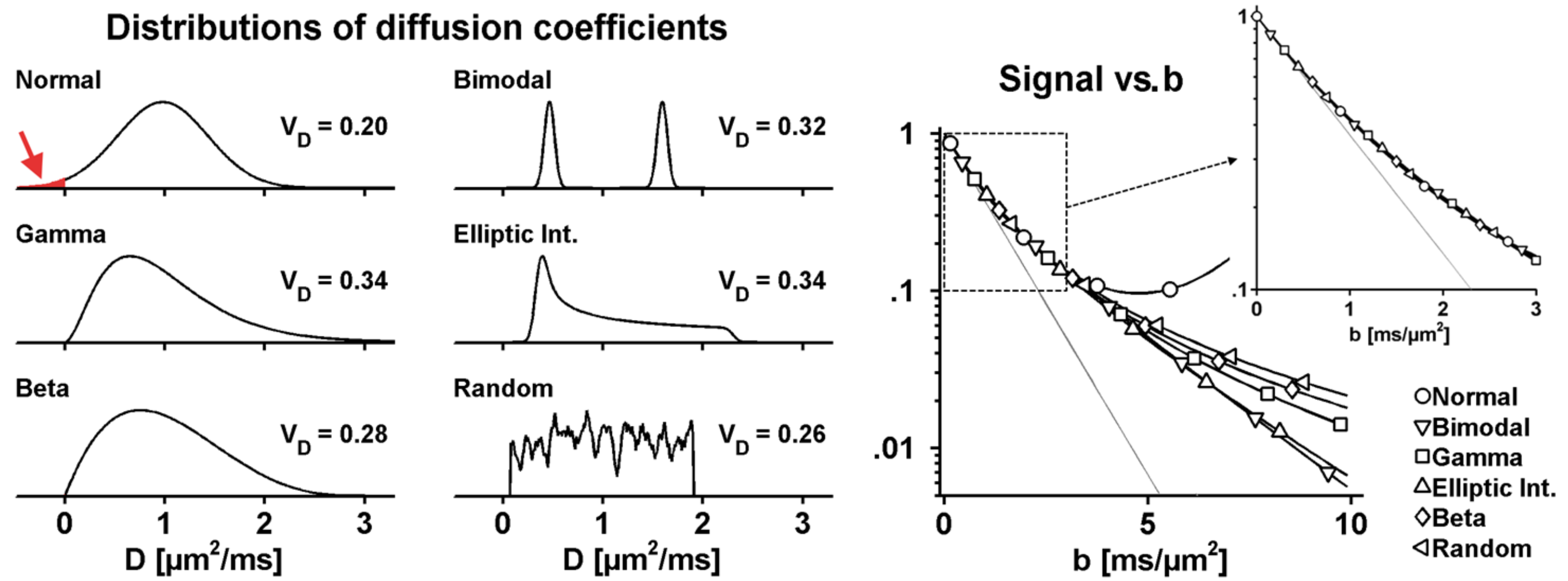}
\caption{Various diffusivity distributions $\mathcal{P}(D)$ that yield similar diffusion-weighted signals. The signals are computed from Eq.~\eqref{Eq_signal_tensor_distribution} with a powder-averaged signal Eq.~\eqref{Eq_powder-average}, aiming at similar $\mathrm{MD}$ and variance. All diffusivity distributions render similar signal curves for moderate diffusion encoding strengths ($b < 3 \; \mathrm{ms}/\mu\mathrm{m}^2 \equiv 3000\; \mathrm{s}/\mathrm{mm}^2$). At strong diffusion encoding strength (up to $b = 10\; \mathrm{ms}/\mu\mathrm{m}^2$), the signal curves diverge, especially the normal distribution since it contains negative diffusivity values (red arrow). The maximal signal difference for the remaining distributions is approximately $2\%$ at $b = 10\; \mathrm{ms}/\mu\mathrm{m}^2$. Figure drawn from Ref.~\cite{SzczepankiewiczThesis:2016}.}
\label{Figure_Models_Filip_thesis}
\end{center}
\end{figure}

\begin{sidenote}{title = A limited freedom of choice for $\mathcal{P}(D)$}
From Fig.~\ref{Figure_Models_Filip_thesis}, one sees that choosing very different functional forms to approximate a diffusivity distribution $\mathcal{P}(D)$ has no real consequences on the signal versus $b$-value data within a reasonable range of $b$-values. However, the choices present significant discrepancy outside of this range. Therefore, statistical approaches should remain within a clinical range of $b$-values to be trustworthy.
\end{sidenote}
\bigskip

\subsubsection{Gamma distribution fitting}
\label{Sec_Gamma_distribution}

 One of the earliest functional form proposed was the log-normal distribution in 1985 \cite{Callaghan_Pinder:1985}. This distribution has probability density
\begin{equation}
\mathcal{P}_\text{log-norm.}(D, \mu, \sigma) = \frac{1}{\sqrt{(2\pi\sigma^2)}}\, \exp^{- (\ln D - \mu)^2/(2\sigma^2)}\, ,
\end{equation}
where $\mu$ and $\sigma$ are two shape parameters (the logarithm ensures that $D>0$). However, even though it seemed promising, this probability density has no analytically tractable Laplace transform, which makes the computation of the signal Eq.~\eqref{Eq_Laplace_1D} quite challenging.\\

To solve this problem, the Gamma distribution function has been introduced in the 2010s \cite{Jensen_Helpern:2010, Roding:2012, Roding:2015, Williamson:2016}:
\begin{equation}
\cbeq{
\mathcal{P}_\Gamma (D, \kappa, \psi) =  \frac{D^{\kappa - 1}}{\psi^\kappa\, \Gamma(\kappa)}\, \exp^{-D/\psi}
} \qquad \text{with} \qquad (\kappa, \psi) \in (\mathbb{R}_+^*)^2 \; ,
\label{Eq_Gamma_distribution_simple}
\end{equation}
where $\kappa$ is the ``shape" parameter (unitless), $\psi$ is the ``scale" parameter (diffusivity units), and 
\begin{equation}
\Gamma \; : \; z \longmapsto \int_0^{+\infty}\! t^{z-1}\exp^{-t}\, \mathrm{d}t
\end{equation}
is the so-called Gamma function. Not only does this distribution satisfy all of the aforementioned requirements for a convenient functional form to approximate $\mathcal{P}(D)$, it also possesses an analytically tractable Laplace transform, Indeed, injecting this probability distribution in Eq.~\eqref{Eq_Laplace_1D} yields
\begin{align}
\frac{\overline{\mathcal{S}}}{\mathcal{S}_0} & = \int_0^{+\infty} \! \mathcal{P}_\Gamma (D, \kappa, \psi)\, \exp^{-bD} \, \mathrm{d}D \nonumber \\
 & = \frac{1}{\psi^\kappa\, \Gamma(\kappa)} \int_0^{+\infty} \! D^{\kappa - 1} \, \exp^{-(1 + b\psi) D/\psi}\, \mathrm{d}D \nonumber \\
 & = \frac{1}{\psi^\kappa\, \Gamma(\kappa)} \left( \frac{\psi}{1+ b\psi} \right)^{\kappa} \underbrace{\int_0^{+\infty} \! t^{\kappa - 1} \, \exp^{-t}\, \mathrm{d}t}_{\Gamma(\kappa)} \nonumber \\ 
 & = \left( 1+ b\psi \right)^{-\kappa}\, .
 \label{Eq_gamma_Laplace_simple}
\end{align}
Using the fascinating properties of the Gamma function, one can show that
\begin{align}
\langle D \rangle & = \int_0^{+\infty} \! D\, \mathcal{P}_\Gamma (D, \kappa, \psi) \, \mathrm{d}D = \kappa\psi \, , \label{Eq_gamma_expectation_simple} \\
\mu_2 & = \int_0^{+\infty} \! (D-\langle D \rangle)^2\, \mathcal{P}_\Gamma (D, \kappa, \psi) \, \mathrm{d}D = \kappa\psi^2\, ,
\end{align}
which in turn gives
\begin{align}
\kappa = \frac{\langle D \rangle^2}{\mu_2}
\qquad \text{and} \qquad
\psi = \frac{\mu_2}{\langle D \rangle}\, .
\end{align}
Replacing the shape and scale parameters in Eq.~\eqref{Eq_gamma_Laplace_simple}, one finally obtains 
\begin{equation}
\cbeq{
\frac{\overline{\mathcal{S}}}{\mathcal{S}_0} = \left(1+ b\, \frac{\mu_2}{\langle D \rangle} \right)^{-\langle D \rangle^2/\mu_2}
}\; .
\label{Eq_signal_Gamma_mu2}
\end{equation}
As shown in Fig.~\ref{Figure_Gamma_cumulant}, this new approach mitigates the overshooting of the cumulant approach.\\

Introduced in Refs.~\cite{Lasic:2014, Szczepankiewicz:2015, Szczepankiewicz:2016}, the diffusional variance decomposition (DIVIDE) simply uses the functional form Eq.~\eqref{Eq_signal_Gamma_mu2} with the total variance Eq.~\eqref{Eq_second_moment_f_variance}, yielding for axisymmetric diffusion encoding
\begin{equation}
\cbeq{
\frac{\overline{\mathcal{S}}}{\mathcal{S}_0} = \left(1+ b\, \frac{V_\mathrm{I} + b_\Delta^2 V_\mathrm{A}}{\mathrm{MD}} \right)^{-\mathrm{MD}^2/(V_\mathrm{I} + b_\Delta^2 V_\mathrm{A})} 
}\; ,
\label{Eq_Signal_DIVIDE}
\end{equation}
where $\mathrm{MD} = \langle D \rangle$, $V_{\mathrm{I}} = \mu_2^\text{sph} = V_\text{iso}$ and $V_{\mathrm{A}} = \mu_2^\text{lin} - \mu_2^\text{sph} = (4/5)\langle D_\text{aniso}^2\rangle$ (see Eq.~\eqref{Eq_diff_second_moments}).\\

\begin{figure}[h!]
\begin{center}
\includegraphics[width=\textwidth]{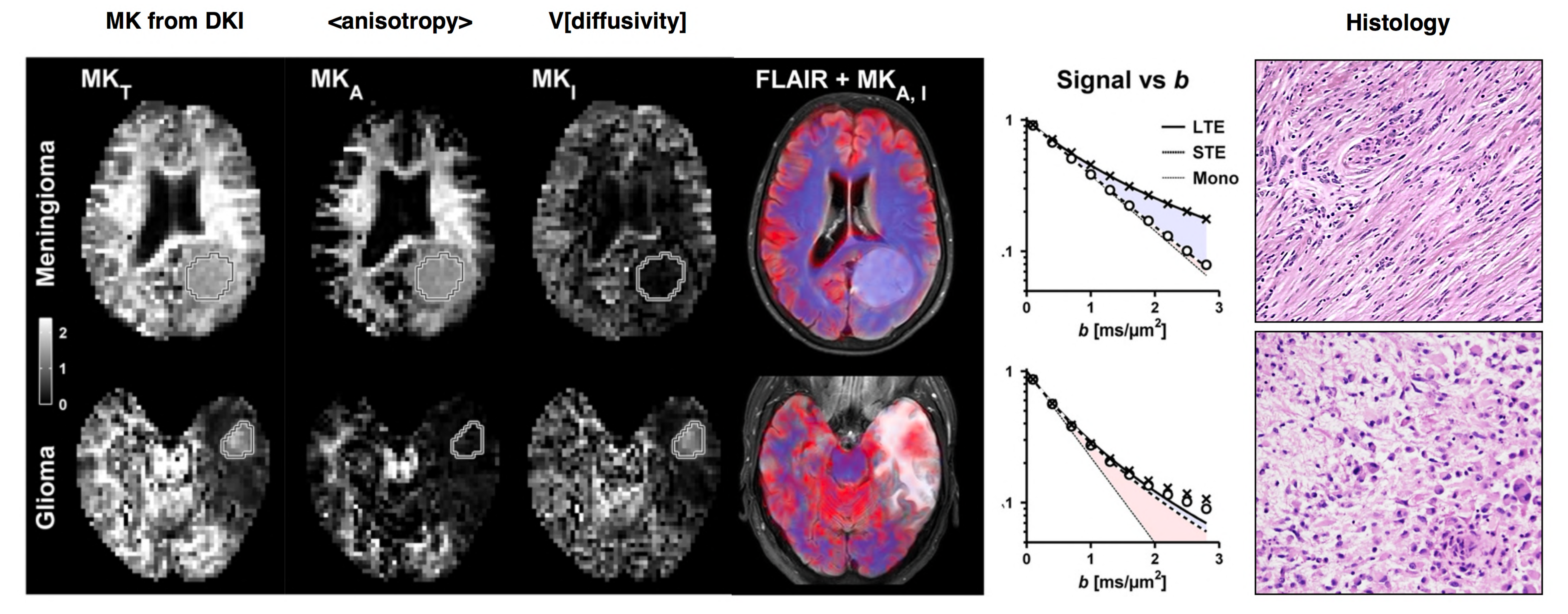}
\caption{\textbf{Left:} Total mean kurtosis $\mathrm{MK}_\text{T}$ (similar to DKI), anisotropic mean kurtosis $\mathrm{MK}_\text{A}$ (normalized anisotropy), isotropic mean kurtosis $\mathrm{MK}_\text{I}$ (normalized isotropic variance) and fluid-attenuated inversion recovery (FLAIR) image (black and white) where $\mathrm{MK}_\text{A}$ and $\mathrm{MK}_\text{I}$ are superimposed in blue and red, respectively. \textbf{Middle:} powder-averaged signals versus $b$-value measured in both tissues. \textbf{Right:} Histology of both tissues. Figure slightly modified from Ref.~\cite{Szczepankiewicz:2016}.}
\label{Figure_DIVIDE}
\end{center}
\end{figure}

In Ref.~\cite{Szczepankiewicz:2015}, the authors target a very specific problem: distinguishing between two kinds of tumors, namely meningiomas and glioblastomas. Indeed, both tumor types exhibit a low FA, which renders DTI unable to distinguish between them. In contrast, $\mu\mathrm{FA}$ is high in meningiomas and low in glioblastomas, indicating that meningiomas contain disordered anisotropic structures, while glioblastomas do not. This interpretation was confirmed on the one hand by histological examination, and on the other hand through the mean kurtoses $\mathrm{MK}_\text{A}=3V_\text{A}/\langle D\rangle^2$ (meningiomas) and $\mathrm{MK}_\text{I}3V_\text{I}/\langle D\rangle^2$ (glioblastomas) \cite{Szczepankiewicz:2016}. This work is illustrated in Fig.~\ref{Figure_DIVIDE}. Moreover, the clinical feasibility of DIVIDE has been recently established \cite{Szczepankiewicz_DIVIDE:2019}.

\vspace*{\fill}

\subsection{Covariance tensor approximation}

An alternative to the previous statistical approach is to generalize the cumulant expansion Eq.~\ref{Eq_cumulant_expansion} to non powder-averaged signals as \cite{Westin:2016} 
\begin{equation}
\ln\! \left( \frac{\mathcal{S}}{\mathcal{S}_0} \right) \underset{\mathrm{Tr}(\mathbf{b})\to 0}{\simeq} -\mathbf{b}:\langle \mathbf{D} \rangle + \frac{1}{2}\, \mathbf{b}^{\otimes 2}: \mathbb{C} - \cdots \, ,
\end{equation}
with the outer tensor product $\otimes$ such that $\mathbf{b}^{\otimes 2} = \mathbf{b}\otimes \mathbf{b}$ and the 6$\times$6 covariance tensor $\mathbb{C} = \langle \mathbf{D}^{\otimes 2}\rangle - \langle \mathbf{D}\rangle^{\otimes 2}$ in Voigt notation. The covariance tensor approximation consists in only keeping the first two terms of this generalized expansion, which is equivalent to considering a normal distribution of diffusion tensors. This equivalency explains why such approach allows for negative isotropic diffusivities, anisotropy parameters and various variances.\\

Aforementioned metrics can then be expressed in terms of tensor inner and outer products involving the voxel-scale averaged diffusion tensor and the covariance tensor. Indeed, introducing the isotropic tensors $\mathbf{E}_{\mathrm{iso}} =\mathbf{I}_3/3$ and $\mathbb{E}_{\mathrm{iso}} = \mathbf{I}_6/3$, where $\mathbf{I}_n$ is the $\mathit{n}\!$ $\times$ $\! \mathit{n}$ identity matrix, and the Voigt notation in which a 3$\times$3 symmetric tensor $\bm{\Lambda}$ writes as the column vector $(\lambda_{\mathit{xx}}\; \lambda_{\mathit{yy}}\; \lambda_{\mathit{zz}}\; \sqrt{2}\,\lambda_{\mathit{yz}}\; \sqrt{2}\,\lambda_{\mathit{xz}}\; \sqrt{2}\,\lambda_{\mathit{xy}})^\mathrm{T}$, one builds the bulk modulus tensor
\begin{equation}
\mathbb{E}_\mathrm{bulk} = \mathbf{E}_{\mathrm{iso}}^{\otimes 2} = \frac{1}{9}
\begin{pmatrix}
1 & 1 & 1 & 0 & 0 & 0 \\
1 & 1 & 1 & 0 & 0 & 0 \\
1 & 1 & 1 & 0 & 0 & 0 \\
0 & 0 & 0 & 0 & 0 & 0 \\ 
0 & 0 & 0 & 0 & 0 & 0 \\
0 & 0 & 0 & 0 & 0 & 0
\end{pmatrix} 
\end{equation}
and the shear modulus tensor
\begin{equation}
\mathbb{E}_\mathrm{shear} = \mathbb{E}_{\mathrm{iso}}-\mathbb{E}_\mathrm{bulk} = \frac{1}{9}
\begin{pmatrix}
2 & -1 & -1 & 0 & 0 & 0 \\
-1 & 2 & -1 & 0 & 0 & 0 \\
-1 & -1 & 2 & 0 & 0 & 0 \\
0 & 0 & 0 & 3 & 0 & 0 \\
0 & 0 & 0 & 0 & 3 & 0 \\ 
0 & 0 & 0 & 0 & 0 & 3
\end{pmatrix}\, ,
\end{equation}
by analogy with the stress tensor in mechanics, and obtains
\begin{align}
\langle\mathit{D}_{\mathrm{iso}}\rangle & = \langle \mathbf{D}\rangle : \mathbf{E}_{\mathrm{iso}} \, , \label{Eq_Cov_expectation_Diso} \\
V_{\mathrm{iso}}  & = \mathbb{C}:\mathbb{E}_\mathrm{bulk} \, , \label{Eq_Cov_variance_Diso} \\
\langle\mathit{D}_\mathrm{aniso}^2\rangle  & = \frac{(\mathbb{C}+\langle\mathbf{D}\rangle^{\otimes 2}):\mathbb{E}_\mathrm{shear}}{2} = \frac{\langle\mathbf{D}^{\otimes 2}\rangle:\mathbb{E}_\mathrm{shear}}{2} \, . \label{Eq_Cov_expectation_Daniso}
\end{align}
While $\langle \mathbf{D}\rangle$ possesses 6 independent elements, $\mathbb{C}$ has 21 independent elements, which makes a total of 27 elements to estimate within this approximation. Consequently, the covariance tensor approximation requires the acquisition of multiple signals over a wide range of $\mathit{b}$-tensor's sizes, shapes, and orientations in order to be reliable. Besides, this approach cannot access higher-order cumulants by construction, like the previous Gamma distribution fitting.

\section{Wrapping up}

The mere combination of linearly and spherically encoded data enables the definitions of a set of metrics on the voxel scale, such as $\mathrm{MD}$ and $\mathrm{FA}$, and on the microscopic (sub-voxel) scale, such as $V_\text{I}=V_\text{iso}$, $V_\text{A}\propto\langle D^2_\text{aniso} \rangle$ and $\mu\mathrm{FA}$. It is then by considering the entire configuration set of these metrics that one extracts the voxel content. This opens the path towards the idea of robust ``virtual biopsy": assessing the brain's microstructures solely through non-invasive interventions.\\

\begin{figure}[h!]
\begin{center}
\includegraphics[width=\textwidth]{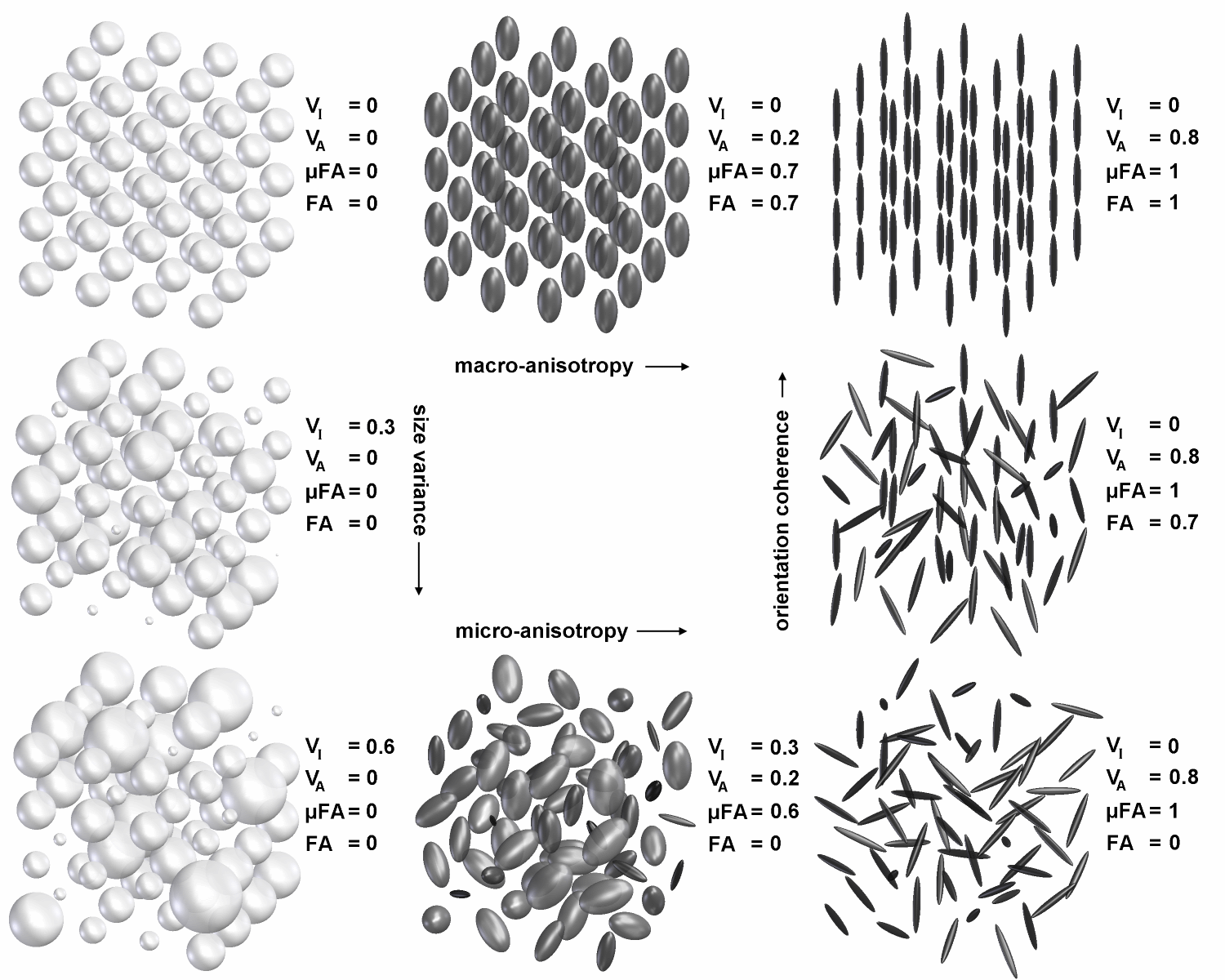}
\caption{Archetypal intra-voxel tensor distributions. The parameters show the isotropic and anisotropic diffusional variances ($V_\text{I}$ and $V_\text{A}$), and the fractional anisotropies on the microscopic and voxel scale ($\mu\mathrm{FA}$ and $\mathrm{FA}$). Figure drawn from Ref.~\cite{SzczepankiewiczThesis:2016}.}
\label{Figure_Tensor_Combinations}
\end{center}
\end{figure}

Fig.~\ref{Figure_Tensor_Combinations} gives a very nice opportunity to test our understanding of the new metrics from Sec.~\ref{Sec_new_metrics}. Let us start from the perfectly homogeneous and isotropic tissue (top left DTD), where all four parameters are zero (no variance in isotropic diffusivities and no anisotropy). One can vary the size of isotropic diffusivities toward the bottom left DTD: the microscopic features of the probed voxel only gain in isotropic variance ($V_\text{I}=V_\text{iso}$) without acquiring any anisotropy whatsoever. One can then increase micro-anisotropy by shaping the different microscopic tensors into randomly oriented thin needles of identical (and very small) radii (bottom right DTD). Now, no isotropic variance is detected, but some anisotropic variance ($V_\text{A}\propto\langle D^2_\text{aniso} \rangle$) emerges from anisotropy. Besides, while the voxel possesses an overall zero $\mathrm{FA}$, its $\mu\mathrm{FA}$ is maximal because all microscopic tensors are fully anisotropic. Finally, one can provide orientational coherence to the probed microstructure by aligning the microscopic needles toward the top right DTD. This coherence only makes the voxel-scale diffusion tensor retain anisotropy, which shows up in the $\mathrm{FA}$. This macro-anisotropy could also have been obtained by simply shaping the top left microscopic isotropic tensors into aligned needles, since $\mu\mathrm{FA} = \mathrm{FA}$ as long as the microscopic tensors remain coherently oriented.\\

\begin{mainpoint}{title = And what about planar encoding?}
Very little work has been done so far regarding the usefulness of planar encoding in microstructural estimation. However, Refs.~\cite{Reisert:2017,Coelho:2019, Coelho_arxiv:2019} indicate that this encoding, associated with linear encoding, can prevent degeneracy and improve precision in parameter estimation. Therefore, much remains to be done, but let us bet that exciting times are coming ahead!
\end{mainpoint}


\end{document}